\documentclass[%
 showpacs,
 reprint,
 groupedaddress,
 showpacs,
 amsmath,amssymb,
 aps,
 prl,
]{revtex4-1}

\usepackage[pdftex]{graphicx, color}% Include figure files
\usepackage{multirow,booktabs}
\usepackage{bm}% bold math
\usepackage{braket}
\usepackage{siunitx}
\usepackage{hyperref}
\usepackage{ulem}

\DeclareMathOperator{\Tr}{Tr}

\DeclareMathOperator{\Log}{Log}
\DeclareMathOperator{\Pf}{Pf}
\newcommand{\BdG}{\mathrm{BdG}}
\newcommand{\Table}[1]{Table~\ref{#1}}
\newcommand{\Eq}[1]{Eq.~\eqref{#1}}
\newcommand{\Eqs}[2]{Eqs.~\eqref{#1}~and~\eqref{#2}}
\newcommand{\Fig}[1]{Fig.~\ref{#1}}
\newcommand{\Ref}[1]{Ref.~\cite{#1}}
\newcommand{\Sec}[1]{Sec.~\ref{#1}}

\graphicspath{{./figs/}}

\begin{document}

\title{$\mathbb{Z}_4$ Topological Superconductivity in UCoGe}
%\thanks{A footnote to the article title}

\author{Akito Daido}
\email[]{daido@scphys.kyoto-u.ac.jp}
\affiliation{%
 Department of Physics, Graduate School of Science, Kyoto University, Kyoto 606-8502, Japan
}%

\author{Tsuneya Yoshida}
%\email[]{}
\affiliation{%
 Department of Physics, Graduate School of Science, Kyoto University, Kyoto 606-8502, Japan
}%

\author{Youichi Yanase}
%\email[]{yanase@scphys.kyoto-u.ac.jp}
\affiliation{%
 Department of Physics, Graduate School of Science, Kyoto University, Kyoto 606-8502, Japan
}%

%\altaffiliation[Also at ]{Physics Department, Kyoto University.}
%\homepage[]{Your web page}

\date{\today}

\begin{abstract}
  Topological nonsymmorphic crystalline superconductivity (TNCS) is an intriguing phase of matter, offering a platform to study the interplay between topology, superconductivity, and nonsymmorphic crystalline symmetries.
  Interestingly, some of TNCS are classified into $\mathbb{Z}_4$ topological phases, which have unique surface states referred to as a \textit{M\"obius strip} or an \textit{hourglass}, and {have not been achieved}
  %can not be achieved
  in symmorphic superconductors.
  However, material realization of $\mathbb{Z}_4$ TNCS has never been known, to the best of our knowledge.
  Here we propose that the paramagnetic superconducting phase of UCoGe under pressure is a promising candidate of $\mathbb{Z}_4$-nontrivial TNCS enriched by glide symmetry.
  We evaluate $\mathbb{Z}_4$ invariants of UCoGe by deriving the formulas relating $\mathbb{Z}_4$ invariants to the topology of Fermi surfaces.
  %We demonstrate $\mathbb{Z}_4$ invariants of UCoGe at Brillouin-zone face are easily evaluated by counting the number of Fermi surfaces.
  %  Applying the formulas to the Fermi surfaces previously obtained by an \textit{ab-initio} calculation,
  Applying the formulas and previous \textit{ab-initio} calculations,
  we clarify that three odd-parity representations, out of four, are $\mathbb{Z}_4$-nontrivial TNCS, while the other is also $\mathbb{Z}_2$-nontrivial TNCS. We also discuss possible $\mathbb{Z}_4$ TNCS in CrAs and related materials.
  %Our findings make a first step to establish $\mathbb{Z}_4$ TNCS as a material phase in nature.
\end{abstract}


% Use showkeys class option if keyword display desired
%\keywords{locally non-centrosymmetric superconductors, staggered Rashba spin-orbit coupling}

\maketitle

%\tableofcontents

%%%%%%%%%% Introduction %%%%%%%%%%
Realization of topological superconductivity (TSC) and Majorana Fermions has been one of the central issues in modern condensed matter physics~\cite{Qi2011,Sato2016review,SatoAndo2017}. Stimulated by the proposal to use Majorana Fermions as qbits of quantum computation~\cite{Kitaev2001}, about two decades of intensive study has revealed the TSC in superconducting heterostructures~\cite{Sau2010,Lutchyn2010,Oreg2010,Alicea2010,Mourik2012,Das2012,Deng2012,Nadj-Perge2014,Zhang2018,Fu2008,Wang2012_TI,Xu2014,Xu2015,Sun2016,Qi2010,Chung2011,Wang2015,He2017,Menard2017} and superconducting topological materials~\cite{Hosur2011,Wang2015_FeSeTe,Xu2016,Zhang2017,Wang2018,Fu2010,Fu2014,Pan2016,Matano2016,Yonezawa2017,Hosur2014,Kobayashi2015}.
%% : nanowire/superconductor\cite{Sato2009_STF,Sato2010_STF,Sau2010,Lutchyn2010,Alicea2010,Mourik2012,Nadj-Perge2014}, topological insulator/SC\cite{Fu2008,}, and quantum anomallous Hall insulator/SC\cite{}.
Even apart from application to quantum computation, TSC in itself is an intriguing topological phase of matter, namely, a new type of unconventional superconductivity.
%% From that respect, several researches proposed realization of Abelian TSC in materials such as $\mathrm{Sr}_2\mathrm{RuO}_4$\cite{Maeno2012,Kallin2016} and so on.\cite{Yoshida2016,Daido2016,honeycomb materials,DIIInatphys}
From that respect, several researches proposed material realization of
spinful chiral superconductivity~\cite{Maeno2012,Kallin2016,Nandkishore2012,Yoshida2016,Daido2016} and
time-reversal symmetric (TRS) TSC in bulk materials~\cite{Fu2010,Fu2014,Matano2016,Yonezawa2017,Pan2016,Hosur2014,Kobayashi2015}.
%including $\mathrm{Sr}_2\mathrm{RuO}_4$.
%It has also been revealed that nodal superconductors (SCs) may have topologically-protected surface states and stability of excitation nodes is ensured by nontrivial topology~\cite{Yada2011,Schnyder2011,Sato2011,Meng-Balents2012,Agterberg2017}.
Now we can say that TSC is recognized as a material phase in nature.

Recently, the concept of TSC is extended in the presence of crystalline symmetries.
Topologically-nontrivial phases whose topological nature is ensured by crystalline symmetries is called topological crystalline superconductivity (TCSC)~\cite{Zhang-Kane-Mele2013,Chiu-Yao-Ryu2013,Morimoto2013,Shiozaki-Sato2014,Chiu-Schnyder2014,Yoshida2015,Shapourian2018,Ueno-Sato2013,Tsutsumi2013,Yanase2016,Yanase2017}, whose material proposal includes $\mathrm{Sr}_2\mathrm{RuO}_4$~\cite{Ueno-Sato2013} and $\mathrm{U}\mathrm{Pt}_3$~\cite{Tsutsumi2013,Yanase2016,Yanase2017}.
TCSC hosts gapless boundary states on a surface preserving the crystalline symmetry, and
such topological crystalline phases may be stable against disorders preserving the relevant symmetries on average~\cite{SatoAndo2017}, as evidenced for weak topological insulators and topological crystalline insulators~\cite{Mong2012,Ringel2012,Fu2012,Fulga2014,Fang2015}.
%statistical recovery of the relevant symmetries may protect them even in the presence of disorders~\cite{Mong2012,Ringel2012,Fu2012,Fulga2014,Fang2015}.
%some of them were show to be stable against impurities preserving the relevant symmetries on average\cite{Mong2012,Ringel2012,Fu2012,Fulga2014,Fang2015}@@.
%% On the contrary to the case of TSC, material proposal of TCSC is seriously lacking. It includes only a few kinds of materials, say, $\mathrm{Sr}_2\mathrm{RuO}_4$\cite{Ueno-Sato2013}, $\mathrm{U}\mathrm{Pt}_3$\cite{Tsutsumi2013,Yanse2016,Yansae2017}, and heavy fermion superlattices\cite{Yoshida2015}.
%% Material proposal of TCSC is important for our understanding of emergence of non-trival topology in superconductors, and therefore, is an urgent issue.
Among various TCSC phases, TCSC in nonsymmorphic crystals
%enriched by nonsymmorphic symmetries
forms a special class, dubbed topological nonsymmorphic crystalline superconductivity (TNCS)~\cite{Varjas2015,Sahoo2016,Wang2016,Shiozaki2016,Yanase2017}.
%%~\cite{Mong2010,Liu2014,Fang2015,Shiozaki2016}@@.%%including TNCIs
In particular, TNCS enriched by glide symmetry possesses surfaces preserving the symmetry, and therefore, has symmetry-protected surface states. {Such topological surface states have been studied in the context of topological insulators, where $4\pi$ periodicity of glide eigenvalues leads to a characteristic structure likened to a \textit{M\"obius strip}~\cite{Shiozaki2015,PYChang2017} or an \textit{hourglass}~\cite{Wang2016_HF}.}
Interestingly, the double-valuedness of eigenvalues may give rise to $\mathbb{Z}_4$ topological phases,
which do not appear in conventional topological periodic tables for symmorphic free fermion systems~\cite{Schnyder2008,Schnyder2009,Kitaev2009,Shiozaki-Sato2014}. Such $\mathbb{Z}_4$ topological phases are a precious platform to study the interplay between topology and nonsymmorphic symmetry, and are worthy of further investigation.

According to the $K$-theory classifications~\cite{Shiozaki2016,Yanase2017}, $\mathbb{Z}_4$ TNCS may be realized in TRS nonsymmorphic superconductors (SCs). In particular, odd-parity superconductivity is preferable in order to achieve nontrivial topology~\cite{Qi2011,Sato2016review,SatoAndo2017}.
%, although famous candidate materials of odd-parity SCs have symmorphic crystal structures\cite{}.
From these perspectives, we identify the high-pressure superconducting phase of UCoGe~\cite{Aoki2014,Huy2007,Hassinger2008,Slooten2009,Bastien2016,Cheung2016,Mineev2017}
%,Cheung2016,Mineev2017}
($S_2$ phase in Ref.~\cite{Slooten2009}) as one of the best platforms to study TNCS.
First of all, crystal structure of UCoGe belongs the space group $Pnma$~\cite{Canepa1996}, which has two glide planes as shown later in detail.
Second, most importantly, UCoGe at ambient pressure is a ferromagnetic superconductor~\cite{Huy2007,Aoki2014}, and therefore, odd-parity superconductivity is strongly suggested.
The $S_2$ phase at high pressure is also expected to be odd-parity superconductivity, since it is continuously connected to the ferromagnetic superconducting phase~\cite{Hassinger2008,Slooten2009,Bastien2016,Cheung2016,Mineev2017}. The observed upper critical field extremely exceeding the Pauli limit~\cite{Slooten2009,Bastien2016} also supports the odd-parity superconductivity.
The time-reversal symmetry is recovered in the $S_2$ phase as evidenced by the vanishing ferromagnetic moment~\cite{Hassinger2008,Slooten2009,Bastien2016,Manago_private}.
%% In particular, $S_2$ has vanishing ferromagnetic moment~\cite{Slooten2009}, and recent NQR experiment consistently reports the absence of internal magnetic field~\cite{Manago_private}@@.
%\footnote{M.~Manago and K.~Ishida, {private communication}.}@@.
%Several thoeretical works also support TRS superconductivity~\cite{Cheung2016,Mineev2017}.@@
%% The $S_2$ phase at high pressure also exhibits upper critical field extremely exceeding the Pauli limit\cite{}, and so is expected to be odd-parity SC.
%% In particular, the $S_2$ phase has vanishing ferromagnetic moment@@, and recent NQR experiment also reports the absence of internal magnetic field\cite{}[private commm.].
Thus, the $S_2$ phase of UCoGe is a promising candidate of nonsymmorphic TRS odd-parity superconductivity, which is hardly known at present.

In this Letter, we propose that UCoGe under pressure may be a $\mathbb{Z}_4$ nontrivial TNCS.
In the context of topological insulators, such $\mathbb{Z}_4$ nontrivial phases have already been proposed~\cite{Wang2016_HF,PYChang2017}, and experimental implication has recently been reported~\cite{Ma2017,Liang2017}.
However, its counterpart in superconductors has remained unknown, although $\mathbb{Z}_2$ nontrivial glide-even TNCS has been proposed in the $A$-phase of $\mathrm{UPt}_3$~\cite{Yanase2017}.
This work is the first material proposal of $\mathbb{Z}_4$ TNCS, to the best of our knowledge.
The present paper is constructed as follows. First, we derive a formula relating the structure of Fermi surfaces (FSs) with $\mathbb{Z}_4$ topological invariants of glide-odd superconductivity, assuming the symmetry of $Pnma$.
We also show similar formulas for $\mathbb{Z}_2$ invariants of glide-even superconductivity.
{The obtained formulas directly predict TNCS for $4$-times-odd-integer sheets of Fermi surfaces, whose topological properties can not be revealed with the similar formulas~\cite{Fu2010,Sato2010_oddparity} predicting TSC in systems with odd-integer sheets of Fermi surfaces.}
Second, we apply the formulas to the FSs obtained by \textit{ab-initio} calculations~\cite{Fujimori2015,Fujimori2016,Czekala2010}, and demonstrate that UCoGe under pressure has nontrivial topological invariants.
Finally, our predictions are confirmed by tight-binding model calculations of surface states.
We also discuss stability of our results against deformation of FSs or against the possible nodal excitations, and show future direction to identify UCoGe as a $\mathbb{Z}_4$ TNCS.

%%%%%%%%%% Formula for Z4 invariant %%%%%%%%%%
\textit{Topological invariants ---}
%% \textcolor{blue}{The crystal space group of UCoGe has been reported as a centrosymmetric nonsymmorphic space group $Pnma$~\cite{Canepa1996}, which includes two glide planes perpendicular to the $x$- and $z$-axis, and three screw rotations along all principal axes.
%% In the following, we derive formulas to give $\mathbb{Z}_4$ and $\mathbb{Z}_2$ invariants of three dimensional (3D) glide-symmetric class DIII systems with additional screw symmetry. Then we demonstrate that the formulas can be significantly simplified by using the symmetry of $Pnma$.}
  %% Glide-odd (-even) class DIII superconductivity is specified by two $\mathbb{Z}_4$ ($\mathbb{Z}_2$) invariants defined on the two glide-invariant planes in the Brillouin zone.
The crystal space group of UCoGe has been reported as a centrosymmetric nonsymmorphic space group $Pnma$~\cite{Canepa1996}, which includes two glide planes and three screw rotations.
%\textcolor{red}
TNCS in UCoGe is specified by the $\mathbb{Z}_4$ and $\mathbb{Z}_2$ invariants of class DIII glide-symmetric systems, which are defined on the two glide-invariant planes in the Brillouin zone~\cite{Shiozaki2016}.
  In the following, we derive formulas to give the $\mathbb{Z}_4$ and $\mathbb{Z}_2$ invariants defined on the Brillouin zone face (ZF), taking additional screw symmetry into account.
  Then we demonstrate that the formulas can be significantly simplified by using the symmetry of $Pnma$.

%Let us start by a system with coaxial glide and screw symmetry,
%\begin{equation}
%  \hat{G}=\Set{M_c|\bm{c}/2+\bm{a}/2},\quad\hat{S}=\Set{C_{2c}|\bm{c}/2+\bm{a}/2},\label{eq:def_Ga}
%\end{equation}
%where $\bm{c}=\hat{c}$ is the primitive lattice translation in the $c$ direction, and $\bm{a}/2$ is a fractional translation perpendicular to $\hat{c}$.
%Here we set the origin to be an inversion center,
%\begin{equation}
%  \hat{I}=\hat{S}^{-1}\hat{G}=\Set{I|\bm{0}}.
%\end{equation}
%%Here $M_z$, $C_{2z}$ and $I$ represent mirror, twofold rotation, and inversion with respect to the origin.
%The space group under consideration is isomorphic to $P2_1/c$, and the discussion below holds for all space groups involving $P2_1/c$ as a translation-equivalent subgroup.
{Let us start by a system with coaxial glide and screw symmetry, $\hat{G}=\Set{M_c|\bm{c}/2+\bm{a}/2}$, $\hat{S}=\Set{C_{2c}|\bm{c}/2+\bm{a}/2},\label{eq:def_Ga}$ where $\bm{c}=\hat{c}$ is the primitive lattice translation in the $c$ direction, and $\bm{a}/2$ is a fractional translation perpendicular to $\hat{c}$.}
Here we set the origin to be an inversion center: $\hat{I}=\hat{S}^{-1}\hat{G}=\Set{I|\bm{0}}$.
The space group under consideration is isomorphic to $P2_1/c$, and the discussion below holds for all space groups involving $P2_1/c$ as a translation-equivalent subgroup.

The key is the following relation,
\begin{equation}
  \hat{G}\hat{I}=\Set{E|\hat{c}+\bm{a}}\hat{I}\hat{G}.
    \label{eq:commu_rel}
\end{equation}
Owing to \Eq{eq:commu_rel}, time-reversal symmetry $\hat{\Theta}$ combined with $\hat{I}$ does not change the glide eigenvalues on the ZF.  To see this, let us focus on glide-invariant planes $k_c=\Gamma_c\equiv0,\,\pi$. The glide operator can be diagonalized on these planes,
\begin{equation}
  \hat{G}\ket{\bm{k}}_\pm=\pm ie^{-i\bm{k}\cdot\bm{a}/2}\ket{\bm{k}}_\pm.
\end{equation}
Using \Eq{eq:commu_rel} we have,
\begin{subequations}\label{eq:Iclosed}
\begin{align}
  \hat{G}\left(\hat{\Theta}\hat{I}\ket{\bm{k}}_\pm\right)
%&=\Set{E|\hat{x}+\hat{z}}\hat{\Theta}\hat{I}\left(\pm ie^{-ik_x/2}\ket{\bm{k}}_\pm\right)\\
  &=\mp e^{-i\Gamma_c}ie^{-ik_a/2}\left(\hat{\Theta}\hat{I}\ket{\bm{k}}_\pm\right)\\
  &=\pm ie^{-ik_a/2}\left(\hat{\Theta}\hat{I}\ket{\bm{k}}_\pm\right),\quad(\Gamma_c=\pi)
\end{align}
\end{subequations}
where $k_a\equiv\bm{k}\cdot\bm{a}$~\cite{Supplement}.
Thus, the symmetry $\hat{\Theta}\hat{I}$ is preserved within each glide eigen-sector.
Note that translation along the $c$-axis $\{E|\hat{c}\}$ in \Eq{eq:commu_rel} is crucial, which is ensured by the coexisting screw symmetry.
The phase factor $e^{-i\Gamma_c}$ disappears when $\hat{S}$ is replaced by an usual rotation $\hat{C}_{2}=\Set{C_{2c}|\bm{a}/2}$.

Equation \eqref{eq:Iclosed} considerably simplifies the expression of the glide topological invariants.
First, we consider topological invariants of glide-odd superconductivity $\{\hat{C},\hat{G}\}=0$, where $\hat{C}$ represents the particle-hole operation.
We choose $\bm{a}$ and $\hat{c}$ as units of crystal translations, and denote a remaining translation unit perpendicular to $\hat{c}$ as $\bm{b}$~\cite{Supplement}.
Glide-odd superconducting phases are classified by the usual 3D winding number $W$ and the two $\mathbb{Z}_4$ topological invariants defined on the two glide-invariant planes $k_c=\Gamma_c$~\cite{Schnyder2008,Shiozaki2016,Yanase2017},
\begin{align}
  \theta_4(\Gamma_c)=2\int_{-\pi}^{\pi}&\frac{dk_b}{\pi i}A_+^{\mathrm{I}}(\pi,k_b,\Gamma_c)\nonumber\\
  &-\int_{0\le k_a\le\pi}\frac{d^2k}{\pi i}F_+(k_a,k_b,\Gamma_c),\label{eq:def_Z4}
%\bm{k}&=(k_a,\,k_b,\,k_c=\Gamma_c).
\end{align}
where $k_b\equiv\bm{k}\cdot\bm{b}$.
%% Here we denoted the glide direction and the normal of the glide plane as $a$ and $c$, respectively. The other crystal translation is denoted as $b$.
%Here we denoted wave numbers corresponding to $\bm{a}$, $\bm{b}$, and $\hat{c}$ as $k_a$, $k_b$, and $k_c$, respectively.
%% Note that Kramers pair is well defined within each glide eigensector on the lines $C_{\mathrm{AII}}(\Gamma_c)=\Set{\bm{k}|(\pi,k_b,\Gamma_c)}$, where $\Theta$ preserves glide eigenvalues.
Here, $\hat{\Theta}$ preserves glide eigenvalues on the lines $C_{\mathrm{AII}}(\Gamma_c)=\Set{\bm{k}|(\pi,k_b,\Gamma_c)}$.
In \Eq{eq:def_Z4}, $A_+^{\mathrm{I}}$ represents the Berry connection of one of the Kramers pair with the positive glide eigenvalue $+ie^{-ik_a/2}$, while $F_+$ is the Berry curvature in the positive glide eigen-sector.

On the ZF $\Gamma_c=\pi$, the Berry curvature vanishes because $\hat{\Theta}\hat{I}$ is preserved in glide eigen-sectors [\Eq{eq:Iclosed}].
Therefore, only the first integral survives in \Eq{eq:def_Z4}. Equation~\eqref{eq:Iclosed} also ensures $\hat{I}$ is closed within each glide eigen-sector on the line $C_{\mathrm{AII}}(\pi)$.
Thus, \Eq{eq:def_Z4} recasts into a Berry phase of a class AII system with inversion symmetry, which has been studied in the context of topological insulators~\cite{Fu-Kane2006,Fu-Kane2007,Qi2011}.
Following Fu and Kane~\cite{Fu-Kane2007}, \Eq{eq:def_Z4} can be rewritten by inversion eigenvalues $\zeta$ of Bogoliubov-de Gennes (BdG) eigenvectors at the two time-reversal invariant momenta (TRIM) $\Gamma_1=(\pi,0,\pi)$ and $\Gamma_2=(\pi,\pi,\pi)$.
Furthermore, $\zeta$ can be rewritten by the inversion eigenvalues of Bloch wave functions, following Refs.~\cite{Fu2010,Sato2010_oddparity}.
Finally, we obtain a formula for the $\mathbb{Z}_4$ invariant~\cite{Supplement},
\begin{gather}
  \theta_4(\pi)=\sum_{i=1,2}\begin{cases}
    M_{+u}^<(\Gamma_i)+M_{-g}^<(\Gamma_i)\\
    M_{u}^<(\Gamma_i),\end{cases}\quad(\text{mod}\ 4)\label{eq:formula1}
  %% \theta_4(\pi)=\sum_{\Gamma=\Gamma_1,\,\Gamma_2}\begin{cases}
  %%   M_{+u}^<(\Gamma)+M_{-g}^<(\Gamma)&(\text{odd parity SC})\\
  %%   M_{u}^<(\Gamma)&(\text{even parity SC}),\end{cases}\label{eq:formula1}
  %% \\\qquad(\text{mod} 4)\nonumber
\end{gather}
where the first and second lines correspond to odd- and even-parity superconductivity, respectively.
Here, $M_{+u}^<(\Gamma_i)\ \bigl(M_{-g}^<(\Gamma_i)\bigr)$ is the number of occupied electron states at $\Gamma_i$ with positive (negative) glide and negative (positive) inversion eigenvalues.
Similarly, $M_{u}^<(\Gamma_i)$ is the number of occupied inversion-odd electron states.
The $\mathbb{Z}_4$ invariant \eqref{eq:formula1} takes either $0$ or $2$
%,since even number of states are occupied
due to Kramers degeneracy.
%\footnote{Strictly speaking, the present result means the \textit{bulk} glide $\mathbb{Z}_4$ invariant at ZF recasts into $\mathbb{Z}_2$ classification (namely, $\theta_4(\pi)=0,2$) in the presence of coexisting screw symmetry.
%In this paper, we call $\theta_4(\pi)=2$ phases as ``$\mathbb{Z}_4$ nontrivial'' simply because the glide $\mathbb{Z}_4$ topological invariant is nontrivial, and corresponding surface states have the $\mathbb{Z}_4=2$ structure.
%On the other hand, note that topological classification of the glide-preserving \textit{surfaces}~\cite{Yanase2017} is unaffected by the screw symmetry, since it does not preserve the surface. In this sense, the surfaces of $\theta_4(\pi)=2$ system with screw symmetry in the bulk gives an example of $\mathbb{Z}_4$ nontrivial superconducting phases of the wallpaper group $pg(+\Theta pg)$@@.}.

Next, we consider glide-even superconductivity $[\hat{C},\hat{G}]=0$.
In this case, topological phases are characterized by four 1D class DIII $\mathbb{Z}_2$ invariants on the lines $C_{\mathrm{AII}}(\Gamma_c)$~\cite{Shiozaki2016,Yanase2017},
\begin{equation}
  \nu_\pm(\Gamma_c)=\int_{-\pi}^{\pi}\frac{dk_b}{\pi i}A_\pm^{\mathrm{I}}(\pi,k_b,\Gamma_c)\quad(\text{mod}\ 2).
\end{equation}
%A discussion parallel to that of previous paraghraph shows that
A parallel discussion leads to~\cite{Supplement}
\begin{gather}
  \nu_\pm(\pi)=\sum_{i=1,2}\begin{cases}
    M_{\pm}^<(\Gamma_i)/2\\
    0,\end{cases}\quad(\text{mod}\ 2)\label{eq:formula2}
\end{gather}
for odd- and even-parity superconductivity, respectively.
Here, $M_\pm^<(\Gamma_i)$ represents the number of electronic occupied states at $\bm{k}=\Gamma_i$ with positive (negative) glide eigenvalue.
These formulas \eqref{eq:formula1} and \eqref{eq:formula2} clarify ``band inversion'' between two TRIM determines topological properties of TNCS in analogy with topological insulators~\cite{Fu-Kane2007}.

%% Equation~\eqref{eq:formula2} can be rewritten by the number of FSs with positive glide eigenvalue counted with sign, defined as
%% \begin{gather}
%%   \nu_\pm(\pi)=\#\text{FS}_\pm/2\equiv\int_{\Gamma_1}^{\Gamma_2}d\bm{k}\cdot\nabla_{\bm{k}}M_\pm^<(\bm{k})/2,\quad (\text{mod}\ 2)
%% \end{gather}
%% for odd-parity SC. This result is consistent with the formulas previously obtained for 1D class DIII invariants\cite{Sato2010_oddparity,Qi2010_DIII}@@@.

\begin{figure}
  \centering
  \includegraphics[height=34mm]{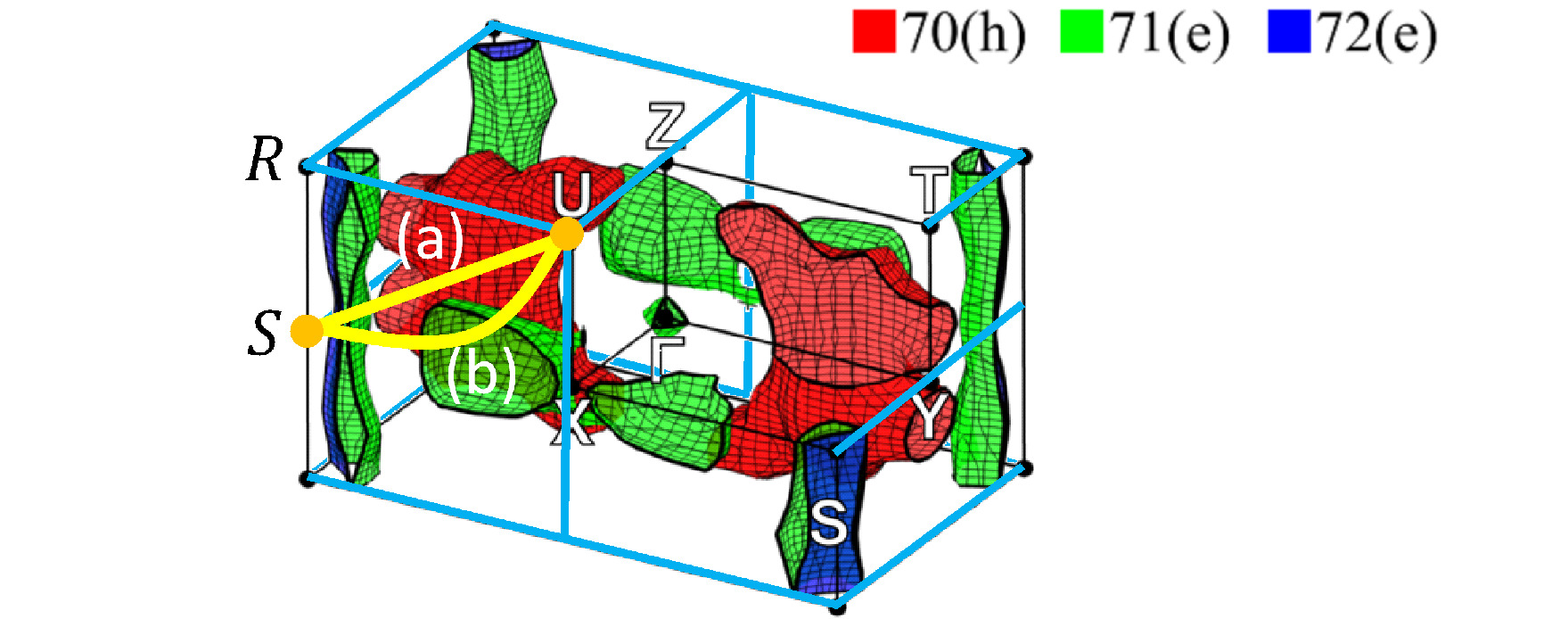}
  \caption{FSs obtained by an \textit{ab-initio} calculation~\cite{Fujimori2015,Fujimori2016}. Lines highlighted by cyan show $\bm{k}$ points where bands are fourfold degenerate owing to nonsymmorphic symmetry~\cite{Bradley-text}. Two yellow lines show paths (a) and (b) connecting $S$ and $U$ points.
    %Copyright from the Physical Society of Japan.}
    %Copyright from American Physical Society.
    \Fig{fig:FSs} adapted with permission from \Ref{Fujimori2015}.
    Copyrighted by the American Physical Society.}
  \label{fig:FSs}
\end{figure}

\begin{table}[thbp]
  \centering
  \caption{Topological invariants of TNCS in the space group $Pnma$. Here, ``IR'' means irreducible representations of $D_{2h}$. $\theta_4^{(a,n)}(\pi)$ are given by modulo four, and $\nu_\pm^{(a,n)}(\pi)$ are by modulo two.}
  \begingroup
  \renewcommand{\arraystretch}{1.5}
  \tabcolsep =2.5mm
  \begin{tabular}{c|l|l}\hline\hline
    \multicolumn{2}{c|}{glide parity \& IR}&topological invariants\\\hline
    \multirow{2}{*}{$\hat{G}_a$}&odd $\ A_u$, $B_{1u}$&$\displaystyle\theta_4^{(a)}(\pi)={\#\text{FS}_{R\to U}}/{2}+\Delta M(R)$\\
    &even $B_{2u}$, $B_{3u}$&$\nu_\pm^{(a)}(\pi)=\#\text{FS}_{R\to U}/4$\\\hline
    \multirow{2}{*}{$\hat{G}_n$}&odd $\ A_u$, $B_{3u}$&$\theta_4^{(n)}(\pi)=\#\text{FS}_{S\to U}/2$\\
    &even $B_{1u}$, $B_{2u}$&$\nu_\pm^{(n)}(\pi)=\#\text{FS}_{S\to U}/4+\Delta M(S)/2$\\\hline\hline
  \end{tabular}
  \endgroup
  \label{tab:formulas}
\end{table}
\begin{table}[htbp]
  \centering
  \caption{Characters of the irreducible representations of $D_{2h}$, into which gap functions of $Pnma$ system are classified~\cite{Sigrist1991,Nomoto2016,Bradley-text}.
    We show all the odd-parity irreducible representations
    %\textcolor{red}{and candidate $d$-vector of UCoGe which may coexist with magnetization in the $\hat{z}$ direction.}
  {and the simplest form of $d$-vector compatible with the pairing symmetry.}}
  \begingroup
  \renewcommand{\arraystretch}{1.2}
  \tabcolsep =2.5mm
  \begin{tabular}{c|cccc|c}\hline\hline
    &$\hat{I}$&$\hat{G}_a$&$\hat{M}_y$&$\hat{G}_n$&$d$-vector\\\hline
    $A_u$&$-1$&$-1$&$-1$&$-1$&$k_x\hat{x},\,k_y\hat{y},\,k_z\hat{z}$\\
    $B_{1u}$&$-1$&$-1$&$+1$&$+1$&$k_y\hat{x},\,k_x\hat{y}$\\
    $B_{2u}$&$-1$&$+1$&$-1$&$+1$&$k_z\hat{x},\,k_x\hat{z}$\\
    $B_{3u}$&$-1$&$+1$&$+1$&$-1$&$k_z\hat{y},\,k_y\hat{z}$\\\hline\hline
    %% $A_u$&$-1$&$-1$&$-1$&$-1$&$k_x\hat{x},\,k_y\hat{y},\,k_z\hat{z}$\\
    %% $B_{1u}$&$-1$&$-1$&$+1$&$+1$&$\sin k_y\hat{x},\,\sin k_x\hat{y}$\\
    %% $B_{2u}$&$-1$&$+1$&$-1$&$+1$&$\sin k_z\hat{x}$\\
    %% $B_{3u}$&$-1$&$+1$&$+1$&$-1$&$\sin k_z\hat{y}$\\\hline\hline
  \end{tabular}
  \endgroup
  \label{tab:IRs}
\end{table}
The formulas \eqref{eq:formula1}~and~\eqref{eq:formula2} for glide topological invariants generally hold in the presence of coexisting screw symmetry. By using the symmetry of $Pnma$, they are further simplified, as shown in \Table{tab:formulas}~\cite{Supplement}.
We classify the odd-parity superconducting states by four irreducible representations of $D_{2h}$ whose symmetries are summarized in \Table{tab:IRs}.
%and show the glide parity which determines the topological invariants.
The parity for glide operation (glide parity) determines the topological invariants, namely, $\mathbb{Z}_2$ or $\mathbb{Z}_4$.
%We classify the pairing symmetry by irreducible representations of $D_{2h}$ and show the glide parity which determines the topological invariants.
%~\cite{Shiozaki2016,Yanase2017}.
%The symmetry of four odd-parity representations is shown in \Table{tab:IRs}.
The space group $Pnma$ includes the $a$-glide and the $n$-glide which are represented by,
%% We classify the pairing symmetries by glide parity and show corresponding topological invariants.
%% All the four odd-parity representations of $D_{2h}$ are shown in \Table{tab:IRs}.
%% %, denoted as $A_u$, $B_{1u}$, $B_{2u}$, and $B_{3u}$.
%% Here,
\begin{equation}
\hat{G}_a=\Set{M_z|\hat{x}/2+\hat{z}/2},\ \hat{G}_n=\Set{M_x|\hat{x}/2+\hat{y}/2+\hat{z}/2},
\end{equation}
respectively. Correspondingly, the topological invariants protected by each glide symmetry are shown. Here,
%denote the $a$-glide and the $n$-glide of $Pnma$, respectively, and
$\#\text{FS}_{\Gamma_1\to\Gamma_2}$ and $\Delta M(\Gamma_i)\ (i=1,2)$ are defined as follows:
\begin{subequations}
\begin{gather}
  \#\text{FS}_{\Gamma_1\to\Gamma_2}\equiv\int_{\Gamma_1}^{\Gamma_2}d\bm{k}\cdot\nabla_{\bm{k}}M^<(\bm{k})\in 4\mathbb{Z},\label{eq:def_numFS}\\
  \Delta M(\Gamma_i)\equiv {M^<_{+u}(\Gamma_i)}-{M^<_{-u}(\Gamma_i)}\in 2\mathbb{Z},\label{eq:def_DeltaN}
\end{gather}
\end{subequations}
where $M^<(\bm{k})$ is the number of occupied states at $\bm{k}$.
The integrand of \Eq{eq:def_numFS} contributes only when the integration path crosses Fermi surfaces, and therefore, $\#\text{FS}_{\Gamma_1\to\Gamma_2}$ can be regarded as the number of Fermi surfaces counted with sign.
In \Eq{eq:def_numFS}, we used the fact $M^<(\Gamma_i)\in 4\mathbb{Z}$ for $\Gamma_i=S,\,U\,R$ [see~\Fig{fig:FSs}], owing to nonsymmorphic band degeneracy~\cite{Bradley-text,Supplement}.
The formulas in \Table{tab:formulas} are one of the main results of this paper.
%, and are derived in \Ref{Supplement} by considering irreducible representations of electronic Bloch functions at each TRIM.

Here we note that $\Delta M$ vanishes when the effective spin-orbit coupling (SOC) is not too large on $S$ and $R$ points. For weak strength of SOC, the system is approximately
% regarded as a spinless system
SU(2) symmetric, and the glide eigenvalues can be interchanged by SU(2) rotation. Then, the glide topological invariants solely depend on the topology of FSs, and the topological conditions are reduced to
%% Here we note that $\Delta M$ is related with the strength of effective spin-orbit coupling (SOC) of bands around Fermi energy.
%% When the system is approximately regarded as spinless, two (spinful) glide eigenvalues can be interchanged owing to SU(2) symmetry, and $\Delta M(\Gamma_i)$ vanishes.
%% Therefore, the glide topological invariants solely depend on FSs, in the case of small or moderate strengh of effective SOC at $S$ and $R$.
%% Then, the topological conditions are reduced to
\begin{equation}
  \#\text{FS}_{R\to U}\in 4(2\mathbb{Z}+1),\label{eq:WeakSOC_Ga}
\end{equation}
for $\theta_4^{(a)}(\pi)=2$ or $\nu_\pm^{(a)}(\pi)=1$, while
\begin{equation}
  \#\text{FS}_{S\to U}\in 4(2\mathbb{Z}+1),\label{eq:WeakSOC_Gn}
\end{equation}
for $\theta_4^{(n)}(\pi)=2$ or $\nu_\pm^{(n)}(\pi)=1$. Conditions~\eqref{eq:WeakSOC_Ga} and \eqref{eq:WeakSOC_Gn} for $\nu_\pm^{(a)}(\pi)$ and $\theta_4^{(n)}(\pi)$ are rigorous, irrespective of the SOC strength.
%% The results obtained in this section show that ``band inversion'' between two TRIM determines topological properties of TNCS in analogy with topological insulators\cite{}.@@
%% These results clarify that ``band inversion'' between two TRIM determines topological properties of TNCS in analogy with topological insulators\cite{}.@@

%%%%%%%%%%%%%%%%%%%% Application to UCoGe %%%%%%%%%%%%%%%%%%%%%%
\textit{Application to UCoGe ---}
We apply the formulas to
UCoGe, whose FSs {in the paramagnetic normal state} have been obtained by \textit{ab-initio} calculations~\cite{Fujimori2015,Fujimori2016}
\footnote{{Although the results are for ambient pressure, pressure effects on the Fermi surfaces are negligible, according to \textit{ab-initio} calculations~\cite{Ishizuka_private} with the lattice constants measured at high pressure~\cite{Adamska2010}.}}.
%In the following, we illustrate the formulas with the $\mathbb{Z}_4$ invariant of $\hat{G}_n$-odd SC.
{Here, we discuss all the odd-parity pairing states, for completeness.}

%Let us apply the formulas to UCoGe.
%The FSs of UCoGe in the paramagnetic normal state have been obtained by  \textit{ab-initio} calculations~\cite{Fujimori2015,Fujimori2016}.
%Although the results are for ambient pressure, they can be used for our purpose, because pressure effects on the results are shown to be negligible (data not shown).

First, we illustrate the formulas for the $\mathbb{Z}_4$ invariant of $\hat{G}_n$-odd superconductivity.
According to \Table{tab:formulas} and \Eq{eq:WeakSOC_Gn}, $\theta_4^{(n)}(\pi)$ is determined by the number of FSs on the line connecting $S$ and $U$ points in the Brillouin zone, and it is independent of the choice of the path by \Eq{eq:def_numFS}.
For instance, we consider two paths (a) and (b) in \Fig{fig:FSs}.
On the path (a), there are two FSs $71$ and $72$, each of which is doubly degenerate due to Kramers theorem. Both of them contribute with the same sign to the integral \eqref{eq:def_numFS}, since they are electron FSs. Therefore, we have $\#\text{FS}_{S\to U}=\pm4$, and conclude $\theta_4^{(n)}(\pi)=2$.
The other path (b) crosses another FS.
However, contribution from this FS vanishes, because of the cancellation of two crossing points.
%since entering and going out from the same FS accompany opposite sign.
Thus, we
%can say that $X$-FS is irrelevant of topological properties, and
consistently obtain nontrivial $\mathbb{Z}_4$ invariant $\theta_4^{(n)}(\pi)=2$.
These discussions imply that the cylindrical FSs $71$ and $72$ play major role for the $\mathbb{Z}_4$ nontrivial TNCS.

%% We note that the path can be chosen on high-symmetry lines such as $S-R-U$. Therefore, if necessary, topological invariants can be more carefully evaluated by using the data of \textit{ab-initio} calculations.

\begin{table}[htbp]
  \centering
  \caption{Summary of the topological invariants of UCoGe obtained by formulas in \Table{tab:formulas} and the \textit{ab-initio} calculations~\cite{Fujimori2015,Fujimori2016}.}
  \begingroup
  \renewcommand{\arraystretch}{1.4}
  \tabcolsep =3mm
  \begin{tabular}{ccc}\hline\hline
    \multirow{2}{*}{IR}&\multicolumn{2}{c}{topological invariants at ZF}\\\cline{2-3}
    &$\hat{G}_a$&$\hat{G}_n$\\\hline
    $A_u$&{$\theta_4^{(a)}(\pi)=2$}&{$\theta_4^{(n)}(\pi)=2$}\\
    $B_{1u}$&{$\theta_4^{(a)}(\pi)=2$}&{$\nu_\pm^{(n)}(\pi)=1$}\\
    $B_{2u}$&{$\nu_\pm^{(a)}(\pi)=1$}&{$\nu_\pm^{(n)}(\pi)=1$}\\
    $B_{3u}$&{$\nu_\pm^{(a)}(\pi)=1$}&{$\theta_4^{(n)}(\pi)=2$}\\\hline\hline
  \end{tabular}
  \endgroup
  \label{tab:results}
\end{table}
%% \begin{table}
%%   \centering
%%   \caption{Topological indices of UCoGe~\cite{Shiozaki2016,Yanase2017,Supplement}. The first and second rows correspond to the two possibilities of the winding number $\mathbb{Z}^W$ of $A_u$ state with an arbitrary integer $n$. Curly and square braces represent glide-odd indices $\{\mathbb{Z}^W,\mathbb{Z}_2^{\mathrm{strong}},\mathbb{Z}_4^{\mathrm{weak}}\}$ and glide-even indices $[\mathbb{Z}_2^{\mathrm{CS}_T},\mathbb{Z}_2^{\mathrm{strong}},\mathbb{Z}_2^{\mathrm{weak}}]$, repectively~\cite{Supplement}. $\mathbb{Z}_2^{\mathrm{strong}}$ represents the strong glide index of TNCS.}
%%   \begingroup
%%   \renewcommand{\arraystretch}{1.5}
%%   \tabcolsep =2.5mm
%%   \begin{tabular}{ccc}\\\hline\hline
%%     &$\hat{G}_a$&$\hat{G}_n$\\\hline
%%     \multirow{2}{*}{$A_u$}&$\{4n,0,2\}$&$\{4n,1,0\}$\\
%%     &$\{4n+2,1,0\}$&$\{4n+2,0,2\}$\\
%%     $B_{1u}$&$\{0,0,2\}$&$[0,1,0]$\\
%%     $B_{2u}$&$[0,0,1]$&$[0,1,0]$\\
%%     $B_{3u}$&$[0,0,1]$&$\{0,1,0\}$\\\hline\hline
%%   \end{tabular}
%%   \endgroup
%%   \label{tab:SIs}
%% \end{table}
In the same way, we can evaluate the other topological invariants of all the odd-parity superconducting states. The results are summarized in \Table{tab:results}.
Here, we take into account $\Delta M(S)=\Delta M(R)=0$ consistent with \textit{ab-initio} calculations revealing small splitting of FS by SOC.
%Note that bands at ZF should be fourfold degenerate in the spinless limit\cite{}.
Actually, it is naturally expected that FS $71$ and $72$ around $S$ and $R$ are nearly degenerate on the ZF where the four-fold degeneracy is ensured in the SU(2)-symmetric limit~\cite{Liang2016}.
Thus, \Fig{fig:FSs} reveals that the splitting of FS by the SOC is small and it does not change the structure of FS, namely, $\Delta M(S)=\Delta M(R)=0$.
%% close to spinless limit, since split of the FSs is small and they may be paired by recovering SU(2) symmetry~\cite{Liang2016}.

Interestingly, \Table{tab:results} shows that
all the candidate odd-parity superconducting states of UCoGe are nontrivial TNCS.
Indeed, $A_u$, $B_{1u}$, and $B_{3u}$ states are $\mathbb{Z}_4$ nontrivial TNCS, while $B_{2u}$ state is $\mathbb{Z}_2$ nontrivial for both $\hat{G}_a$ and $\hat{G}_n$.
%% We note that the other topological invariants can not be determined in the disucussion here.
%% However, we can show that each superconductivity has nontrivial strong glide indices~\cite{Shiozaki2016,Yanase2017} at least for one of $\hat{G}_a$ or $\hat{G}_n$, based on some natural assumptions~\cite{Supplement}. Thus, the $S_2$ phase of UCoGe is identified as strong glide TNCS.
Owing to these nontrivial topological invariants at ZF, the strong index of the $K$-theory for three-dimensional system is also nontrivial. The strong indices obtained under reasonable assumptions are shown in
%\Table{tab:SIs}.
%We show the details in
Supplemental Materials~\cite{Supplement,Yoshida2018}.

%Notably, judging from the current situation, the paring symmetry of UCoGe is unlikely to be the $B_{2u}$ state.
%Indeed, there is an experimental study~\cite{Hattori2012} suggesting $A_u$ representation of $C_{2h}$ for the ferromagnetic superconducting phase, which smoothly deforms into either $A_u$ or $B_{1u}$ states of $D_{2h}$ at the high pressure phase.
%The $A_u$ state is also supported by a group theoretical argument~\cite{Cheung2016}.
%Thus, it immediately follows from \Table{tab:results} that UCoGe is the first material candidate of $\mathbb{Z}_4$ TNCS.
Notably, the pairing symmetry of UCoGe may not be the $B_{2u}$ state.
There is an experimental study~\cite{Hattori2012} suggesting $A_u$ representation of $C_{2h}$ for the ferromagnetic superconducting phase, which smoothly deforms into either $A_u$ or $B_{1u}$ states of $D_{2h}$ at the high pressure phase.
In this case, $\mathbb{Z}_4$ TNCS immediately follows from \Table{tab:results}.
Thus, UCoGe is the first and promising material candidate of $\mathbb{Z}_4$ TNCS, although further experimental effort is required to fully identify the pairing symmetry.

\begin{figure}
  \centering
  \includegraphics[width=100mm]{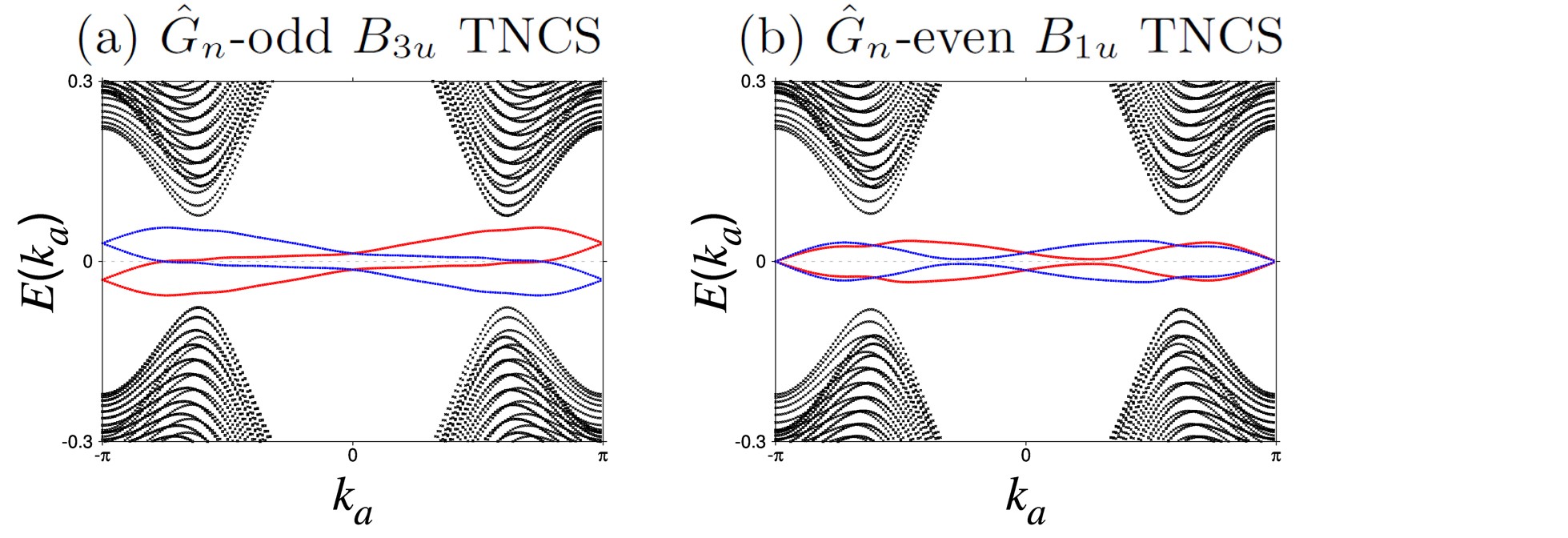}
  \caption{$(0\bar{1}1)$ surface states of (a) glide-odd $B_{3u}$ and (b) glide-even $B_{1u}$ superconducting states at the glide-invariant ZF $k_x=\pi$. Topological surface states with positive (negative) glide eigenvalues are highlighted by red (blue).}
  \label{fig:surface_states}
\end{figure}

Our analytic results are confirmed by a numerical analysis of
%In order to check our predictions, we constructed
a single-orbital tight-binding model reproducing the two cylinder FSs $71$ and $72$
%, and calculated surface spectrum for each odd-parity SC
~\cite{Supplement}.
Calculated surface states are consistent with TNCS as we show $(0\bar{1}1)$ surface states of $\mathbb{Z}_4$ nontrivial $B_{3u}$ state and $\mathbb{Z}_2$ nontrivial $B_{1u}$ state in \Fig{fig:surface_states}.
%(Other surface spectrum are also available in Supplemental materials\cite{Supplement}).
We clearly see a M\"obius structure with the $4\pi$ periodicity.
Such an unconventional structure of surface states is characteristic of $\theta_4=2$ ($\nu_\pm=1$) state, and would provide an important experimental evidence of $\mathbb{Z}_4$ ($\mathbb{Z}_2$) TNCS when it is observed.
Note that analogous state cannot be realized in standard topological phases, while $\theta_4=1$ state shows a conventional helical surface state.
%% \begin{figure}
%%   \centering
%%   \begin{tabular}{cc}
%%     (a) $\hat{G}_n$-odd $B_{3u}$ TNCS&(b) $\hat{G}_n$-even $B_{1u}$ TNCS\\
%%     \includegraphics[width=40mm]{B3u_lifted_Gn.pdf}&\includegraphics[width=40mm]{B1u_lifted_Gn.pdf}
%%   \end{tabular}
%%   \caption{$(0\bar{1}1)$ surface states of (a) glide-odd $B_{3u}$ and (b) glide-even $B_{1u}$ superconducting states at the glide-invariant ZF $k_x=\pi$. Topological surface states with positive (negative) glide eigenvalues are highlighted by red (blue).}
%%   \label{fig:surface_states}
%% \end{figure}

We note that
%% $\hat{G}_a$ and $\hat{G}_n$ combined with lattice translations are also glide operations of $Pnma$, and give the same topological invariants as \Table{tab:results}.Thus,
topological surface states protected by the $a$-glide or the $n$-glide appear not only on the $(010)$ or $(0\bar{1}1)$ surfaces but also on the $(2n,2m+1,0)$ or $(0,2n+1,2m+1)$ surfaces, respectively, with $n$ and $m$ taking arbitrary integer~\cite{Supplement}.
This option may reduce experimental difficulty to observe topological surface states.

\textit{Discussion ---}
First, we stress that our results are stable against adiabatic deformation of FSs unless Lifshitz transition occurs.
This is important for UCoGe since \textit{ab-initio} calculations may be inaccurate for heavy fermion systems.
Even when FSs significantly deviate from \Fig{fig:FSs}, it is easy to judge whether the system is topological or not, using the formulas in \Table{tab:formulas}.
{For example, another \textit{ab-initio} calculation of UCoGe~\cite{Czekala2010} shows two closed FSs enclosing $S$ point, instead of the cylindrical FSs in \Fig{fig:FSs}.
The topological invariants of $\hat{G}_n$ remain nontrivial even in this situation,
%, since four-times-odd-integer number of FSs still exist on the $S-U$ line
%\textcolor{red}{although $\hat{G}_a$-topological invariants are trivial.}
although $\nu_\pm^{(a)}(\pi)$ is trivial. It is revealed by a detailed analysis that $\theta_4^{(a)}(\pi)$ also remains nontrival due to exceptional contribution by $\Delta M(R)$\cite{Supplement}, showing robustness of $\mathbb{Z}_4$ TNCS.

{Second, the TNCS obtained here is robust against possible excitation nodes.
Indeed, suggested line nodes protected by nonsymmorphic symmetries~\cite{Norman1995,Micklitz2009,Kobayashi2016,Nomoto2017,Sumita2018} may not exist in UCoGe due to small splitting of Fermi surfaces~\cite{Supplement}.
Even in their presence, topological surface states remain to exist, because the line nodes do not close band gap within each glide eigen-sector~\cite{Supplement}.}

Next, we discuss other candidates of TNCS on the basis of the obtained formulas.
For example, quasi-linear Dirac semimetals CrAs and CrP~\cite{Wu2014,Kotegawa2014,Niu2017} also crystallize in the space group $Pnma$.
Applying our formulas to the \textit{ab-initio} calculations~\cite{Niu2017,Goh_private}
%\footnote{S.~K.~Goh and W.~C.~Yu, {private communication}.}
, we yield \Table{tab:results} for CrAs and CrP as well, showing TNCS in odd-parity superconducting states.
Importantly, recent experiments point to spin-triplet superconductivity {of $B_{3u}$ state} in the high-pressure phase of CrAs\cite{Kotegawa2015,Guo2018}. Therefore, CrAs family is also a promissing candidate for $\mathbb{Z}_4$ TNCS.
%and related materials are also promising candidates of $\mathbb{Z}_4$ TNCS.
%, altough  topological properties may be hindered to some extent by observed line-nodal behaviour\cite{}.
%% We also propose $\mathrm{CrAs}_{1-x}\mathrm{P}_x$ may be a platform to study TNCS.
%% If a clean sample is obtained, $\mathrm{CrAs}_{(1-x)}\mathrm{P}_x$ may exhibit superconductivity at ambient pressure, and is easier to access in experiments.

Finally, we discuss experimental identification of TNCS.
For an experimental identification of UCoGe as TNCS, the determination of FSs and pairing symmetry are highly desired. By our formulas, the conditions for the TNCS can be examined by these bulk properties.
One of the direct approaches to the TNCS is surface probes such as ARPES. Unfortunately, such measurements are difficult to be done in this case, since TRS superconductivity of UCoGe requires high pressures and low temperatures.
%For future experimental identification of UCoGe as TNCS, the determination of FSs and pairing symmetry are highly desired.
%By our formulas, the conditions for the TNCS can be examined by FSs and pairing symmetries.
%Applying our formulas, we can clarify whether PM superconductivity of UCoGe is TNCS or not.
A possible route to observe gapless surface quasiparticles is to measure magnetic properties of surfaces and their magnetic field angle dependence. Surface magnetization is expected to be strongly dependent on the field direction.
%\textcolor{blue}{, similar to Majorana Isingness of TCSC\cite{Shiozaki2016}}.@@@
A feasible measurement is nuclear magnetic resonance (NMR) which has recently detected nanoscale magnetic properties~\cite{Yamanaka2015}.
%For the same reason, NMR mesuarements may also capture signature of surface states as Knight shift\cite{Yamanaka2015}.
%Another feasible approach is to apply pressure in a \textit{chemical} way and to apply surface probes at ambient pressure. However, such an approach generally takes much time, and is left as an future issue.
Another strategy may be the search of TRS superconducting phase at ambient pressure in related materials. That would enable various experimental techniques to detect topological surface sates.

%%%%%%%%%% Summary and discussion %%%%%%%%%%
%% \textit{Summary ---}
%% {In this Letter, we have investigated TRS TNCS specified by $\mathbb{Z}_4$ topological invariants, focusing on the high-pressure superconducting phase of UCoGe.
%% Formulas connecting glide topological invariants with number of FSs
%% %in the presence of screw symmetry
%% are derived, and further simplified for the space group $Pnma$.
%% Combining the formulas with previous \textit{ab-initio} calculations of UCoGe, we have shown that all the ZF topological invariants take nontrivial values. Thus, UCoGe under high pressure is identified as a promissing candidate of TNCS. In particular, three out of four pairing symmetries are $\mathbb{Z}_4$ nontrivial TNCS, which has never been proposed for real SCs.
%% We also proposed CrAs and related materials as another candidate of $\mathbb{Z}_4$ TNCS. }

%%%%%%%%%% Acknowledgments %%%%%%%%%%
\begin{acknowledgments}
  \textit{Acknowledgments ---}
  The authors are grateful to K.~Shiozaki, J.~Ishizuka, and S.~Sumita for fruitful discussions.
  They also thank S.~Fujimori for the data of band structure calculation used in \Fig{fig:FSs}, and S.~K.~Goh and W.~C.~Yu for their kind offer of band-calculation results for CrAs and CrP.
 This work was supported by Grant-in Aid for Scientific Research on Innovative Areas ``J-Physics'' (15H05884) and ``Topological Materials Science'' (16H00991) from JSPS of Japan, and by JSPS KAKENHI Grants No. JP15K05164, No. JP15H05745, and No. JP17J10588, No. JP18H05842, No. 18H01178, and No. 18H05227.
\end{acknowledgments}

\clearpage

\renewcommand{\thesection}{S\arabic{section}}
\renewcommand{\theequation}{S\arabic{equation}}
\setcounter{equation}{0}
\renewcommand{\thefigure}{S\arabic{figure}}
\setcounter{figure}{0}
\renewcommand{\thetable}{S\arabic{table}}
\setcounter{table}{0}
\makeatletter
\c@secnumdepth = 2
\makeatother

\onecolumngrid
\begin{center}
 {\large \textmd{Supplemental Materials:} \\[0.3em]
 {\bfseries $\mathbb{Z}_4$ Topological Superconductor UCoGe}}
\end{center}

\setcounter{page}{1}
\section{Space group operations and Fourier transformations}
We begin with clarifying the notations for the representation matrices of symmetry operations and two kinds of Fourier transformation.

Let us consider a space group element $\hat{g}=\Set{p|\bm{a}}$ whose operation on real-space coordinates is given by
\begin{equation}
  \hat{g}\bm{x}=p\bm{x}+\bm{a}.
\end{equation}
Here, $p$ and $\bm{a}$ are the point group operation and a translation associated with $\hat{g}$, respectively.
The operation of $\hat{g}$ on the Hilbert space is defined by the following relation,
\begin{equation}
  \hat{g}c^\dagger_{l}(\bm{R}+\bm{r}_n)\hat{g}^{-1}=c^\dagger_{l'}(p\bm{R}+\Delta\bm{R}_n^g+\bm{r}_{n'})D_{n'n}(\hat{g})\mathcal{D}_{l'l}(\hat{g}),
\end{equation}
where $\bm{R},\,\Delta\bm{R}_n^g$ are lattice vectors, $l,l'$ represent internal degrees of freedom, and $n,n'$ specify the sublattices whose position within a unit cell are given by $\bm{r}_n,\bm{r}_{n'}$, respectively. 
We used the following constraint between $\bm{r}_n$, $\bm{r}_{n'}$ and $\Delta\bm{R}_n^g$,
\begin{equation}
  \hat{g}(\bm{r}_n)=p\bm{r}_n+\bm{a}=\bm{r}_{n'}+\Delta\bm{R}_n^g,
\end{equation}
since $\hat{g}$ is a symmetry of the crystal.
Note that $D_{n'n}(\hat{g}){=}\delta_{n',\hat{g}(n)}$ represents the permutation of sublattices, and $\mathcal{D}_{l'l}(\hat{g})$ is a (generally double-valued) representation associated with internal degrees of freedom.
  
In this paper, we use two definitions of Fourier transformation.
The first one is the Fourier transformation compatible with the periodicity of Brillouin zone,
\begin{subequations}
\begin{gather}
  c^\dagger_{\bm{k}ln}\equiv\frac{1}{\sqrt{V}}\sum_{\bm{R}}e^{i\bm{k}\cdot\bm{R}}c^\dagger_l(\bm{R}+\bm{r}_n),\label{eq:periodic}\\
c^\dagger_{\bm{k}+\bm{G}ln}=c^\dagger_{\bm{k}ln},  
\end{gather}
\end{subequations}
with any reciprocal lattice vector $\bm{G}$.
We refer to this basis as periodic basis, where the $\mathbb{Z}_4$ invariant is compactly defined~\cite{Shiozaki2016_SM}.
In the following, we use this basis, unless otherwise specified.
The other is the %\textcolor{red}{site-dependent Fourier transformation},
so-called L\"owdin orbital,
\begin{subequations}
\begin{gather}
  c^\dagger_{ln}(\bm{k})\equiv\frac{1}{\sqrt{V}}\sum_{\bm{R}}e^{i\bm{k}\cdot(\bm{R}+\bm{r}_n)}c^\dagger_l(\bm{R}+\bm{r}_n)=c^\dagger_{\bm{k}ln'}V_{n'n}(\bm{k}),\quad V_{n'n}(\bm{k})=e^{i\bm{k}\cdot\bm{r}_n}\delta_{n'n},\label{eq:Lowdin}\\
  c^\dagger_{ln}(\bm{k}+\bm{G})=c^\dagger_{ln'}(\bm{k})V_{n'n}(\bm{G}).
\end{gather}
\end{subequations}
The operation of $\hat{g}$ is the most simplified in this basis,
\begin{subequations}
\begin{align}
  \hat{g}c^\dagger_{ln}(\bm{k})\hat{g}^{-1}&=\frac{1}{\sqrt{V}}\sum_{\bm{R}}e^{i\bm{k}\cdot(\bm{R}+\bm{r}_n)}c^\dagger_{l'}(p\bm{R}+\Delta\bm{R}_n^g+\bm{r}_{n'})D_{n'n}(\hat{g})\mathcal{D}_{l'l}(\hat{g})\\
  &=\frac{1}{\sqrt{V}}\sum_{\bm{R}'=p\bm{R}+\Delta\bm{R}_n^g}e^{i\bm{k}\cdot(p^{-1}(\bm{R}'-\Delta\bm{R}_n^g)+\bm{r}_n)}c^\dagger_{l'}(\bm{R}'+\bm{r}_{n'})D_{n'n}(\hat{g})\mathcal{D}_{l'l}(\hat{g})\\
  &=\frac{1}{\sqrt{V}}\sum_{\bm{R}'}e^{ip\bm{k}\cdot(\bm{R}'-\Delta\bm{R}_n^g+p\bm{r}_n)}c^\dagger_{l'}(\bm{R}'+\bm{r}_{n'})D_{n'n}(\hat{g})\mathcal{D}_{l'l}(\hat{g})\\
  &=\frac{1}{\sqrt{V}}\sum_{\bm{R}'}e^{ip\bm{k}\cdot(\bm{R}'-\bm{a}+\bm{r}_{n'})}c^\dagger_{l'}(\bm{R}'+\bm{r}_{n'})D_{n'n}(\hat{g})\mathcal{D}_{l'l}(\hat{g})\\
  &=e^{-ip\bm{k}\cdot\bm{a}}c^\dagger_{l'n'}(p\bm{k})D_{n'n}(\hat{g})\mathcal{D}_{l'l}(\hat{g})\equiv c^\dagger_{l'n'}(p\bm{k})U_{(l'n')(ln)}^g(\bm{k}),
\end{align}
\end{subequations}
and for this reason, we use the basis in \Sec{sec:extension}.
Accordingly, the transformation of $c^\dagger_{\bm{k}ln}$ follows as
\begin{subequations}
\begin{gather}
  \hat{g}c^\dagger_{\bm{k}ln}\hat{g}^{-1}=c^\dagger_{p\bm{k}l'n'}\tilde{U}^g_{(l'n')(ln)}(\bm{k}),\\
  \tilde{U}^g_{(l'n')(ln)}(\bm{k})=(V(p\bm{k})U^g(\bm{k})V(\bm{k})^\dagger )_{(l'n')(ln)}=e^{i(p\bm{k}\cdot\bm{r}_{n'}-p\bm{k}\cdot\bm{a}-\bm{k}\cdot\bm{r}_n)}D_{n'n}(\hat{g})\mathcal{D}_{l'l}(\hat{g})=e^{-ip\bm{k}\cdot\Delta\bm{R}_n^g}D_{n'n}(\hat{g})\mathcal{D}_{l'l}(\hat{g}).
\end{gather}
\end{subequations}
Thus, the representation matrix of a space group operation $\hat{g}$ in the periodic basis is given by $\tilde{U}^g(\bm{k})$.

For clarity, we define the particle-hole symmetry of the BdG Hamiltonian as follows:
\begin{gather}
  CH_{\mathrm{BdG}}(\bm{k})C^{-1}=-H_{\mathrm{BdG}}(-\bm{k}),
\end{gather}
where $C$ includes complex conjugation.
Note that the symbol $\hat{C}$ (and $\hat{G}$, etc.) used in the main text represents an operator acting on creation/annihilation operators, while $C$ acts on BdG Hamiltonian.
Using $C$ and $\hat{C}$, glide- even/odd superconductivity is expressed by,
\begin{equation}
C\tilde{U}^G_{\mathrm{BdG}}(\bm{k})=\pm\tilde{U}^G_{\mathrm{BdG}}(-\bm{k})C,\quad \text{or equivalently,}\quad \hat{C}\hat{G}=\pm\hat{G}\hat{C}.
\end{equation}
Here, $\tilde{U}^{G}_\BdG(\bm{k})$ represents the glide operation extended to the Nambu space.
Note that it implicitly includes the U(1) gauge rotation for glide-odd superconductivity~\cite{SatoAndo2017_SM}.
%is implicitly repraced by that followed by U(1) transformation, for glide-odd SC~\cite{SatoAndo2017_SM}.
We adopt the same rules of notation for other space group operations as well.

\section{Derivation of the formulas (5) and (9)}
\label{sec:derivation_formulas}
The formulas (5) and (9) hold in a system with coexisting glide and screw symmetry.
The space group generated by glide and screw operations with a common axis is called $P2_1/c$. The following results derived in this section are valid for all the space groups which include $P2_1/c$ as a translation-equivalent subgroup.

In the following, we derive Eqs.~(5) and (9),
{ under the assumption that the representation matrix of inversion symmetry $\tilde{U}^I(\bm{k})$ can be taken $\bm{k}$-independent. Such a choice of $\tilde{U}^I(\bm{k})$ is possible when all the atoms do not have inversion as their site symmetry (see \Sec{sec:proof_Reps}). In particular, UCoGe and CrAs satisfy the condition, and the following discussion is sufficient for these candidate materials. Later in \Sec{sec:extension}, we extend the proof of formulas to $\bm{k}$-dependent inversion operation.}

First, we fix the notations. We set lattice translation vectors $\bm{a},\bm{b}$, and $\hat{c}$ so as to simplify the glide operator,
\begin{equation}
  \hat{G}=\Set{M_c|\hat{c}/2+\bm{a}/2}.
\end{equation}
Here $\bm{a}$ is orthogonal to $\hat{c}$, the primitive translation vector in the $c$ direction.
In the case of the $a$-glide $\hat{G}_a$ of $Pnma$, $\hat{c}=\hat{z}$ and $\bm{a}=\hat{x}$, while
$\hat{c}=\hat{x}$ and $\bm{a}=\hat{y}+\hat{z}$ for the $n$-glide $\hat{G}_n$.
Then, inversion and screw operations are given by
\begin{equation}
  \hat{S}=\Set{C_{2c}|\hat{c}/2+\bm{a}/2},\quad
  \hat{I}=\Set{I|\bm{0}}.
\end{equation}
We take $\bm{b}$ as another lattice translation vector in the plane normal to $\hat{c}$.
It should be noticed that $\bm{a}$ and $\bm{b}$ are not necessarily taken to be orthogonal.
For example, it is convenient to take $\bm{b}=\hat{z}$ for $\hat{G}_n$ of $Pnma$.
Actually, this choice of crystal translation vectors does not give rise to Brillouin zone folding, while the orthogonal vector $\bm{b}'=\hat{y}-\hat{z}$ does, since $|\hat{c}\cdot\bm{a}\times\bm{b}'|>|\hat{c}\cdot\bm{a}\times\bm{b}|$.
Corresponding wave numbers are given in an usual way,
\begin{subequations}
\begin{gather}
  \bm{k}=k_a\bm{\alpha}+k_b\bm{\beta}+k_c\hat{c},\\
  \bm{\alpha}=\frac{\bm{b}\times\hat{c}}{\bm{a}\cdot\bm{b}\times\hat{c}},\quad
  \bm{\beta}=\frac{\bm{a}\times\hat{c}}{\bm{b}\cdot\bm{a}\times\hat{c}}.
\end{gather}
\end{subequations}
The first BZ is given by $-\pi<k_a,k_b,k_c\le\pi$.

For example, when we adopt the above choice of $\bm{a},\bm{b}$ for $\hat{G}_n$,
we have
\begin{subequations}
\begin{gather}
  \bm{\alpha}=\hat{y},\quad \bm{\beta}=\hat{z}-\hat{y},\quad \hat{c}=\hat{x},\\
  k_a=k_y+k_z,\quad k_b=k_z,\quad k_c=k_x.
\end{gather}
\end{subequations}
The BdG Hamiltonian $H_{\mathrm{BdG}}(\bm{k})$ in terms of $(k_a,k_b,k_c)$, denoted by $\tilde{H}_{\mathrm{BdG}}(k_a,k_b,k_c)$, is readily obtained from $H_{\mathrm{BdG}}(k_x,k_y,k_z)$,
\begin{equation}
  \tilde{H}_{\mathrm{BdG}}(k_a,k_b,k_c)=H_{\mathrm{BdG}}(k_a\bm{\alpha}+k_b\bm{\beta}+k_c\hat{c})=H_{\mathrm{BdG}}(k_x=k_c,k_y=k_a-k_b,k_z=k_b).
  %,\quad (-\pi\le k_a,k_b,k_c\le\pi).
\end{equation}
Then, calculations can be done in an effective cubic BZ $(-\pi< k_a,k_b,k_c\le\pi)$.

In the following, we consider the glide-invariant planes $k_c=\Gamma_c$, where the BdG Hamiltonian and the glide operator can be diagonalized at the same time,
\begin{subequations}
\begin{gather}
    H_{\mathrm{BdG}}(\bm{k})\ket{u_{n\bm{k}\pm}}=E_{n}^\pm(\bm{k})\ket{u_{n\bm{k}\pm}},\\
    \tilde{U}^G_{\mathrm{BdG}}(\bm{k})\ket{u_{n\bm{k}\pm}}={\pm ie^{-ik_a/2}}\ket{u_{n\bm{k}\pm}},\quad \bm{k}=(k_a,k_b,\Gamma_c),    
\end{gather}
\end{subequations}
where $\tilde{U}^G_{\mathrm{BdG}}(\bm{k})$ is the representation matrix of $\hat{G}$.
We impose periodic constraint on the Bloch function, $\ket{{u}_{n\bm{k}+\bm{G}\pm}}=\ket{{u}_{n\bm{k}\pm}}$.
The subscript $n$ is the index of the occupied states of the BdG Hamiltonian with glide eigenvalue {$\pm ie^{-ik_a/2}$}.
It is convenient to define the list of the occupied eigenvectors as
\begin{equation}
  \Psi_{\bm{k}\pm}\equiv\left(\ket{u_{1\bm{k}\pm}},\,\ket{u_{2\bm{k}\pm}},\,\cdots,\,\ket{u_{2N\bm{k}\pm}}\right).
\end{equation}
Here, the number of occupied states with each glide eigenvalue is an even integer $2N$, since $\hat{\Theta}$ preserves glide eigenvalues $\pm ie^{-ik_a/2}=\pm1$ at the lines $C_{\mathrm{AII}}(\Gamma_c)=\Set{\bm{k}|(\pi,k_b,\Gamma_c)}$ \cite{Shiozaki2016_SM}.
%, while flips the eigenvalues elsewhere. 
We can define Kramers pairs on $C_{\mathrm{AII}}(\Gamma_c)$, labeled by I and II, within each glide eigen-sector. We denote the list of states for the Kramers pairs as $\Psi_{\bm{k}\pm}^{(\mathrm{I})}$ and $\Psi_{\bm{k}\pm}^{(\mathrm{II})}$, and assign the band index $n=2\alpha-1$ for $\ket{u_{n\bm{k}\pm}}$ in $\Psi_{\bm{k}\pm}^{(\mathrm{I})}$ and $n=2\alpha$ for $\ket{u_{n\bm{k}\pm}}$ in $\Psi_{\bm{k}\pm}^{(\mathrm{II})}$, with $1\le\alpha\le N$.
Then, we have
\begin{subequations}
\begin{gather}
    \Psi_{\bm{k}\pm}^{(\mathrm{I})}\equiv\left(\ket{u_{1\bm{k}\pm}},\,\ket{u_{3\bm{k}\pm}},\,\cdots,\,\ket{u_{2N-1\bm{k}\pm}}\right),\quad   \Psi_{\bm{k}\pm}^{(\mathrm{II})}\equiv\left(\ket{u_{2\bm{k}\pm}},\,\ket{u_{4\bm{k}\pm}},\,\cdots,\,\ket{u_{2N\bm{k}\pm}}\right),\\
  \Psi_{\bm{k}\pm}^{(\mathrm{I})}=\left[\Theta\Psi_{-\bm{k}\pm}^{(\mathrm{II})}\right]U_\chi(\bm{k}),\quad \bm{k}=(\pi,k_b,\Gamma_c),
\end{gather}
\end{subequations}
where $U_\chi(\bm{k})$ is a $N\times N$ unitary matrix.

%% For clarity, we define the particle-hole symmetry of the BdG Hamiltonian as follows:
%% \begin{gather}
%%   CH_{\mathrm{BdG}}(\bm{k})C^{-1}=-H_{\mathrm{BdG}}(-\bm{k}),
%% \end{gather}
%% where $C$ includes complex conjugation.
%% Note that the symbol $\hat{C}$ (and $\hat{G}$, etc.) used in the main text represents an operator acting on creation/anihilation operators, while $C$ acts on BdG Hamiltonian.
%% Using $C$ and $\hat{C}$, glide even/odd SC is expressed by,
%% \begin{equation}
%% C\tilde{U}^G_{\mathrm{BdG}}(\bm{k})=\pm\tilde{U}^G_{\mathrm{BdG}}(-\bm{k})C,\quad \text{or equivalently,}\quad \hat{C}\hat{G}=\pm\hat{G}\hat{C}.
%% \end{equation}
%% Here, the glide operation $\hat{G}$ for glide-odd SC implicitly includes the U(1) gauge rotation~\cite{SatoAndo2017_SM}.
%% %is implicitly repraced by that followed by U(1) transformation, for glide-odd SC~\cite{SatoAndo2017_SM}.
%% We adopt the same rules of notation for other space group operations as well.

\subsection{Glide-odd superconductivity}
\label{subsec:glide_oddSC}
We here derive Eq.~(5).
As shown in Eq.~(4), the $\mathbb{Z}_4$ topological invariant for glide-odd superconductivity is given by~\cite{Shiozaki2016_SM}
\begin{equation}
  \theta_4(\Gamma_c)=2\int_{-\pi}^{\pi}\frac{dk_b}{\pi i}A_+^{\mathrm{I}}(\pi,k_b,\Gamma_c)\nonumber\\
  -\int_{0\le k_a\le\pi}\frac{d^2k}{\pi i}F_+(k_a,k_b,\Gamma_c),\label{eq:def_Z4_SM}
%\bm{k}&=(k_a,\,k_b,\,k_c=\Gamma_c).
\end{equation}
which is defined in the periodic basis. The first term, defined by
%{$+$@@the Berry phase of half the Kramers pair defined by (Fu-Kane2006)}
%\begin{subequations}
%\begin{gather}
\begin{equation}
  A_+^{\mathrm{I}}(\bm{k})\equiv\Tr\left[{\Psi_{\bm{k}+}^{(\mathrm{I})}}^\dagger\partial_{k_b}\Psi_{\bm{k}+}^{(\mathrm{I})}\right],\quad\bm{k}=(k_a=\pi,k_b,k_c=\Gamma_c),
    %% A_+^I(\bm{k})\equiv\sum_{\alpha=1}^{N}\,{}_+\!\!\braket{\psi_{2\alpha-1}(\bm{k})|\partial_{k_b}|\psi_{2\alpha-1}(\bm{k})}_+,\\
%\end{gather}
  %\end{subequations}
\end{equation}
represents the Berry connection of one of the Kramers pair with positive glide eigenvalue, while the second term
%% \begin{subequations}
%% \begin{gather}
\begin{equation}
  F_+(\bm{k})\equiv\sum_{i,j=a,b}\epsilon_{ij}\partial_{k_i}\Tr\left[\Psi^\dagger_{\bm{k}+}\partial_{k_j}\Psi_{\bm{k}+}\right],\quad\bm{k}=(k_a,k_b,k_c=\Gamma_c),\label{eq:Berrycurvature}
%% \end{gather}
    %% \end{subequations}
\end{equation}
represents the Berry curvature in the positive glide eigen-sector.
%% Here, $2\alpha$ ($2\alpha-1$) runs over the occupied states of BdG Hamiltonian with positive glide eigenvalue, about half the Kramers pair.
%The first integral in \Eq{eq:def_Z4_SM} is well-defined, since $\hat{\Theta}$ is closed within each glide eigen-sector on the integration line.%%  Actually, glide eigenvalues are real on this line, and
%% \begin{gather}
%%   \hat{G}\ket{\bm{k}}_\pm=\pm\ket{\bm{k}}_\pm,\\
%%   \hat{G}\left(\hat{\Theta}\ket{\bm{k}}_\pm\right)=\hat{\Theta}\left(\pm\ket{\bm{k}}_\pm\right)=\pm\left(\hat{\Theta}\ket{\bm{k}}_\pm\right).
%% \end{gather}
%In other words, the positive glide eigen-sector on the integration line $C_{\mathrm{AII}}(\Gamma_c)$ belongs to class AII~\cite{Shiozaki2016_SM}.

As shown in the main text, $\hat{\Theta}\hat{I}$ is closed within each glide eigen-sector at ZF, in the presence of screw symmetry. It is well known that the Berry curvature vanishes in the presence of $\hat{\Theta}\hat{I}$ symmetry, and therefore, the topological invariant recasts into the Berry phase of 1D class AII system,
\begin{equation}
  \theta_4(\pi)=2\int_{-\pi}^{\pi}\frac{dk_b}{\pi i}A_+^{\mathrm{I}}(\pi,k_b,\Gamma_c=\pi)\in2\mathbb{Z}.
\end{equation}
Note that inversion operation also closes within the positive glide eigen-sector on $C_{\mathrm{AII}}(\pi)=\Set{\bm{k}|(\pi,k_b,\pi)}$.
Thus, we can use the Fu-Kane formula~\cite{Fu-Kane2007_SM} to evaluate the integral,
\begin{subequations}
\begin{gather}
  (-1)^{\theta_4(\pi)/2}=\prod_{\alpha=1}^N\zeta_{2\alpha-1}^+(\Gamma_1)\zeta_{2\alpha-1}^+(\Gamma_2),\label{eq:FuKane}\\
  \Gamma_1=(k_a=\pi,k_b=0,k_c=\pi),\quad\Gamma_2=(k_a=\pi,k_b=\pi,k_c=\pi).
\end{gather}
\end{subequations}
%where product about $2\alpha$ runs over one of the Kramers pair out of all the $2N$ occupied states with positive glide eigenvalue.
Here, $\zeta_n^\pm(\Gamma_i)=\pm1$ represents the inversion eigenvalues of occupied BdG eigen-states at TRIM,
\begin{equation}
  \tilde{U}^I_{\mathrm{BdG}}(\Gamma_i)\ket{u_{n\Gamma_i\pm}}=\zeta_n^\pm(\Gamma_i)\ket{u_{n\Gamma_i\pm}}.
\end{equation}

The inversion eigenvalues of Bogoliubov quasiparticles $\zeta_n^\pm(\Gamma_i)$ can be rewritten by eigenvalues of electron Bloch wave functions, by following Refs.~\cite{Fu2010_SM,Sato2010_oddparity_SM}.
Except for accidental cases, FSs are off TRIM, and therefore we can take the limit of vanishing order parameter at TRIM without gap closing.
Then, the wave function of BdG quasiparticles $\ket{u_{n\Gamma_i\pm}}$ is expressed by the electronic Bloch wave functions $\ket{v_{m\Gamma_i\pm}}$ as follows. Let us define
%% Note that we can specify energy and eigenvalues of $\hat{G}$ and $\hat{I}$ at the same time, since \Eq{eq:comm_rel_SM} reads $[\hat{G},\hat{I}]=0$ at $\Gamma_i$. We define
%% \begin{gather}
%%     H_{\mathrm{BdG}}(\Gamma_i)\ket{\psi_\alpha(\Gamma_i)}_\pm=E_{\alpha,\pm}(\Gamma_i)\ket{\psi_\alpha(\Gamma_i)}_\pm,\\
%%     \hat{I}\ket{\psi_\alpha(\Gamma_i)}_\pm=\zeta_{\alpha,\pm}(\Gamma_i)\ket{\psi_\alpha(\Gamma_i)}_\pm,\\
%%     \hat{G}\ket{\psi_\alpha(\Gamma_i)}_\pm=\pm\ket{\psi_\alpha(\Gamma_i)}_\pm,
%% \end{gather}
%% and
\begin{subequations}
\begin{gather}
  H_N(\Gamma_i)\ket{v_{m\Gamma_i\pm}}=\epsilon_{m}^\pm(\Gamma_i)\ket{v_{m\Gamma_i\pm}},\\
\tilde{U}^{G}(\Gamma_i)\ket{v_{m\Gamma_i\pm}}=\pm\ket{v_{\Gamma_im\pm}},\\
\tilde{U}^{I}(\Gamma_i)\ket{v_{m\Gamma_i\pm}}=\omega_{m}^\pm(\Gamma_i)\ket{v_{m\Gamma_i\pm}},
\end{gather}
\end{subequations}
where $H_N(\Gamma_i)$ is the Hamiltonian in the normal state and $\omega_m^\pm(\Gamma_i)=\pm1$ is the inversion eigenvalues.
%and $\tilde{U}^{g}(\bm{k})$ is the representation matrix of space group element $g$ in the normal state.
%% Ocuupied states of BdG Hamiltonian with positive glide eigenvalue is given by $\ket{\psi_{\alpha,+}(\Gamma_i)}$ such that $E_{\alpha,+}(\Gamma_i)<0$.
In the weak coupling limit, the occupied states of BdG Hamiltonian $\ket{u_{n\Gamma_i+}}$ $(E_{n}^+(\Gamma_i)<0)$ is given by the Bloch wave functions,
%\begin{subequations}
\begin{gather}
  \ket{u_{\Gamma_in+}}=\begin{cases}\left(\ket{v_{m\Gamma_i+}},0\right)^T&(\epsilon_{m}^+(\Gamma_i)<0),\\
    {C}\left(\ket{v_{m\Gamma_i-}},0\right)^T&
      (\epsilon_{m}^-(\Gamma_i)>0),\end{cases}%\quad(E_{n}^+(\Gamma_i)<0)
\end{gather}
%\end{subequations}
which are obtained from the anticommutation relation $\{\hat{C},\hat{G}\}=0$ characteristic of glide-odd superconductivity.
%since we are considering glide-odd superconductivity $\{\hat{C},\hat{G}\}=0$.
Inversion eigenvalues $\zeta_n^+(\Gamma_i)$ and $\omega_m^\pm(\Gamma_i)$ are related by
\begin{equation}
  \zeta_{n}^+(\Gamma_i)=\begin{cases}\omega_{m}^+(\Gamma_i)&(\epsilon_{m}^+(\Gamma_i)<0)\\
\eta_I\omega_{m}^-(\Gamma_i)&(\epsilon_{m}^-(\Gamma_i)>0),\end{cases}
\end{equation}
where $\eta_I=\pm1$ specifies the inversion-even or -odd superconductivity,
\begin{equation}
  \hat{C}\hat{I}=\eta_I\hat{I}\hat{C}.
\end{equation}

Note that Kramers pair at $\Gamma_i$, which shares the same glide eigenvalue, also possesses the same inversion eigenvalue, since $[\hat{\Theta},\hat{I}]=0$ and $\hat{I}^2=\hat{E}$.
Thus, we rewrite \Eq{eq:FuKane} as
\begin{gather}
  (-1)^{\theta_4(\pi)/2}=\prod_{i=1,2}(-1)^{M_{+u}^<(\Gamma_i)/2}(-1)^{M_{-u}^>(\Gamma_i)/2}\eta_I^{M_-^>(\Gamma_i)/2},\label{eq:formula_temp1}
\end{gather}
where $M_{+u}^<(\Gamma_i)\in2\mathbb{Z}$ $\bigl(M_{-u}^>(\Gamma_i)\in2\mathbb{Z}\bigr)$ represents number of occupied (unoccupied) states with positive (negative) glide and negative inversion eigenvalues.
$M_-^>(\Gamma_i)\in2\mathbb{Z}$ is the number of unoccupied eigenstates with negative glide eigenvalue.
The total number of simultaneous eigenstates with each eigenvalue, defined by
\begin{equation}
  M_{\pm u}(\Gamma_i)\equiv M_{\pm u}^<(\Gamma_i)+M_{\pm u}^>(\Gamma_i)\quad\text{and}\quad M_{\pm g}(\Gamma_i)\equiv M_{\pm g}^<(\Gamma_i)+M_{\pm g}^>(\Gamma_i),
\end{equation}
is equivalent between $\Gamma_1$ and $\Gamma_2$, {when $\tilde{U}^I(\bm{k})$ is $\bm{k}$-independent}. That is,
\begin{equation}
  M_{\pm u}(\Gamma_1)=M_{\pm u}(\Gamma_2),\quad M_{\pm g}(\Gamma_1)=M_{\pm g}(\Gamma_2).\label{eq:switcheigen}
\end{equation}
We prove \Eq{eq:switcheigen} in \Sec{sec:proof_Reps}.
%% The relations \Eq{eq:switcheigen} may be natural, by considering a limit of the atomic insulator with chemical potential $\mu\to-\infty$, where $M_{\pm u}(\Gamma_i)^<=M_{\pm g}^<(\Gamma_i)=0$. This limiting procedure does not alter $M_{\pm u}(\Gamma_i)$ and $M_{\pm g}(\Gamma_i)$, since they are independet of the parameters of the Hamiltonian. Naive expectation is that the atomic insulator is topologically trivial, and then, relations similar to \Eq{eq:switcheigen} hold.
Equation~\eqref{eq:switcheigen} means that the superscripts ``$>$'' and ``$<$'' can be switched in \Eq{eq:formula_temp1}, since the factor $(-1)^{M_{\pm u(g)}(\Gamma_i)/2}$ cancels between $i=1$ and $i=2$.
Thus, we obtain Eq.~(5), by using the definitions
\begin{subequations}
\begin{gather}
  M_-^<(\Gamma_i)\equiv M_{-g}^<(\Gamma_i)+M_{-u}^<(\Gamma_i),\\
  M_{u}^<(\Gamma_i)\equiv M_{+u}^<(\Gamma_i)+M_{-u}^<(\Gamma_i).
\end{gather}
\end{subequations}
\subsection{Glide-even superconductivity}
\label{subsec:glide_oddSC}
Topological invariants of glide-even superconductivity are given by four 1D class DIII invariants on the $C_{\mathrm{AII}}(\Gamma_c)$,
\begin{equation}
  \nu_\pm(\Gamma_c)=\int_{-\pi}^{\pi}\frac{dk_b}{\pi i}A_\pm^{\mathrm{I}}(\pi,k_b,\Gamma_c)\quad(\text{mod}\ 2),
\end{equation}
since each glide eigen-sector on $C_{\mathrm{AII}}(\Gamma_c)$ preserves particle-hole symmetry as well as time-reversal symmetry~\cite{Shiozaki2016_SM}.
%% Actually, we can show $\hat{C}$ is also closed within glide eigen-sector,
%% \begin{gather}
%%   \hat{G}\ket{\bm{k}}_\pm=\pm\ket{\bm{k}}_\pm,\\
%%   \hat{G}\left(\hat{C}\ket{\bm{k}}_\pm\right)=\hat{C}\left(\pm\ket{\bm{k}}_\pm\right)=\pm\left(\hat{C}\ket{\bm{k}}_\pm\right),
%% \end{gather}
%% since $[\hat{G},\hat{C}]=0$.
From Eq.~(3), we have inversion symmetry in each glide eigen-sector also in this case. Thus, topological invariants can be again regarded as the Berry phase of inversion symmetric class AII system. Therefore, discussion parallel to the previous section holds, with a slight modification
\begin{equation}
  \ket{u_{n\Gamma_i\pm}}=\begin{cases}\left(\ket{v_{m\Gamma_i\pm}},0\right)^T&(\epsilon_{m}^\pm(\Gamma_i)<0),\\
    {C}\left(\ket{v_{m\Gamma_i\pm}},0\right)^T&
      (\epsilon_{m}^\pm(\Gamma_i)>0),\end{cases}\quad(\text{for }E_{n}^\pm(\Gamma_i)<0)
\label{eq:eq25}
\end{equation}
since $[\hat{G},\hat{C}]=0$.
Using \Eq{eq:eq25}, we obtain Eq.~(9),
\begin{gather}
  (-1)^{\nu_\pm(\pi)}=\prod_{i=1,2}(-1)^{M_{\pm u}^<(\Gamma_i)/2}(-1)^{M_{\pm u}^>(\Gamma_i)/2}\eta_I^{M_{\pm}^>(\Gamma_i)/2}=\prod_{i=1,2}(-1)^{M_{\pm u}(\Gamma_i)/2}\eta_I^{M_{\pm}^>(\Gamma_i)/2}=\prod_{i=1,2}\eta_I^{M_{\pm}^<(\Gamma_i)/2}.\label{eq:formula2_temp}
\end{gather}
We see that even-parity SCs are trivial.

Equation~(9) for odd-parity SCs can be rewritten by the number of FSs with positive/negative glide eigenvalue counted with sign,  
\begin{gather}
  \nu_\pm(\pi)=\#\text{FS}_\pm/2\equiv\int_{\Gamma_1}^{\Gamma_2}d\bm{k}\cdot\nabla_{\bm{k}}M_\pm^<(\bm{k})/2.\quad (\text{mod}\ 2)
\end{gather}
This formula is consistent with the results previously obtained for 1D class DIII invariants~\cite{Fu2010_SM,Sato2010_oddparity_SM,Qi2010_DIII_SM}.

\section{Glide topological invariants in the space group $Pnma$}
\label{sec:topo_Pnma}
In this section, we simplify Eqs.~(5) and (9) using the symmetry of the space group $Pnma$.
First, we specify the TRIM $\Gamma_1$ and $\Gamma_2$ for each glide symmetry $\hat{G}_a$ and $\hat{G}_n$.
They are the TRIM where eigenvalues of $\hat{G}_a$ and $\hat{G}_n$, given by $\pm ie^{-ik_x/2}$ and $\pm ie^{-ik_y/2-ik_z/2}$, respectively, take real values. Thus, we have,
\begin{equation}
  \Gamma_1=(k_x=\pi,k_y=0,k_z=\pi)=U,\quad\Gamma_2=(k_x=\pi,k_y=\pi,k_z=\pi)=R,
\end{equation}
for $\hat{G}_a$, and
\begin{equation}
  \Gamma_1=(k_x=\pi,k_y=\pi,k_z=0)=S,\quad\Gamma_2=(k_x=\pi,k_y=0,k_z=\pi)=U,
\end{equation}
for $\hat{G}_n$. In the following, we decompose the irreducible representations of the little group $\mathcal{M}^{\bm{k}}$ at $\bm{k}=U,R,S$ into simultaneous eigenstates of $\hat{I}$ and $\hat{G}=\hat{G}_a,\hat{G}_n$.
In other words, we derive compatibility relations between irreducible representations of the little group $\mathcal{M}^{\bm{k}}$ $(\bm{k}=S,U,R)$ and those of its subgroup $\mathcal{M}\equiv\{\hat{E},\hat{I},\hat{G},\hat{G}\hat{I}\}$.
Then, relations between number of occupied states with certain glide/inversion eigenvalues are obtained, and the formulas for the topological invariants are simplified.
For clarity, we call $\mathcal{M}$ with $\hat{G}=\hat{G}_a$, $\hat{G}_n$ as $\mathcal{M}_a$, $\mathcal{M}_n$.

We list up symmetry operations of $Pnma$ and derive their (anti-)commutation relations for latter use.
$Pnma$ consists of identity, inversion, mirror reflection in $y$ direction, two glide reflections and three screw rotations, in addition to translations. They are represented by,
\begin{subequations}
\begin{gather}
  \hat{E}=\Set{E|\bm{0}},\\
  \hat{I}=\Set{I|\bm{0}},\\
  \hat{G}_a=\Set{M_z|\hat{x}/2+\hat{z}/2},\\
  \hat{M}_y=\Set{M_y|\hat{y}/2},\\
  \hat{G}_n=\Set{M_x|\hat{x}/2+\hat{y}/2+\hat{z}/2},\\
  \hat{S}_z=\Set{C_{2z}|\hat{x}/2+\hat{z}/2},\\
  \hat{S}_y=\Set{C_{2y}|\hat{y}/2},\\
  \hat{S}_x=\Set{C_{2x}|\hat{x}/2+\hat{y}/2+\hat{z}/2}.
\end{gather}
\end{subequations}
Generators of $\mathcal{M}^{\bm{k}}$ $(\bm{k}=U,R,S)$ are $\hat{I},\hat{M}_y$, and $\hat{G}_a$ (with added $\hat{\bar{E}}$ of double group, strictly speaking).
We may also use another set, $\hat{I}$, $\hat{M}_y$, $\hat{G}_n$ and $\hat{\bar{E}}$, for convenience.
The following relations hold,
\begin{subequations}
\begin{gather}
  \hat{G}_a\hat{I}=\Set{E|\hat{x}+\hat{z}}\hat{I}\hat{G}_a,\\
  \hat{G}_n\hat{I}=\Set{E|\hat{x}+\hat{y}+\hat{z}}\hat{I}\hat{G}_n,\\
  \hat{M}_y\hat{I}=\Set{E|\hat{y}}\hat{I}\hat{M}_y,\\
  \hat{M}_y\hat{G}_a=-\hat{G}_a\hat{M}_y,\\
  \hat{M}_y\hat{G}_n=-\Set{E|-\hat{y}}\hat{G}_n\hat{M}_y,
%  \\\hat{G}_a\hat{G}_n=-\Set{E|-\hat{z}+\hat{x}}\hat{G}_n\hat{G}_a,  
\end{gather}
\end{subequations}
where $-1$ corresponds to $\hat{\bar{E}}$.
Note that
\begin{equation}
  [\hat{I},\hat{G}_{a}]=0,\quad\hat{I}^2=\hat{G}_a^2=\hat{E},
\end{equation}
for $U,R$ and
\begin{equation}
  [\hat{I},\hat{G}_n]=0,\quad\hat{I}^2=\hat{G}_n^2=\hat{E},
\end{equation}
for $S,U$. Thus, $\mathcal{M}_{a}$ at $U,R$ and $\mathcal{M}_n$ at $S,U$ are Abelian group and they are isomorphic with each other. Below, we represent these four isomorphic groups as $\mathcal{M}$, for simplicity. (Note that $\mathcal{M}$ closes without $\hat{\bar{E}}$).
There are four 1D representations of $\mathcal{M}$ as listed in \Table{tab:M_IRs}, each of which corresponds to simultaneous eigenstate of $\hat{I}$ and $\hat{G}$.
\begin{table}
  \centering
  \caption{Irreducible representations of $\mathcal{M}$. 
%$(\hat{G}=\hat{G}_a,\,\hat{G}_n)$. %($\mathcal{M}_a$ and $\mathcal{M}_n$ are isomorphic.)
  }
  \begingroup
  \renewcommand{\arraystretch}{1.2}
  \tabcolsep =2.5mm
  \begin{tabular}{c|ccc}\hline\hline
    &$\hat{E}$&$\hat{I}$&$\hat{G}$\\\hline
    $\Gamma_{+g}$&$+1$&$+1$&$+1$\\\hline
    $\Gamma_{-g}$&$+1$&$+1$&$-1$\\\hline
    $\Gamma_{+u}$&$+1$&$-1$&$+1$\\\hline
    $\Gamma_{-u}$&$+1$&$-1$&$-1$\\\hline\hline
  \end{tabular}
  \endgroup
  \label{tab:M_IRs}
\end{table}

\subsection{Irreducible representations at $R$}
\label{subsec:irrep_of_R}
We first analyze irreducible representations at $R$. The following anticommutation relations hold,
\begin{subequations}
\begin{gather}
  \{\hat{M}_y,\hat{I}\}=0,\\
  \{\hat{M}_y,\hat{G}_a\}=0.
%  ,\\\{\hat{G}_n,\hat{I}\}=0,\\
%  \{\hat{G}_n,\hat{G}_a\}=0,\\
%        [\hat{M}_y,\hat{G}_n]=0.
\end{gather}
\end{subequations}
We start from the group $\mathcal{M}_a$, and add $\hat{M}_y$ (and $\hat{\bar{E}}$, strictly speaking) to obtain the little group $\mathcal{M}^{R}$.
Considering the above anticommutation relations, we obtain irreducible representations of $\mathcal{M}^R$ as shown in \Table{tab:IR_R}.
Note that $\hat{M}_y$ flips both inversion and $\hat{G}_a$ eigenvalues, and therefore, $\Gamma_{\pm g}$ and $\Gamma_{\mp u}$ are paired up. Thus, we obtain 2D irreducible representations $E_1$ and $E_2$.
\begin{table}
  \centering
  \begingroup
  \renewcommand{\arraystretch}{1.5}
  \tabcolsep =2.5mm
  \caption{Irreducible representations of $\mathcal{M}^R$ and compatibility relations with those of $\mathcal{M}_a$.}
  \label{tab:IR_R}
  \begin{tabular}{c||c|ccc}\hline\hline
    {irreps. of $\mathcal{M}^R$}&irreps. of $\mathcal{M}_a$&$\hat{I}$&$\hat{G}_a$&$\hat{M}_y$\\\hline
    \multirow{2}{*}{$E_1$}&$\Gamma_{+g}$&$+1$&$+1$&\multirow{2}{*}{$\displaystyle\begin{pmatrix}0&1\\-1&0\end{pmatrix}$}\\\cline{2-4}
    &$\Gamma_{-u}$&$-1$&$-1$&\\\hline
    \multirow{2}{*}{$E_2$}&$\Gamma_{+u}$&$-1$&$+1$&\multirow{2}{*}{$\displaystyle\begin{pmatrix}0&1\\-1&0\end{pmatrix}$}\\\cline{2-4}
    &$\Gamma_{-g}$&$+1$&$-1$&\\\hline\hline
  \end{tabular}
  \endgroup
\end{table}
When we add time-reversal symmetry, each representation is simply doubled by Kramers degeneracy, to give irreducible co-representations $E_1'=2E_1$ and $E_2'=2E_2$ (Type (b) of Wigner's theorem~\cite{Bradley_SM}). For example, representation space of $E_1'$ is spanned by bases of $E_1$ and their Kramers partners,
%Actually, all states in $E_1'$ ($E_2'$) are orthogonal.
%Let us consider $\ket{\Gamma_{+g}}$ and $\ket{\Gamma_{-u}}$ in $E_1$.
%The bases of $E_1'$ are given by
\begin{equation}
  \ket{\Gamma_{+g}},\   \ket{\Gamma_{-u}},\   \hat{\Theta}\ket{\Gamma_{+g}},\   \hat{\Theta}\ket{\Gamma_{-u}}.
\end{equation}
%We can easily see that they are orthogonal to each other. 
%Actually,
The latter two bases again form $E_1$, since $\hat{\Theta}$ preserves inversion and glide eigenvalues.
Note that $\ket{\Gamma_{+g}}$ are orthogonal to $\ket{\Gamma_{-u}}$ and $\hat{\Theta}\ket{\Gamma_{-u}}$ with different eigenvalues, and also to $\hat{\Theta}\ket{\Gamma_{+g}}$ owing to Kramers theorem.
%Note that $\hat{\Theta}$ preserves inversion and glide eigenvalues.
In this way, all the states are orthogonal to each other.

Table~\ref{tab:IR_R} tells us that $\ket{\Gamma_{+g}}$ always accompanies $\ket{\Gamma_{-u}}$ at R, for instance.
Thus, we obtain the following relations for the number of occupied states,
\begin{subequations}
\begin{gather}
  M^<_{\pm g}(R)=M^<_{\mp u}(R),\label{eq:formula_Ra}\\
  M^<_{+u}(R)+M^<_{-g}(R)=2M^<_{+u}(R),\label{eq:formula_Rb}\\
  %M_u^<(R)=M_{+u}^<(R)+M_{-u}^<(R)=M_{-g}^<(R)+M_{+g}^<(R)=M_g^<(R)=M^<(R)/2,\label{eq:formula_R2}\\
  M_u^<(R)=M_g^<(R)=M^<(R)/2,\label{eq:formula_Rc}\\
  %  M_+^<(R)=M_{+u}^<(R)+M_{+g}^<(R)=M_{+u}^<(R)+M_{-u}^<(R)=M_u^<(R)=M^<(R)/2.\label{eq:formula_R3}
  M_+^<(R)=M_-^<(R)=M^<(R)/2.\label{eq:formula_Rd}
\end{gather}
\label{eq:formula_R}
\end{subequations}
We also conclude
\begin{equation}
  M^<_{\pm g}(R)\in 2\mathbb{Z},\quad M^<_{\pm u}(R)\in 2\mathbb{Z},
  \label{eq:KP_num}
\end{equation}
from the results of irreducible co-representations.
(Equation~\eqref{eq:KP_num} reproduces the general result $M_{\pm u,g}^<(\Gamma_i)\in2\mathbb{Z}$ in the previous section.)

\subsection{Irreducible representations at $S$}
\label{subsec:irrep_of_S}
Next, we analyze irreducible representations at $S$.
(Anti-)commutation relations are given by,
\begin{subequations}
\begin{gather}
  \{\hat{M}_y,\hat{I}\}=0,\\
  [\hat{M}_y,\hat{G}_n]=0,
\end{gather}
\end{subequations}
and we obtain compatibility relations between $\mathcal{M}^S$ and $\mathcal{M}_n$ in \Table{tab:IR_S}.
\begin{table}
  \centering
  \begingroup
  \renewcommand{\arraystretch}{1.5}
  \tabcolsep =2.5mm
  \caption{Irreducible representations of $\mathcal{M}^S$ and compatible relations with those of $\mathcal{M}_n$.}
  \label{tab:IR_S}
  \begin{tabular}{c||c|ccc}\hline\hline
    {irreps. of $\mathcal{M}^S$}&irreps. of $\mathcal{M}_n$&$\hat{I}$&$\hat{G}_n$&$\hat{M}_y$\\\hline
    \multirow{2}{*}{$E_+$}&$\Gamma_{+g}$&$+1$&$+1$&\multirow{2}{*}{$\displaystyle\begin{pmatrix}0&1\\-1&0\end{pmatrix}$}\\\cline{2-4}
    &$\Gamma_{+u}$&$-1$&$+1$&\\\hline
    \multirow{2}{*}{$E_-$}&$\Gamma_{-g}$&$+1$&$-1$&\multirow{2}{*}{$\displaystyle\begin{pmatrix}0&1\\-1&0\end{pmatrix}$}\\\cline{2-4}
    &$\Gamma_{-u}$&$-1$&$-1$&\\\hline\hline
  \end{tabular}
  \endgroup
\end{table}
Irreducible co-representations are given by $E_\pm'=2E_\pm$.
Thus, we obtain
\begin{subequations}
\begin{gather}
    M^<_{\pm g}(S)=M^<_{\pm u}(S),\label{eq:formula_Sa}\\
    %  M^<_{+u}(S)+M^<_{-g}(S)=M^<_{+u}(S)+M^<_{-u}(S)=M^<_{u}(S),\label{eq:formula_S1}\\
    M^<_{+u}(S)+M^<_{-g}(S)=M^<_{u}(S),\label{eq:formula_Sb}\\
    %M_u^<(S)=M_{+u}^<(S)+M_{-u}^<(S)=M_{+g}^<(S)+M_{-g}^<(S)=M_g^<(S)=M^<(S)/2,\\\label{eq:formula_S2}
    M_u^<(S)=M_g^<(S)=M^<(S)/2,\label{eq:formula_Sc}\\
    M^<_{\pm g}(S)\in 2\mathbb{Z},\quad M^<_{\pm u}(S)\in 2\mathbb{Z}\label{eq:formula_Sd}.
\end{gather}
\label{eq:formula_S}
\end{subequations}
\subsection{Irreducible representations at $U$}
\label{subsec:irrep_of_U}
Finally, we analyze irreducible representations at $U$.
(Anti-)commutation relations are given by
\begin{subequations}
\begin{gather}
  [\hat{M}_y,\hat{I}]=0,\\
  \{\hat{M}_y,\hat{G}_n\}=0,\\
  \{\hat{M}_y,\hat{G}_a\}=0,
\end{gather}
\end{subequations}
and we obtain compatibility relations between $\mathcal{M}^U$ and $\mathcal{M}_{a}\,(\mathcal{M}_n)$ in \Table{tab:IR_Ua} (\Table{tab:IR_Un}).
An identical result is obtained for both $\mathcal{M}_a$ and $\mathcal{M}_n$.
Irreducible co-representations are given by $E_g'=2E_g$ and $E_u'=2E_u$.
Thus, we obtain for both $\hat{G}_a$ and $\hat{G}_n$,
\begin{subequations}
\begin{gather}
    M^<_{+g}(U)=M^<_{-g}(U),\quad M^<_{+u}(U)=M^<_{-u}(U),\label{eq:formula_Ua}\\
  %% M^<_{+u}(U)+M^<_{-g}(U)=M^<_{+u}(U)+M^<_{+g}(U)=M^<_{+}(U),\label{eq:formula_U1}\\
  %% M_u^<(U)=M_{+u}^<(U)+M_{-u}^<(U)=2M_{+u}^<(U),\label{eq:formula_U2}\\
    %% M^<_{+}(U)=M^<_{+u}(U)+M^<_{+g}(U)=M^<_{-u}(U)+M^<_{-g}(U)=M^<_{-}(U)=M^<(U)/2,\\\label{eq:formula_U3}
    M^<_{+u}(U)+M^<_{-g}(U)=M^<_{+}(U),\label{eq:formula_Ub}\\
    M_u^<(U)=2M_{+u}^<(U),\label{eq:formula_Uc}\\
    M^<_{+}(U)=M^<_{-}(U)=M^<(U)/2,\label{eq:formula_Ud}\\
  M^<_{\pm g}(U)\in 2\mathbb{Z},\quad M^<_{\pm u}(U)\in 2\mathbb{Z}.\label{eq:formula_Ue}
\end{gather}
\label{eq:formula_U}
\end{subequations}

\begin{table}
  \centering
  \begingroup
  \renewcommand{\arraystretch}{1.5}
  \tabcolsep =2.5mm
  \caption{Irreducible representations of $\mathcal{M}^U$ and compatible relations with those of $\mathcal{M}_a$.}
  \label{tab:IR_Ua}
  \begin{tabular}{c||c|ccc}\hline\hline
    {irreps. of $\mathcal{M}^U$}&irreps. of $\mathcal{M}_a$&$\hat{I}$&$\hat{G}_a$&$\hat{M}_y$\\\hline
    \multirow{2}{*}{$E_g$}&$\Gamma_{+g}$&$+1$&$+1$&\multirow{2}{*}{$\displaystyle\begin{pmatrix}0&1\\-1&0\end{pmatrix}$}\\\cline{2-4}
    &$\Gamma_{-g}$&$+1$&$-1$&\\\hline
    \multirow{2}{*}{$E_u$}&$\Gamma_{+u}$&$-1$&$+1$&\multirow{2}{*}{$\displaystyle\begin{pmatrix}0&1\\-1&0\end{pmatrix}$}\\\cline{2-4}
    &$\Gamma_{-u}$&$-1$&$-1$&\\\hline\hline
  \end{tabular}
  \endgroup
\end{table}
\begin{table}
  \centering
  \begingroup
  \renewcommand{\arraystretch}{1.5}
  \tabcolsep =2.5mm
  \caption{Irreducible representations of $\mathcal{M}^U$ and compatibility relations with those of $\mathcal{M}_n$. The same results have been obtained for $\mathcal{M}_a$ in \Table{tab:IR_Ua}.}
  \label{tab:IR_Un}
  \begin{tabular}{c||c|ccc}\hline\hline
    {irreps. of $\mathcal{M}^U$}&irreps. of $\mathcal{M}_n$&$\hat{I}$&$\hat{G}_n$&$\hat{M}_y$\\\hline
    \multirow{2}{*}{$E_g$}&$\Gamma_{+g}$&$+1$&$+1$&\multirow{2}{*}{$\displaystyle\begin{pmatrix}0&1\\-1&0\end{pmatrix}$}\\\cline{2-4}
    &$\Gamma_{-g}$&$+1$&$-1$&\\\hline
    \multirow{2}{*}{$E_u$}&$\Gamma_{+u}$&$-1$&$+1$&\multirow{2}{*}{$\displaystyle\begin{pmatrix}0&1\\-1&0\end{pmatrix}$}\\\cline{2-4}
    &$\Gamma_{-u}$&$-1$&$-1$&\\\hline\hline
  \end{tabular}
  \endgroup
\end{table}

\subsection{Topological invariants for $\hat{G}_a$ of $Pnma$}
Using the results obtained in this section, we simplify the formulas of topological invariants, Eq.~(5) and Eq.~(9), in the case of $\hat{G}_a$.
First we analyze glide-odd $\mathbb{Z}_4$ invariant.
Applying Eqs.~\eqref{eq:formula_Rb}, \eqref{eq:formula_Rc}, \eqref{eq:formula_Ub} and \eqref{eq:formula_Uc} to Eq.~(5), we obtain
\begin{equation}
  \theta_4^{(a)}(\pi)=\begin{cases}M_+^<(U)\\M^<(R)/2\end{cases},\quad (\text{mod}\ 4)\label{eq:Z4temp_a}
\end{equation}
for inversion-odd and -even superconductivity, respectively.
From \Eqs{eq:formula_Rc}{eq:formula_Ud}, the first line of \Eq{eq:Z4temp_a} is rewritten by
%\begin{subequations}
\begin{gather}
  M_+^<(U)=M^<(U)/2=\#\text{FS}_{R\to U}/2+M^<(R)/2=\#\text{FS}_{R\to U}/2+M^<_u(R)=\#\text{FS}_{R\to U}/2+2M^<_{u-}(R)+\Delta M(R).\label{eq:Z4temp_b}
%  M^<(R)/2=M^<_u(R)=2M^<_{u-}(R)+\Delta M(R).
\end{gather}
%\end{subequations}
For odd-parity superconductivity, this equation modulo four gives the first row of Table~I.
%% The second line of \Eq{eq:Z4temp_a} may be rewritten as
%% \begin{equation}
%%   M^<(R)/2=M^<(U)/2+\#\text{FS}_{U\to R}/2=M_+^<(U)+\#\text{FS}_{U\to R}/2,
%% \end{equation}
%% by using \Eq{eq:formula_U3}.
{Even-parity superconductivity ($B_{1g}$) may also become $\mathbb{Z}_4$ nontrivial TNCS, when excitation is gapful and the filling condition $M^<(R)\in 4(2\mathbb{Z}+1)$ is satisfied.}

Next, we analyze glide-even $\mathbb{Z}_2$ invariant.
Applying \Eqs{eq:formula_R}{eq:formula_U} to Eq.~(9) for odd-parity superconductivity, we have
\begin{equation}
  \nu_\pm^{(a)}(\pi)=M_\pm^<(R)/2+M_\pm^<(U)/2=M_\pm^<(R)+(M^<(U)-M^<(R))/4=\#\text{FS}_{R\to U}/4\quad(\text{mod}\ 2).
\end{equation}
This gives the second line of Table~I.

\subsection{Topological invariants for $\hat{G}_n$ of $Pnma$}
We simplify Eqs.~(5) and (9) for the case of $\hat{G}_n$.
First we consider $\mathbb{Z}_4$ invariant of glide-odd superconductivity.
Using Eqs.~\eqref{eq:formula_Sb}, \eqref{eq:formula_Sc}, \eqref{eq:formula_Ub}, \eqref{eq:formula_Uc} and \eqref{eq:formula_Ud}, we obtain
\begin{equation}
  \theta_4^{(n)}(\pi)=\begin{cases}M^<(S)/2+M^<(U)/2\\
  M^<(S)/2\end{cases},\quad (\text{mod}\ 4)
\end{equation}
for inversion-odd and -even superconductivity, respectively.
The first line can be expressed as
\begin{equation}
  M^<(S)/2+M^<(U)/2=\#\text{FS}_{S\to U}/2+M^<(S).
\end{equation}
This equation modulo four gives the third row of Table~I, since $M^<(S)=2M_u^<(S)\in4\mathbb{Z}$ by \Eq{eq:formula_Sc}.
%% For even parity superconductivity,
%% \begin{equation}
%%   M^<(S)/2=\#\text{FS}_{S\to U}/2+M^<(U)/2.
%% \end{equation}
Even-parity superconductivity ($B_{3g}$) may also become nontrivial $\mathbb{Z}_4$ TNCS when the filling condition $M^<(S)\in 4(2\mathbb{Z}+1)$ is realized.

The $\mathbb{Z}_2$ topological invariants of glide-even odd-parity superconductivity are evaluated as
\begin{equation}
\nu_\pm^{(n)}(\pi)=M^<_\pm(S)/2+M^<(U)/4=\#\text{FS}_{S\to U}/4+M^<_\pm(S)/2+M^<(S)/4.
\end{equation}
Here, the latter two terms are rewritten as
\begin{equation}
  M^<_\pm(S)/2+M^<(S)/4=M^<_\pm(S)+(M^<_\mp(S)-M^<_\pm(S))/4=\Delta M(S)/2,\quad(\text{mod}\ 2)
\end{equation}
where we used $M^<_\pm(S)=M^<_{\pm u}(S)+M^<_{\pm g}(S)=2M^<_{\pm u}(S)\in2\mathbb{Z}$.
Thus, we obtain the last row of Table~I.

\subsection{$\theta_4^{(a)}(\pi)$ for $S$-Fermi surfaces of Ref.~\cite{Samsel_SM}}
In this section, we evaluate $\theta_4^{(a)}(\pi)$ of possible $S$-Fermi surfaces.
As noted in the main text, the $S$-Fermi surfaces obtained in a band structure calculation~\cite{Samsel_SM} are interesting situation,
because for $\theta_4^{(a)}(\pi)$ the correction $\Delta M(R)$ due to large SOC plays an important role.
Let us focus on Fig. 2(a) of Ref.~\cite{Samsel_SM}.
At the $R$ point, the splitting of bands $251$ and $255$, both of which are four-fold degenerate, is so large that the band $255$ does not cross the Fermi energy on the $U$-$R$ line, as opposed to the results of Refs.~\cite{Fujimori2015_SM,Fujimori2016_SM}.
In the SU(2) limit, the bands at $R$ are eight-fold degenerate, and therefore, the bands $251$ and $255$ are paired up in this limit.
Thus, this is the situation we referred to as effective SOC is large, and $\Delta M(R)$ is expected to have finite contribution.

%A direct way to evaluate $\theta_4^{(a)}(\pi)$ in application to band structure calculations is to use 
An alternative expression of $\theta_4^{(a)}(\pi)$ [see Eqs.~\eqref{eq:Z4temp_a} and \eqref{eq:Z4temp_b}] enables a direct evaluation in application to band structure calculations:
\begin{equation}
\theta_4^{(a)}(\pi)=M^<(U)/2.
\end{equation}
Since $M^<(U)=252$ in Ref.~\cite{Samsel_SM}, we readily obtain $\theta_4^{(a)}(\pi)=2$.
Thus, $\mathbb{Z}_4$ invariants are also nontrivial for $S$-Fermi surfaces of Ref.~\cite{Samsel_SM}.

\section{Proof of Equation~(\ref{eq:switcheigen})}
\label{sec:proof_Reps}
%We here prove the relation \Eq{eq:switcheigen} in order to complete the derivation of Eqs.~(5) and (9).
In this section, we first show that $\tilde{U}(\bm{k})$ can be taken $\bm{k}$-independent when all the atoms do not have inversion as their site symmetry.
Then, we prove \Eq{eq:switcheigen} when $\tilde{U}(\bm{k})$ is taken $\bm{k}$-independent, to complete the derivation of Eqs.~(5) and (9).

%@@@@@@@First, we consider the character of $\hat{I}$.
%Note that we can always choose $\tilde{U}^I(\bm{k})$ so as not to depend on $\bm{k}$.
Let us suppose that the site symmetry of all the atoms is noncentrosymmetric.
%consider the situation where all the atoms don't have inversion as their site symmetry.
We take the origin of the lattice to be an inversion center, and write the set of all the atoms within an unit cell by $B=\Set{\bm{r}_n\in(\text{unit cell})}$.
Note that we can choose the unit cell so as to satisfy $B\equiv\Set{\bm{r}_n,-\bm{r}_n|\bm{r}_n\in B/2}$, where $B/2$ is a subset of $B$ containing half the atoms in the unit cell.
This can be understood by an inductive way:
Let us start by taking an arbitrary atom $\bm{r}_1\in B$. Then, $\hat{I}\bm{r}_1=-\bm{r}_1$ must be translationally inequivalent to $\bm{r}_1$, since the site symmetry lacks $\hat{I}$. We name this atom as $\bm{r}_2\equiv-\bm{r}_1$. Next, we take another atom, if any, which is translationally inequivalent to both $\bm{r}_1$ and $\bm{r}_2$, and name it $\bm{r}_3$. We obtain $\bm{r}_4\equiv-\bm{r}_3$, which is translationally inequivalent to all the $\bm{r}_1$, $\bm{r}_2$, and $\bm{r}_3$. Thus, we obtain $B/2=\Set{\bm{r}_1,\,\bm{r}_3,\,\cdots}$, by repeating the same procedure.
Note that $\Delta\bm{R}^I_n=0$ holds for all $\bm{r}_n\in B$ for this choice of $B$. Thus, $\tilde{U}^I(\bm{k})=\tilde{U}^I(\bm{0})$ is satisfied.

{In the following, we show \Eq{eq:switcheigen} by assuming $\tilde{U}^I(\bm{k})=\tilde{U}^I(\bm{0})$.
For this purpose, we adopt the tight-binding representation of the space group operations.
%at $\bm{k}=\Gamma_i\ (i=1,2)$. 
%For clarity, here we briefly repeat the discussion of \Sec{sec:topo_Pnma} in the tight-binding basis. %, before making detailed discussion.
Let us consider the TRIM $\Gamma_i$ $(i=1,2)$.
At these points, the representation matrices satisfy $[\tilde{U}^{I}(\Gamma_i),\tilde{U}^{G}(\Gamma_i)]=0$ and $\tilde{U}^I(\Gamma_i)^2=\tilde{U}^{G}(\Gamma_i)^2=\tilde{U}^E(\Gamma_i)$. Therefore, $\gamma(\Gamma_i)\equiv\Set{\tilde{U}^{E}(\Gamma_i),\,\tilde{U}^{I}(\Gamma_i),\,\tilde{U}^{{G}}(\Gamma_i),\,\tilde{U}^{{G}I}(\Gamma_i)}$ %$(i=1,2)$
is a representation of the Abelian group $\mathcal{M}$, whose
%, the Abelian subgroup of the little group $\mathcal{M}^{\Gamma_i}$.
irreducible representations have been summarized in \Table{tab:M_IRs}.
%Here, $\gamma(\Gamma_i)$ is the tight-binding representation of $\mathcal{M}^a$ and $\mathcal{M}^n$ for $G=G_a$ and $G=G_n$, respectively.
%As $\gamma(\Gamma_1)$ and $\gamma(\Gamma_2)$ are specified by a common group multiplication table, they are isomorphic to $\mathcal{M}$@@. 
%Thus, the argument of $\mathcal{M}(\Gamma_i)$ is omitted below. 
It can be proven that the two representations $\gamma(\Gamma_1)$ and $\gamma(\Gamma_2)$ are identical when $\tilde{U}^I(\bm{k})=\tilde{U}^I(\bm{0})$.
Then, it follows that the irreducible decomposition of $\gamma(\Gamma_i)$
\begin{equation}
  \gamma(\Gamma_i)=M_{+u}(\Gamma_i)\Gamma_{+u}+M_{-u}(\Gamma_i)\Gamma_{-u}+M_{+g}(\Gamma_i)\Gamma_{+g}+M_{-g}(\Gamma_i)\Gamma_{-g},
\end{equation}
are also identical, and \Eq{eq:switcheigen} is proved.}

We prove the equivalence of $\gamma(\Gamma_1)$ and $\gamma(\Gamma_2)$ below. It is sufficient to examine the characters of $\tilde{U}^E(\Gamma_i)$, $\tilde{U}^{I}(\Gamma_i)$, $\tilde{U}^{G}(\Gamma_i)$, and $\tilde{U}^{GI}(\Gamma_i)$.
First, we obtain
\begin{gather}
  \chi_{\gamma(\Gamma_1)}(E)=\Tr\tilde{U}^E(\bm{0})=\chi_{\gamma(\Gamma_2)}(E),\\
  \chi_{\gamma(\Gamma_1)}(I)=\Tr\tilde{U}^I(\bm{0})=\chi_{\gamma(\Gamma_2)}(I),
\end{gather}
since $\tilde{U}^E(\bm{k})$ and $\tilde{U}^I(\bm{k})$ are $\bm{k}$-independent.
Next, we consider the characters of $\hat{G}$ and $\hat{S}\equiv\hat{G}\hat{I}$.
Note that $\hat{S}$ is the screw symmetry in our setup.
Since they are nonsymmorphic symmetries, the permutation matrices $D({G})$ and $D({S})$ are off-diagonal.
Actually, the assumption
\begin{equation}
  \hat{g}\bm{r}_n=^\exists\bm{R}_0+\bm{r}_n,
\quad (\hat{g}=\Set{p_c|\bm{a}/2+\hat{c}/2}=\hat{G},\,\hat{S}),%\quad p_c=M_c,\,C_{2c}),
\end{equation}
with $\bm{R}_0$ being a lattice translation vector,
leads to a contradiction:
\begin{subequations}
\begin{gather}
  \bm{R}_0+\bm{r}_n=\hat{g}\bm{r}_n=\Set{p_c|\bm{a}/2+\hat{c}/2}\bm{r}_n=p_c\bm{r}_n+\bm{a}/2+\hat{c}/2,\\
  \bm{a}/2+\hat{c}/2=(\bm{r}_n-p_c\bm{r}_n)+\bm{R}_0.
\end{gather}
\end{subequations}
The $a$-component ($c$-component) is fractional in the left-hand side, although it is integer in the right-hand side for $\hat{G}$ ($\hat{S}$).
%The lefthand side of the second line is fractional, while the righthand side is integral, in terms of $\bm{a}$ or $\hat{c}$ for $\hat{G}$ or $\hat{S}$, respectively.
Therefore, we have
\begin{equation}
  \chi_{\gamma(\Gamma_1)}(\hat{g})=0=\chi_{\gamma(\Gamma_2)}(\hat{g})\quad (\hat{g}=\hat{G},\,\hat{S}).\\
\end{equation}
Thus, $\gamma(\Gamma_1)$ and $\gamma(\Gamma_2)$ are identical, and \Eq{eq:switcheigen} holds.
% the representation of the Abelian group $\mathcal{M}$ is equivalent between $\Gamma_1$ and $\Gamma_2$, and \Eq{eq:switcheigen} holds.

\section{Extension of the formulas to general inversion operation}
%\section{Extension of the formulas to the case of $\bm{k}$-dependent inversion}
\label{sec:extension}
In this section, we show that the formulas derived in the previous sections remain valid even when $\tilde{U}^{{I}}(\bm{k})$ is $\bm{k}$-dependent.
The derivation in the previous sections using the periodic basis \Eq{eq:periodic} can not be directly applied in such a case. This is because $\bm{k}$-derivative of $\tilde{U}^{I}(\bm{k})$ makes finite contribution.
To solve the difficulty, we rewrite \Eq{eq:def_Z4_SM} in the L\"owdin basis \Eq{eq:Lowdin}, in which inversion operation is $\bm{k}$-independent.
%We clarify the
\subsection{Glide-odd $\mathbb{Z}_4$ invariant in the L\"owdin basis}
First, we rewrite the $\mathbb{Z}_4$ invariant in the L\"owdin basis, without assuming the presence of the inversion symmetry.
In this section, we distinguish the state vectors, Berry connection, and Berry curvature in the periodic basis from those in the L\"owdin basis by their subscript and argument, e.g. $A_{\bm{k}}$ represents Berry connection in the periodic basis, while $A(\bm{k})$ represents Berry connection in the L\"owdin basis.
Here, we rewrite $F_+(\bm{k})$ defined in \Eq{eq:Berrycurvature} as $F_{\bm{k}+}$, in order to emphasize it is defined via periodic basis. \textit{Instead, we use $F_+(\bm{k})$ to represent the Berry curvature in the L\"owdin basis.}
%% We define the list of occupied states in the periodic basis by
%% \begin{equation}
%%   \Psi_{\bm{k}+}=(\ket{\psi_{1\bm{k}}}_+, \ket{\psi_{2\bm{k}}}_+, \cdots, \ket{\psi_{2N\bm{k}}}_+), 
%% \end{equation}
%% where the subscript $+$ represents positive glide eigenvalues,
%% \begin{equation}
%%   \tilde{U}^G_{\BdG}(\bm{k})\Psi_{\bm{k}+}=\Psi_{\bm{k}+}(+ie^{-ik_a/2}).
%% \end{equation}
The list of occupied states in the L\"owdin basis is given by,
\begin{equation}
  \Psi_+(\bm{k})=V(\bm{k})^\dagger\Psi_{\bm{k}+},\quad \Psi_+(\bm{k}+\bm{G})=V(\bm{G})^\dagger\Psi_+(\bm{k}).
\end{equation}
Here, $\Psi_+(\bm{k})$ satisfy the eigen-equation of the glide operator,
\begin{equation}
  {U}^G_{\BdG}(\bm{k})\Psi_{+}(\bm{k})=\Psi_+(\bm{k}+\bm{G}_c)\cdot ie^{-ik_a/2},\quad \bm{G}_c\equiv M_c\bm{k}-\bm{k}=-2\Gamma_c\hat{c},
\end{equation}
and equivalently,
\begin{subequations}
\begin{gather}
  \bar{U}^G_{\BdG}(\bm{k})\Psi_{+}(\bm{k})=\Psi_+(\bm{k})\cdot ie^{-ik_a/2},\\
  \bar{U}^G_\BdG(\bm{k})\equiv V(\bm{G}_c)U^G_\BdG(\bm{k})=V(\bm{k})^\dagger\tilde{U}^G_\BdG(\bm{k})V(\bm{k}).
\end{gather}
\end{subequations}
Hereafter, we denote $(k_a,k_b,\Gamma_c)$ as $\bm{k}$.

The Berry connections in each basis are defined by,
\begin{subequations}
\begin{gather}
  \bm{A}_{\bm{k}+}\equiv\Tr\left[\Psi^\dagger_{\bm{k}+}\partial_{\bm{k}}\Psi_{\bm{k}+}\right],\\
  \bm{A}_{+}(\bm{k})\equiv\Tr\left[\Psi^\dagger_{+}(\bm{k})\partial_{\bm{k}}\Psi_{+}(\bm{k})\right],
\end{gather}
\end{subequations}
which are related to each other by
\begin{subequations}
\begin{gather}
  \bm{A}_+(\bm{k})=\bm{A}_{\bm{k}+}-i\bm{r}_{+}^<(\bm{k}),\\
  i\bm{r}_+^<(\bm{k})\equiv i\Tr\left[\Psi_{\bm{k}+}^\dagger \hat{\bm{r}}\Psi_{\bm{k}+}\right],\quad (\hat{\bm{r}})_{nn'}\equiv -i(V(\bm{k})^\dagger\partial_{\bm{k}}V(\bm{k}))_{nn'}=\bm{r}_n\delta_{nn'}.
\end{gather}
\end{subequations}
Accordingly,
\begin{subequations}
  \begin{gather}
    F_+(\bm{k})=F_{\bm{k}+}-i\hat{c}\cdot\nabla_{\bm{k}}\times\bm{r}^<_+(\bm{k}),
  \end{gather}
\end{subequations}
holds for the Berry curvature
%\begin{equation}
  $F_{\bm{k}+}=\epsilon_{ij}\partial_i(\bm{A}_{\bm{k}+})_j,$ and $F_{+}(\bm{k})=\epsilon_{ij}\partial_i(\bm{A}_{+}(\bm{k}))_j.$
%\end{equation}

The $\mathbb{Z}_4$ invariant $\theta_4(\Gamma_c)$ is rewritten as \cite{Fu-Kane2006_SM}
\begin{equation}
  \theta_4(\Gamma_c)=2\left\{\int_{0}^\pi\frac{dk_b}{\pi i}\left.\bm{\beta}\cdot\bm{A}_{\bm{k}+}\right|_{k_a=\pi}+\frac{1}{\pi i}\Log\left[\frac{\Pf[w_{\Gamma_2+}]}{\Pf[w_{\Gamma_1+}]}\right]\right\}-\int_{0\le k_a\le\pi}\frac{d^2k}{\pi i}F_{\bm{k}+},\label{eq:def_Z4_SM2}
%\bm{k}&=(k_a,\,k_b,\,k_c=\Gamma_c).
\end{equation}
where
%\begin{equation}
  $w_{\bm{k}+}\equiv\Psi_{-\bm{k}+}^\dagger\Theta\Psi_{\bm{k}+}.$
%\end{equation}
When we define similar quantity $\theta^L(\Gamma_c)$ in the L\"owdin basis as
\begin{subequations}
\begin{gather}
  \theta^L(\Gamma_c)=2\left\{\int_{0}^\pi\frac{dk_b}{\pi i}\left.\bm{\beta}\cdot\bm{A}_{+}(\bm{k})\right|_{k_a=\pi}+\frac{1}{\pi i}\Log\left[\frac{\Pf[w_{+}(\Gamma_2)]}{\Pf[w_{+}(\Gamma_1)]}\right]\right\}-\int_{0\le k_a\le\pi}\frac{d^2k}{\pi i}F_{+}(\bm{k}),\label{eq:def_thetaL_SM}\\
  w_{+}(\bm{k})\equiv\Psi_{+}(-\bm{k})^\dagger\Theta\Psi_{+}(\bm{k})=w_{\bm{k}+},
\end{gather}
\end{subequations}
the $\mathbb{Z}_4$ invariant recasts into
\begin{equation}
  \theta_4(\Gamma_c)=\theta^L(\Gamma_c)+\delta r_b,
  \end{equation}
where $\delta r_b$ is given by
\begin{equation}
  \delta r_b\equiv2\int_0^\pi\frac{dk_b}{\pi}\left.\bm{\beta}\cdot\bm{r}_+^<(\bm{k})\right|_{k_a=\pi}
   -\int_{0\le k_a\le\pi}\frac{d^2k}{\pi}\hat{c}\cdot\nabla_{\bm{k}}\times\bm{r}^<_+(\bm{k}).\label{eq:deltar}
\end{equation}
This term gives the correction to $\theta^L(\Gamma_c)$ purely from atomic positions within an unit cell, as we see below.

Let us simplify the expression of $\delta{r}_b$.
Note that
\begin{equation}
  \Psi_{\bm{k}'+}\Psi_{\bm{k}'+}^\dagger=\Theta\Psi_{-\bm{k}'+}(\Theta\Psi_{-\bm{k}'+})^\dagger,
\end{equation}
since $\Theta$ preserves glide eigenvalues on the line $C_{\mathrm{AII}}(\Gamma_c)=\Set{\bm{k}'|\bm{k}'\equiv(\pi,k_b,\Gamma_c)}$.
It follows that $\bm{r}_{+}^<(\bm{k}')=\bm{r}_{+}^<(-\bm{k}')$.
Thus, the first term in \Eq{eq:deltar} can be rewritten as
\begin{equation}
  2\int_0^\pi\frac{dk_b}{\pi}\bm{\beta}\cdot\bm{r}_+^<(\bm{k}')=\int_{-\pi}^\pi\frac{dk_b}{\pi}\bm{\beta}\cdot\bm{r}_+^<(\bm{k}').
\end{equation}
By using Stokes' theorem, we obtain
\begin{equation}
  \delta r_b=\int_{-\pi}^\pi\frac{dk_b}{\pi}\bm{\beta}\cdot\bm{r}_+^<(\bm{k}''),\quad \bm{k}''\equiv(0,k_b,\Gamma_c).
\end{equation}
Owing to the particle-hole symmetry, which preserves glide eigenvalues on the line $\Set{\bm{k}''}$, we have
\begin{equation}
  \Psi_{\bm{k}''+}\Psi_{\bm{k}''+}^\dagger+C\Psi_{-\bm{k}''+}(C\Psi_{-\bm{k}''+})^\dagger=P_+(\bm{k}'').
\end{equation}
Here, $P_+(\bm{k})\equiv(\tilde{U}^G_\BdG(\bm{k})+ie^{-ik_a/2})/(2ie^{-ik_a/2})$ is the projection operator onto positive glide eigen-space. It follows that
\begin{subequations}
\begin{align}
  \bm{r}_{+}^<(\bm{k}'')&=\Tr\left[\Psi_{\bm{k}''+}\Psi_{\bm{k}''+}^\dagger\hat{\bm{r}}\right]\\
  &=-\Tr\left[C\Psi_{-\bm{k}''+}(C\Psi_{-\bm{k}''+})^\dagger\hat{\bm{r}}\right]+\Tr\left[P_+(\bm{k})\hat{\bm{r}}\right]\\
  &=-\bm{r}_+^<(-\bm{k}'')+\Tr\left[P_+(\bm{k}'')\hat{\bm{r}}\right].
\end{align}
\end{subequations}
Furthermore,
\begin{equation}
  \Tr\left[P_+(\bm{k})\hat{\bm{r}}\right]=\frac{1}{2}\Tr[\hat{\bm{r}}]=2n_{\mathrm{orb}}\sum_n\bm{r}_n,
  \label{eq:tr_temp}
\end{equation}
since $\Tr[\tilde{U}^G_\BdG(\bm{k})\hat{\bm{r}}]\propto\Tr[s_c]=0$, with $s_c$ being the Pauli matrix of the Kramers degrees of freedom. Note that $\Tr[\cdot]$ in \Eq{eq:tr_temp} is taken in the space spanned by Nambu, Kramers, sublattices, and local non-Kramers (that is, orbital) degrees of freedom whose number is $n_{\mathrm{orb}}\ge1$.
Thus, we obtain the final expression,
\begin{equation}
  \delta r_b=\int_{-\pi}^\pi\frac{dk_b}{2\pi}\bm{\beta}\cdot\left[\bm{r}_+^<(\bm{k}'')+\bm{r}_+^<(-\bm{k}'')\right]=2n_{\mathrm{orb}}\bm{\beta}\cdot\sum_n\bm{r}_n.
\end{equation}
\subsection{Glide-odd $\mathbb{Z}_4$ invariant in the presence of ${k}$-dependent inversion symmetry}
Here, we evaluate the $\mathbb{Z}_4$ invariant on the ZF $\theta_4(\pi)=\theta^L(\pi)+\delta r_b$ in the presence of inversion symmetry.
A discussion parallel to $\bm{k}$-independent inversion operator can be used to evaluate $\theta^L(\pi)$, since inversion operation is $\bm{k}$-independent in the L\"owdin basis when we place the origin at an inversion center:
\begin{equation}
  U^I_\BdG\Psi_+(\bm{k})=\Psi_+(-\bm{k})U_\chi^I(\bm{k}),
\end{equation}
where $U_\chi^I(\bm{k})$ is a gauge-transformation matrix, and we have
\begin{equation}
  U^I_\BdG\Psi_+(\Gamma_i)=\Psi_+(-\Gamma_i)\hat{\zeta},
\end{equation}
where $\hat{\zeta}$ is a diagonal matrix with components $\pm1$.
It should be noticed that $\hat{\zeta}$ represents the eigenvalues of $\tilde{U}^I_\BdG(\Gamma_i)$, not of ${U}^I_\BdG$,
since 
\begin{equation}
  \bar{U}^I_\BdG(\Gamma_i)\Psi_+(\Gamma_i)=\Psi_+(\Gamma_i)\hat{\zeta},\quad
  \bar{U}^I_\BdG(\Gamma_i)=V(-2\Gamma_i)U^I_\BdG=V(\Gamma_i)^\dagger\tilde{U}^I_\BdG(\Gamma_i)V(\Gamma_i),
\end{equation}
and therefore, $\bar{U}^I_\BdG(\Gamma_i)$ is unitary-equivalent to $\tilde{U}^I_\BdG(\Gamma_i)$, not to $U_\BdG^I(\Gamma_i)$.
Thus, $\theta^L(\pi)$ can be rewritten as
\begin{equation}
  (-1)^{\theta^L(\pi)/2}=\left(\prod_{i=1,2}(-1)^{M_{+u}^<(\Gamma_i)/2+M_{-u}^<(\Gamma_i)/2}\eta_I^{M_-^<(\Gamma_i)/2}\right)\cdot\left(\prod_{i=1,2}(-1)^{M_{-u}(\Gamma_i)/2}\eta_I^{M_-(\Gamma_i)/2}\right),
  \label{eq:thetaL}
\end{equation}
by \Eq{eq:formula_temp1}.
The latter product is the extra factor coming from the $k$-dependence of $\tilde{U}^I(\bm{k})$, and can not be ignored in general.
This term may take the value $-1$ when atoms are placed at an inversion center on the face of the unit cell, e.g. Wyckoff position $4a$ in $Pnma$.
However, we see below that $\delta r_b$ cancels out this factor, and the formulas in the main text still hold for $k$-dependent inversion cases. {(This cancellation is physically reasonable, since $\theta_4(\pi)=0$ holds for an atomic superconductor $\mu\to-\infty$.)}

Note that the atoms within an unit cell can be classified into the following two groups.
The first group, named $A$, includes the atoms having inversion as their site symmetry,
while otherwise the atoms are classified into the group $B$.
We can define the subset $B/2$ of $B$, so as to satisfy $B=\Set{\pm\bm{r}_n|\bm{r}_n\in B/2}$ (see \Sec{sec:proof_Reps}).
%(by retaking the unit cell, if necessary).
Thus, contribution to $\delta r_b$ only comes from the atoms in the group $A$.
Similarly, we can define the subset $A/2$ of $A$ so that $A=\Set{\bm{r}_n,\,\hat{G}(\bm{r}_{n})|\bm{r}_n\in A/2}$,
since $\hat{G}$ must exchange sublattices.
%We choose $A_{1/2}$ such that $\bm{r}_{G(n)}=G(\bm{r}_n)$.@@@@
%These specific choices of atoms in the unitcell does not lose generality, since $\theta_4(\Gamma_c)$ is independent of such choices.
In our setup,
\begin{equation}
  \hat{G}(\bm{r}_n)=\hat{M}_c\bm{r}_n+\bm{a}/2+\hat{c}/2,
\end{equation}
and we have $\bm{\beta}\cdot(\bm{r}_n+\hat{G}(\bm{r}_n))=2\bm{r}_n\cdot\bm{\beta}$.
Using $\hat{I}(\bm{r}_n)=-\bm{r}_n=\bm{r}_n+\Delta\bm{R}^I_n$ for $\bm{r}_n\in A$, {and} $\pi\bm{\beta}=\bm{\Gamma}_2-\bm{\Gamma_1}$, we find
\begin{subequations}
\begin{align}
  \delta r_b&=4n_{\mathrm{orb}}\bm{\beta}\cdot\sum_{n\in A/2}\bm{r}_n\\
  &=2n_{\mathrm{orb}}\bm{\beta}\cdot\sum_{n\in A/2}(-\Delta\bm{R}_n^I)\\
  &=\sum_{l:\mathrm{orb}}\,\left(2\sum_{n\in A/2}(-\bm{\beta}\cdot\Delta\bm{R}_n^I)\right)\\
  &=\sum_{l:\mathrm{orb}}p_l\,\left(2\sum_{n\in A/2}(-\bm{\beta}\cdot\Delta\bm{R}_n^I)\right)\quad (\text{mod}\ 4)\label{eq:mod4_localorb}\\
  &=2\sum_{l:\mathrm{orb}}p_l\sum_{n\in A/2}\left(-\frac{\bm{\Gamma}_2\cdot\Delta\bm{R}_n^I}{\pi}+\frac{\bm{\Gamma}_1\cdot\Delta\bm{R}_n^I}{\pi}\right)\\
  &=2\sum_{l:\mathrm{orb}}p_l\sum_{n\in A/2}\left(\frac{e^{i\bm{\Gamma}_2\cdot\Delta\bm{R}_n^I}-1}{2}-\frac{e^{i\bm{\Gamma}_1\cdot\Delta\bm{R}_n^I}-1}{2}\right)\quad (\text{mod}\ 4)\\
  &=\sum_{l:\mathrm{orb}}p_l\sum_{n\in A/2}\left({e^{i\bm{\Gamma}_2\cdot\Delta\bm{R}_n^I}}-{e^{i\bm{\Gamma}_1\cdot\Delta\bm{R}_n^I}}\right),
\end{align}
\end{subequations}
where $p_l=\pm1$ is the parity of the local orbitals (Note that atoms in the group $A$ are locally centrosymmetric).
Here, an equality $e^{i\Gamma_i\cdot\Delta\bm{R}_{G(n)}^I}=e^{i\Gamma_i\cdot\Delta\bm{R}_n^I}$ holds,
since
\begin{subequations}
\begin{gather}
  \hat{I}\hat{G}\bm{r}_n=\hat{G}\bm{r}_n+\Delta\bm{R}_{G(n)}^I,\\
  \hat{I}\hat{G}\bm{r}_n=\Set{E|-\bm{a}-\hat{c}}\hat{G}(\bm{r}_n+\Delta\bm{R}_n^I)=\hat{G}\bm{r}_n+\hat{M}_c\Delta\bm{R}_n^I-\bm{a}-\hat{c},\\
  \bm{\Gamma}_i\cdot\Delta\bm{R}_{G(n)}^I=(\hat{M}_c\bm{\Gamma}_i)\cdot\Delta\bm{R}_{n}^I+\bm{\Gamma}_i\cdot(-\bm{a}-\hat{c})=\bm{\Gamma}_i\cdot\Delta\bm{R}_n^I+\bm{G}_c\cdot\Delta\bm{R}_n^I-2\pi=\bm{\Gamma}_i\cdot\Delta\bm{R}_n^I\quad(\text{mod }2\pi).
\end{gather}
\end{subequations}
Thus,
\begin{subequations}
\begin{align}
  \delta r_b&=\frac{1}{2}\sum_{l:\mathrm{orb}}p_l\sum_{n\in A}\left({e^{i\bm{\Gamma}_2\cdot\Delta\bm{R}_n^I}}-{e^{i\bm{\Gamma}_1\cdot\Delta\bm{R}_n^I}}\right)\\
  &=\frac{1}{4}(\chi_{\Gamma_2}(\hat{I})-\chi_{\Gamma_1}(\hat{I}))\label{eq:character_I}\\
  &=\frac{1}{4}\bigl[(M_{+g}(\Gamma_2)+M_{-g}(\Gamma_2)-M_{+u}(\Gamma_2)-M_{-u}(\Gamma_2)\\
    &\qquad\qquad-(M_{+g}(\Gamma_1)+M_{-g}(\Gamma_1)-M_{+u}(\Gamma_1)-M_{-u}(\Gamma_1))\bigr].
\end{align}
\end{subequations}
In the second line, we used $\tilde{U}^I(\bm{k})_{nls,nls}=e^{i\bm{k}\cdot\Delta\bm{R}^I_n}\,p_l$ in the subspace spanned by the $A$-atoms, and $\tilde{U}^I(\bm{k})_{nls,nls}=0$ for $B$-atoms.
%[{red}{Extension of \Eqs{eq:mod4_localorb}{eq:character_I} to the case with parity-mixed local orbitals is trivial, since two orbitals are paired up by $\hat{I}$ and $\tilde{U}^I(\bm{k})$ is off-diagonal in that subspace.}
Note that the rotation of basis for Kramers degrees of freedom changes only glide eigenvalues, and thus $M_{+g,u}(\Gamma_i)=M_{-g,u}(\Gamma_i)=M_{g,u}(\Gamma_i)/2$.
In addition, $M_g(\Gamma_1)+M_u(\Gamma_1)=M_g(\Gamma_2)+M_u(\Gamma_2)$.
Therefore,
\begin{subequations}
\begin{align}
  \delta r_b&=\frac{1}{2}(-M_u(\Gamma_2)+M_u(\Gamma_1))=-M_{-u}(\Gamma_2)+M_{-u}(\Gamma_1)\\
  &=\frac{1}{2}(M_g(\Gamma_2)-M_g(\Gamma_1))=M_{-g}(\Gamma_2)-M_{-g}(\Gamma_1).
\end{align}
\end{subequations}
Thus, $\delta r_b$ modulo four cancels the correction in \Eq{eq:thetaL}, and the formulas in the main text remain valid even when $\bm{k}$-dependence of $\tilde{U}^I(\bm{k})$ is inevitable.

\subsection{Glide-even $\mathbb{Z}_2$ invariant in the L\"owdin basis}
%We skip the detail of the calculation, since discussion similar to the previous subsection holds, by using TRS and particle-hole symmetry of the line $C_{\mathrm{AII}}(\Gamma_c)$.
A calculation similar to the previous subsection reveals the $\mathbb{Z}_2$ invariant of glide-even SCs. By using TRS and particle-hole symmetry on the line $C_{\mathrm{AII}}(\Gamma_c)$, we obtain
\begin{subequations}
\begin{gather}
  \nu_\pm(\Gamma_c)=\nu_\pm^L(\Gamma_c)+\delta r_b/2,\\
  \nu_\pm^L(\Gamma_c)=\int_0^\pi\frac{dk_b}{\pi i}\bm{\beta}\cdot\bm{A}_\pm(\bm{k}')+\frac{1}{\pi i}\Log\left[\frac{\Pf[w_\pm(\Gamma_2)]}{\Pf[w_\pm(\Gamma_1)]}\right].
\end{gather}
\end{subequations}
\subsection{Glide-even $\mathbb{Z}_2$ invariant in the presence of $k$-dependent inversion symmetry}
We show the formulas for the $\mathbb{Z}_2$ invariants remain valid even when the $\bm{k}$-dependence of $\tilde{U}^I(\bm{k})$ is inevitable.
We can easily evaluate $\nu_\pm^L(\pi)$ by using \Eq{eq:formula2_temp},
\begin{equation}
  \nu_\pm^L(\pi)=\left(\prod_{i=1,2}\eta_I^{M_\pm^<(\Gamma_i)/2}\right)\cdot\left(\prod_{i=1,2}(-1)^{M_{\pm u}(\Gamma_i)/2}\eta_I^{M_\pm(\Gamma_i)}\right).
\end{equation}
The latter product cancels with the correction $\delta r_b/2=(-M_{\pm u}(\Gamma_2)+ M_{\pm u}(\Gamma_1))/2=(M_{\pm g}(\Gamma_2)-M_{\pm g}(\Gamma_1))/2$. Thus, the formulas in the main text have been proven.
%remain valid even when the $\bm{k}$-dependence of $\tilde{U}^I(\bm{k})$ can not be removed.

\section{A tight-binding Hamiltonian for $\mathrm{UCoGe}$}
\label{sec:modelcalc}
In this section, we give a single-orbital tight-binding Hamiltonian for UCoGe.
This is a minimal model of $\mathbb{Z}_4$ and $\mathbb{Z}_2$ nontrivial TNCS in the space group $Pnma$. 
%The model is constructed within nearest-neighbor hopping.

\subsection{Normal-part Hamiltonian and model parameters}
First, we show spin-independent part of the tight-binding Hamiltonian.
Figure~\ref{fig:UCoGeUC} shows uranium atoms in a unit cell.
There are four uranium atoms labeled by $(a1,a2,b1,b2)$. Atoms $(a1,a2)$ and $(b1,b2)$ are placed on the planes $y=-1/4,1/4$, respectively, and form zigzag chains in the $x$ direction. We call these chains as chain $a$ and chain $b$.
Within nearest neighbor coupling, hopping part of the Hamiltonian is given by
\begin{subequations}
\begin{align}
  \hat{H}_{\mathrm{hop}}&=\sum_{\bm{k}}\bm{C}^\dagger_{\bm{k}}\,H_{\mathrm{hop}}(\bm{k})\,\bm{C}_{\bm{k}},\\
  \bm{C}_{\bm{k}}&=\left(c_{\bm{k}a1\uparrow},\,c_{\bm{k}a1\downarrow},\,c_{\bm{k}a2\uparrow},\,c_{\bm{k}a2\downarrow},
  %\right.\\&\qquad\qquad\left.
  c_{\bm{k}b1\uparrow},\,c_{\bm{k}b1\downarrow},\,c_{\bm{k}b2\uparrow},\,c_{\bm{k}b2\downarrow}\right)^T,\\
  &H_{\mathrm{hop}}(\bm{k})=s_0\otimes\begin{pmatrix}H_a(\bm{k})&H_{ab}(\bm{k})\\H_{ab}(\bm{k})^\dagger&H_b(\bm{k})\end{pmatrix}_\eta,\\
  &H_a(\bm{k})=\begin{pmatrix}\xi(\bm{k})&\xi_{12}(\bm{k})\\\xi_{12}(\bm{k})^*&\xi(\bm{k})\end{pmatrix}_\sigma,\\
  &H_b(\bm{k})=H_a(\bm{k})^T,\\
  &H_{ab}(\bm{k})=\begin{pmatrix}v_1(\bm{k})&0\\0&v_2(\bm{k})\end{pmatrix}_\sigma,
\end{align}
\end{subequations}
where $s$ shows spin, while $\sigma$ and $\eta$ represent sublattice degrees of freedom corresponding to $(1,2)$ and $(a,b)$, respectively.
Here, $H_a$ and $H_b$ represent the intra-chain hopping within the chain $a$ and the chain $b$, while $H_{ab}$ represents the inter-chain hopping between the chains $a$ and $b$. $H_a$, $H_b$, and $H_{ab}$ are given by
\begin{subequations}
\begin{gather}
  \xi(\bm{k})=2t_1'\cos k_x+2t_2\cos k_y+2t_3\cos k_z-\mu,\\
  \xi_{12}(\bm{k})=t_1(1+e^{-ik_x}),\\
  v_1(\bm{k})=e^{-ik_x}(1+e^{-ik_y})(t_{ab}+e^{ik_z}t_{ab}'),\\
  v_2(\bm{k})=e^{ik_z}(1+e^{-ik_y})(t_{ab}+e^{-ik_z}t_{ab}').
\end{gather}
\end{subequations}
\begin{figure}
  \centering
  \includegraphics[height=60mm]{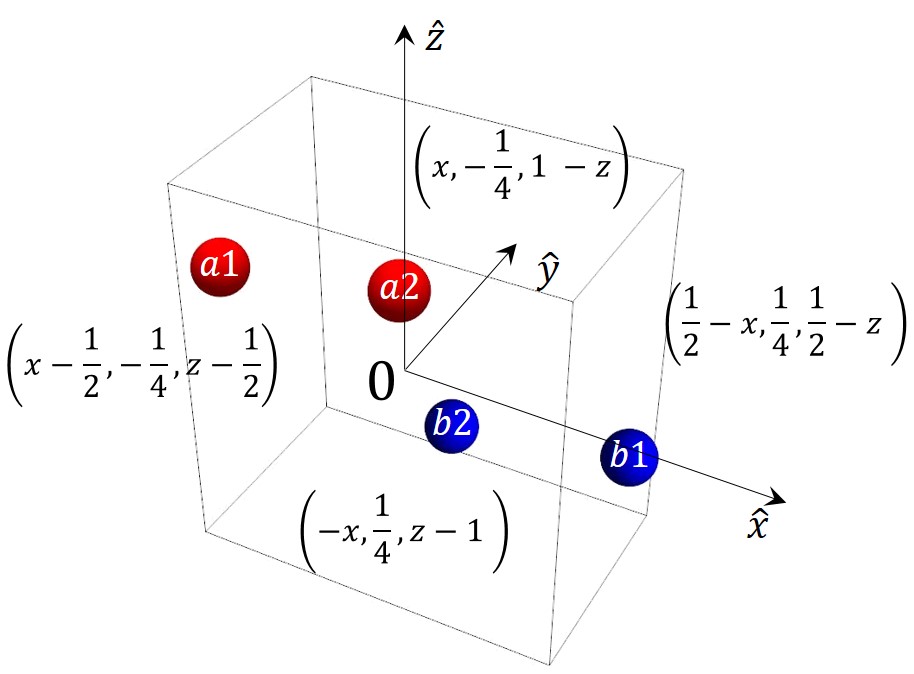}
  \caption{Uranium atoms in a unit cell of UCoGe. Atoms $(a1,a2)$ on a plane $y=-1/4$ are highlighted by red, while $(b1,b2)$ on $y=1/4$ by blue. All the four atoms are placed at an equivalent Wyckoff position $4c$. Position of the atoms is parametrized by $(x,z)=(0.0101,0.7075)$~\cite{Canepa1996_SM}, where lattice constants are normalized to be unity.}
  \label{fig:UCoGeUC}
\end{figure}

Next, we introduce spin-dependent part of the tight-binding Hamiltonian.
%% Here we call them antisymmetric spin-orbit coupling (ASOC), although the system is globally inversion symmetric and its uniform component vanishes.
%% ASOC is calssified into the two types, where one is intra-sublattice, and the other is inter-sublattice.
For simplicity of the model, we take only intra-sublattice spin-orbit coupling into account.
In accordance with the $C_s$ local symmetry of the uranium atoms, we obtain spin-orbit coupling term,
\begin{subequations}
\begin{align}
  H_{\mathrm{SOC}}(\bm{k})&=\alpha(\delta_\alpha\sin k_x\,s_y-\sin k_y\,s_x)\sigma_z\eta_z\\
  &+\beta(\sin k_y\,s_z+\delta_\beta \sin k_z\,s_y)\eta_z,\\
  &=\bm{g}_\alpha(\bm{k})\cdot\bm{s}\sigma_z\eta_z+\bm{g}_\beta(\bm{k})\cdot\bm{s}\eta_z,
\end{align}
\end{subequations}
where
\begin{subequations}
\begin{gather}
  \bm{g}_\alpha(\bm{k})=\alpha(-\sin k_y,\delta_\alpha\sin k_x,0)^T,\\
  \bm{g}_\beta(\bm{k})=\beta(0,\delta_\beta\sin k_z,\sin k_y)^T,\\
  \alpha,\,\delta_\alpha,\,\beta,\,\delta_\beta\in\mathbb{R}.
\end{gather}
\end{subequations}
Here, $s_\mu$, $\sigma_\mu$, and $\eta_\mu$ are the Pauli matrices for spin, sublattice $(1,2)$, and sublattice $(a,b)$, respectively.
Thus, {the matrix representation of} normal-part Hamiltonian is given by the sum of these two parts:
\begin{gather}
  H(\bm{k})=H_{\mathrm{hop}}(\bm{k})+H_{\mathrm{SOC}}(\bm{k}).\label{eq:TBmodel_normal}
\end{gather}

We adopt the following parameters to mimic the cylinder FSs $71$ and $72$ in Fig.~1,
\begin{equation}
  (t_1,t_2,t_3,t_{ab},t_{ab}',\mu,t_1',\alpha,\delta_\alpha,\beta,\delta_\beta)=(1, 0.2, 0.1, 0.5, 0.1, 0.55, 0.1, 0.3, 0.5, 0.3, 0.5).\label{eq:param}
\end{equation}
FSs of the model \Eqs{eq:TBmodel_normal}{eq:param} are depicted in \Fig{fig:FSs_model}.
\begin{figure}
  \centering
%  \begin{tabular}{ccc}
%    (a) $k_z=0$ &(b) $k_z=\pi/2$ &(c) $k_z=\pi$ \\
    %  \includegraphics[height=40mm]{FS_kz0.pdf}&\includegraphics[height=40mm]{FS_kz025Pi.pdf}&\includegraphics[height=40mm]{FS_kzPi.pdf}\end{tabular}
    \includegraphics[height=40mm]{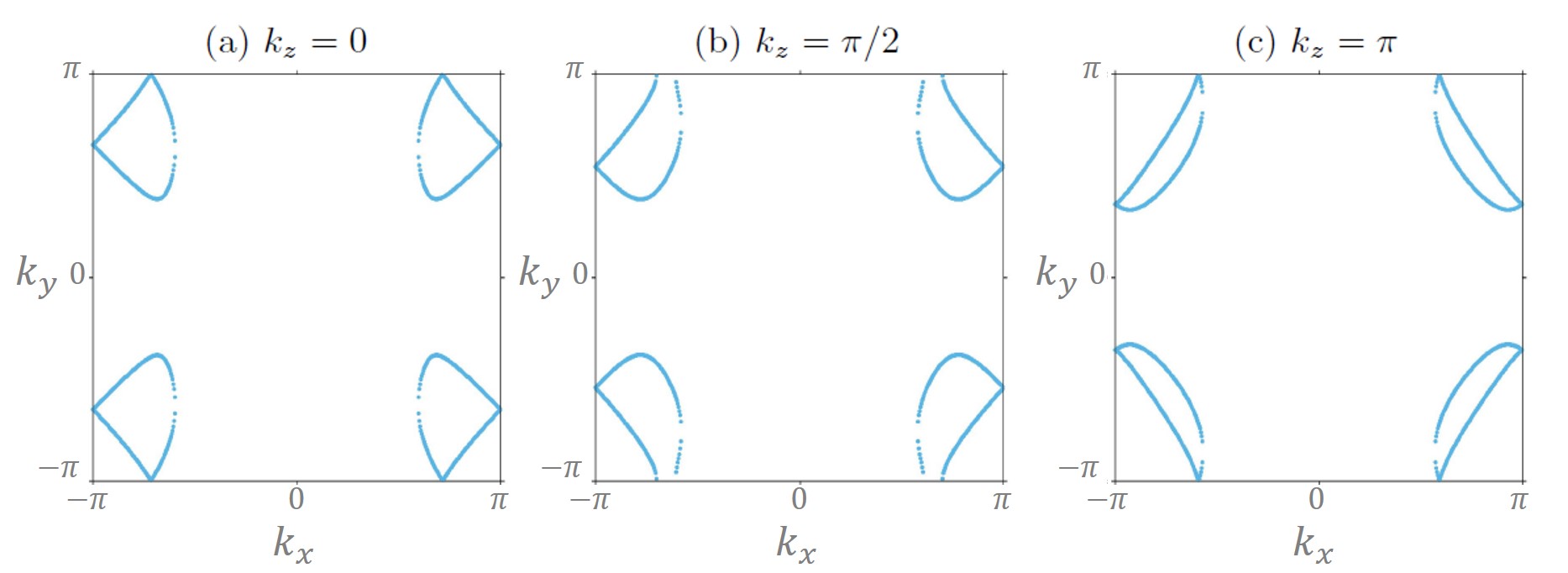}
  \caption{FS of the model. The panels (a,b,c) show the $k_z=0,\,\pi/2,$ and $\pi$ slices of the FS, respectively.}
  \label{fig:FSs_model}
\end{figure}

\subsection{Symmetry operations}
Here, we write down the representation matrices of symmetry operations in $Pnma$.
%We place the origin of the lattice at an inversion center $(1/2,1/2,1/2)$ in \Fig{fig:UCoGeUC}.
For brevity, we show only a set of generators $\hat{I}$, $\hat{G}_a$ and $\hat{G}_n$,
\begin{subequations}
\begin{gather}
  \tilde{U}^I(\bm{k})=\eta_x,\\
  \tilde{U}^{G_a}(\bm{k})=-is_z\otimes\begin{pmatrix}\begin{pmatrix}0&e^{-ik_x}\\1&0\end{pmatrix}_\sigma &0\\0&\begin{pmatrix}0&e^{ik_z}\\e^{-ik_x+ik_z}&0\end{pmatrix}_\sigma\end{pmatrix}_\eta,\\
  \tilde{U}^{G_n}(\bm{k})=-is_x\otimes\begin{pmatrix}0&\begin{pmatrix}0&e^{ik_x-ik_y}\\e^{-ik_y}&0\end{pmatrix}_\sigma\\\begin{pmatrix}0&e^{-ik_z}\\e^{ik_x-ik_z}&0\end{pmatrix}_\sigma&0\end{pmatrix}_\eta.
\end{gather}
\end{subequations}
Other operations are obtained by their combination.
Symmetry of the Hamiltonian $H(\bm{k})$ are written as
\begin{equation}
  \tilde{U}^g(\bm{k})H(\bm{k})\left[\tilde{U}^g(\bm{k})\right]^\dagger=H(p\bm{k}),
  \label{eq:sym_Pnma}
\end{equation}
for $\hat{g}=\Set{p|\bm{a}}\in Pnma$.
  
\subsection{BdG Hamiltonian}
The BdG Hamiltonian is given by
%% \begin{gather}
%%   \hat{H}=\sum_{\bm{k}}(\bm{C}_{\bm{k}}^\dagger,\,\bm{C}_{-\bm{k}}^T)H_{BdG}(\bm{k})\begin{pmatrix}\bm{C}_{\bm{k}}\\\bm{C}_{-\bm{k}}^*\end{pmatrix},\\
%%     H_{BdG}(\bm{k})=\begin{pmatrix}H(\bm{k})&\Delta(\bm{k})\\\Delta(\bm{k})^\dagger&-H(-\bm{k})^T\end{pmatrix}_\tau,
%% \end{gather}
%% \begin{gather}
%%   \hat{H}=\sum_{\bm{k}}(\bm{C}_{\bm{k}}^\dagger,\,\bm{C}_{-\bm{k}}^T(-is_y))H_{\mathrm{BdG}}(\bm{k})\begin{pmatrix}\bm{C}_{\bm{k}}\\is_y\bm{C}_{-\bm{k}}^*\end{pmatrix},\\
%%     H_{\mathrm{BdG}}(\bm{k})=H(\bm{k})\otimes\tau_z+\tilde{\Delta}(\bm{k})\otimes\tau_x,
%% \end{gather}
\begin{subequations}
\begin{gather}
  \hat{H}=\frac{1}{2}\sum_{\bm{k}}\Phi_{\bm{k}}^\dagger H_{\mathrm{BdG}}(\bm{k})\Phi_{\bm{k}},\\
    \Phi_{\bm{k}}^\dagger\equiv (\bm{C}_{\bm{k}}^\dagger,\,\bm{C}_{-\bm{k}}^T(-is_y)),\quad H_{\mathrm{BdG}}(\bm{k})=H(\bm{k})\otimes\tau_z+\tilde{\Delta}(\bm{k})\otimes\tau_x,
\end{gather}
\end{subequations}
where $\tau_\mu$ represents the Pauli matrix in Nambu space.
%In the first line, we expressed Hermitian conjugate of creation/anihilation operators by the superscript $*$.
Here, $\tilde{\Delta}(\bm{k})$ is the superconducting order parameter in the current basis,
which is connected to the usual definition of $\hat{\Delta}(\bm{k})$ in \Ref{Nomoto2016_SM} by $\tilde{\Delta}(\bm{k})=\hat{\Delta}(\bm{k})(-is_y)$.
The time-reversal symmetry of the system imposes Hermitian property on $\tilde{\Delta}(\bm{k})$,
%and can be taken as a Hermintian matrix owing to TRS,
\begin{subequations}
\begin{gather}
  \Theta H_{\mathrm{BdG}}(\bm{k})\Theta^{-1}=H_{\mathrm{BdG}}(-\bm{k}),\quad \Theta=is_yK,\\
  \tilde{\Delta}(\bm{k})^\dagger=\tilde{\Delta}(\bm{k}),
\end{gather}
\end{subequations}
with complex conjugation $K$.
The BdG Hamiltonian has the particle-hole symmetry
\begin{equation}
  CH_{\mathrm{BdG}}(\bm{k})C^{-1}=-H_{\mathrm{BdG}}(-\bm{k}),\quad C=\tau_ys_yK.
\end{equation}
Symmetries of $Pnma$ in the normal state \Eq{eq:sym_Pnma} are extended to the superconducting state as follows.
Let us take $\hat{g}=\Set{p|\bm{a}}\in Pnma$. Then, BdG Hamiltonian of $\hat{g}$-even (-odd) superconductivity, where $\tilde{U}^g(\bm{k})\tilde{\Delta}(\bm{k}){\tilde{U}^g(\bm{k})}^\dagger=\pm\tilde{\Delta}(p\bm{k})$, preserves the symmetry
\begin{subequations}
\begin{gather}
  \tilde{U}^g_{\mathrm{BdG}}(\bm{k})H_{\mathrm{BdG}}(\bm{k}){\tilde{U}^g_{\mathrm{BdG}}(\bm{k})}^\dagger=H_{\mathrm{BdG}}(p\bm{k}),\\
  \tilde{U}^g_{\mathrm{BdG}}(\bm{k})\equiv \tilde{U}^g(\bm{k})\otimes\begin{pmatrix}1&0\\0&\pm1\end{pmatrix}_\tau.
\end{gather}
\end{subequations}
Here, the U(1) gauge rotation is combined with the crystal symmetry. Then, we have
\begin{equation}
\quad C\tilde{U}^g_{\mathrm{BdG}}(\bm{k})=\pm \tilde{U}^g_{\mathrm{BdG}}(-\bm{k})C.
\end{equation}

Superconducting order parameters are given by the basis functions of the point group $D_{2h}$, which is associated with the space group $Pnma$~\cite{Nomoto2016_SM}.
We assume a simple form of $\tilde{\Delta}(\bm{k})$ in \Table{tab:OPs_model} for $A_u$, $B_{1u}$, $B_{2u}$, and $B_{3u}$ states. Note that topological properties are insensitive to specific $\bm{k}$-dependence of $\tilde{\Delta}(\bm{k})$. 
Magnitude of the order parameters is taken as $d_0=0.5\,t_1$, for visibility of the figures of surface spectrum.
The $\delta$-term in $B_{2u}$ state was incorporated so as to remove accidental excitation nodes at $k_z=\pi$.
\begin{table}
  \centering
  \caption{List of order parameters used in the numerical calculation. We take $\delta=1$ in $B_{2u}$.}
  \label{tab:OPs_model}
  \begingroup
  \renewcommand{\arraystretch}{1.5}
  \tabcolsep =2.5mm
  \begin{tabular}{c|c}\hline\hline
    irrep.&$\tilde{\Delta}(\bm{k})/d_0$\\\hline
    $A_u$&$\sin k_xs_x+\sin k_ys_y+\sin k_zs_x\sigma_z$\\
    $B_{1u}$&$\sin k_xs_y+\sin k_ys_x+\sin k_zs_y\sigma_z$\\
    $B_{2u}$&$\sin k_xs_x\sigma_z+\sin k_ys_y\sigma_z+\sin k_zs_x+\delta\left\{(1+\cos k_x)s_x\sigma_y\eta_z-\sin k_xs_x\sigma_x\right\}/2$\\
    $B_{3u}$&$\sin k_xs_y\sigma_z+\sin k_ys_x\sigma_z+\sin k_zs_y$\\\hline\hline
  \end{tabular}
  \endgroup
\end{table}
\subsection{Accidental point nodes in the model of $B_{2u}$ and $B_{3u}$ states}
{The model presented in this section shows accidental point nodes in $B_{2u}$ and $B_{3u}$ states, while $A_u$ and $B_{1u}$ states are gapful.}
{Actually, both $B_{2u}$ and $B_{3u}$ states have point nodes on the $k_z=0$ plane, which are protected by the glide-winding number~\cite{Yanase2017_SM}.
}
%% \textcolor{red}{The $B_{2u}$ state has point nodes on the $k_z=0$ plane, which are protected by the glide-winding number~\cite{Yanase2017_SM}.
%% Similary, the $B_{3u}$ state has point nodes on the $k_z=0$ plane protected by the glide-winding number, and accompanies flat-band surface states in an appropriate surface direction.
%% }
%{
%However, they are not protected by the symmetry, and can be removed by tuning the $k$-dependence of the order parameters, that is, not protected by symmetry.
%They accompany flat-band surface states in the appropriate surface directions, in accordance with the bulk-boundary correspondence.}
{Note that, however, these point nodes are not protected by symmetry.
  Therefore, their existence depends on the detailed $k$-dependence of the order parameters.
  In particular, they can be pair-annihilated by tuning order parameters, since all the Fermi surfaces are connected (\Fig{fig:FSs_model}~(a)).
  Thus, they may be the artifact of the model.
  For this reason, we do not discuss these point nodes in the following part of the paper.
%It needs more experimental information about the gap functions to determine nodal structures.
}

%The $B_{3u}$ state also has point nodes on the $k_y=\pi$ plane, which are due to the winding number enriched by the accidental symmetry $\eta_z$ of $H_{\mathrm{BdG}}(k_y=\pi)$.
%The Hamiltonian can be block-diagonalized simultaneously by $\hat{M}_y$ and $\eta_z$, where the chiral symmetry $i\Theta C$ is closed within each eigen-sector. Thus, the point nodes are protected by the winding number of the class AIII.
{
  The $B_{3u}$ state also has point nodes on the $k_y=\pi$ plane due to the accidental symmetry $\eta_z$ of $H_{\mathrm{BdG}}(k_y=\pi)$.
  They are protected by the winding number of class AIII defined within each eigen-sector of Hamiltonian simultaneously block-diagonalized by $\hat{M}_y$ and $\eta_z$.}
{
  Note that the mirror winding number~\cite{Zhang-Kane-Mele2013_SM} of $H_{\mathrm{BdG}}(k_y=\pi)$ vanishes owing to the $\hat{\Theta}\hat{I}$ symmetry.
Thus, the point nodes on $k_y=\pi$ are also the artifact of the model, and vanish in realistic situations. The presence of these point nodes is not harmful for the glide topological invariants and  topological surface states.}
%These point nodes are placed on the low-symmetry $k$-points within each plane.
%For these reasons, we don't discuss the point nodes in the following part.
\section{Calculated surface states of glide TNCS}
\subsection{Surface states of $\hat{G}_a$-TNCS}
In this section, we show results of the model calculations of surface states protected by $\hat{G}_a$-topological invariants.
We adopt a slab geometry, where lattice sites are given by
\begin{equation}
  \Set{\bm{R}=n\hat{x}+m\hat{y}+l\hat{z}|\ (n,l)\in\mathbb{Z}^2,\quad0\le m\le L_y}.
\end{equation}
Translation operators satisfy
\begin{equation}
  \hat{T}_x^{L_x}=\hat{1},\quad\hat{T}_z^{L_z}=\hat{1}.
\end{equation}
It is easily confirmed that the glide symmetry $\hat{G}_a$ is preserved on the surfaces $\Set{\bm{R}|m=0}$ and $\Set{\bm{R}|m=L_y}$.
\subsubsection{$\hat{G}_a$-odd TNCS}
First we show the results of $\hat{G}_a$-odd superconductivity $A_u$ and $B_{1u}$.
Figures~\ref{fig:Ga_odd}(a) and \ref{fig:Ga_odd}(b) show the surface states of $A_u$ and $B_{1u}$ at $k_z=\pi$, respectively.
\begin{figure}
  \centering
%  \begin{tabular}{cccc}
%    (a) $A_u$ &(b) $B_{1u}$ &(c) $A_u(+\Delta H)$ &(d) $B_{1u}(+\Delta H)$ \\
%    \includegraphics[width=45mm]{Au_Ga.pdf}&\includegraphics[width=45mm]{B1u_Ga.pdf}&\includegraphics[width=45mm]{Au_lifted_Ga.pdf}&\includegraphics[width=45mm]{B1u_lifted_Ga.pdf}\\
%  \end{tabular}
  \includegraphics[width=180mm]{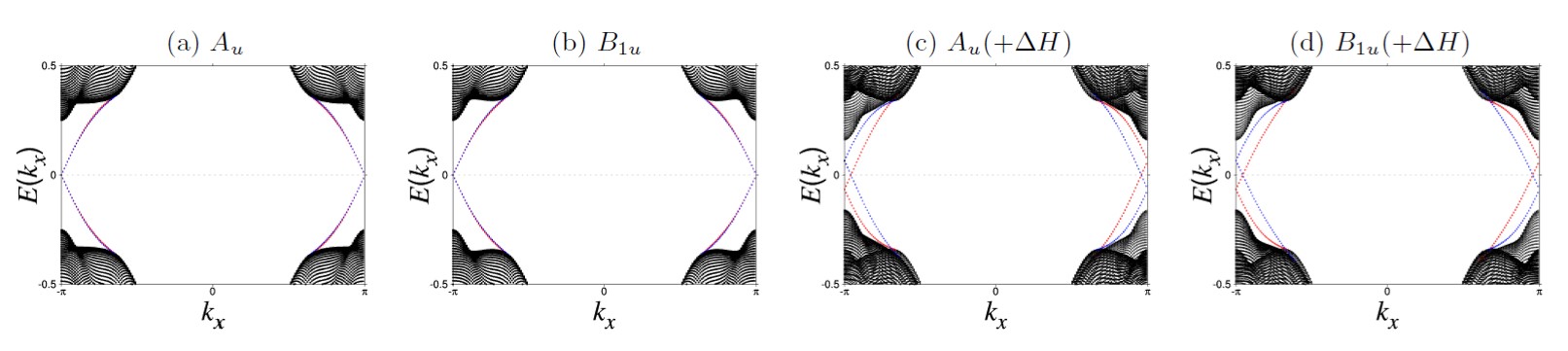}
  \caption{$(010)$ surface states of $\hat{G}_a$-odd superconductivity with $k_z=\pi$. We take $L_x=200$ and $L_y=50$.
  Surface states with positive- and negative-glide eigenvalues are highlighted by red and blue, respectively.}
  \label{fig:Ga_odd}
\end{figure}
Surface states with positive and negative glide eigenvalues are highlighted by red and blue, respectively.
Topological surface states consistent with nontrivial $\mathbb{Z}_4$ invariant $\theta_4^{(a)}(\pi)=2$ appear in the figures.

Although zero-energy surface states are pinned to $k_x=\pi$, these are the artifact of the model.
Actually, there is an emergent internal symmetry $[H_{\mathrm{BdG}}(\pi,k_y,\pi),\sigma_z]=0$ for both $A_u$ and $B_{1u}$ states. The fourfold degeneracy at $k_x=k_z=\pi$ follows from the anticommutation relation $\{\tilde{U}^{G_a}_{\mathrm{BdG}}(\bm{k}),\sigma_z\}=0$, by making use of the discussion in \Ref{Niu2017_SM}.
Here, $\sigma_z$ is an accidental symmetry which is not included in $Pnma$, and therefore, should be broken in reality.
%% Indeed, the accidental degeneracy of surface states at $k_x=\pi$ is lifted by adding the following term to the BdG Hamiltonian,
Indeed, the pinning can be removed by adding the following term to $H_{\mathrm{BdG}}(k_x,k_y,\pi)$,
\begin{equation}
  \Delta H(\bm{k})=\epsilon(\tilde{U}^{G_a}(\bm{k})+\tilde{U}^{G_a}(\bm{k})^\dagger)\tau_z,\quad k_z=\pi,\label{eq:additional_term}
\end{equation}
which is compatible with the $\hat{G}_a$, $\hat{\Theta}$, and $\hat{C}$ symmetries.
%% %@@@-------Old version---------
%% Although zero-energy surface states are pinned to $k_x=\pi$, this is an artificial result of the model.
%% Actually, there is an emergent symmetry $\sigma_z$ of $H_{\mathrm{BdG}}(\pi,k_y,\pi)$ for both $A_u$ and $B_{1u}$ states, which anticommutes with $\hat{G}_a$.
%% Here, $\sigma_z$ is an accidental symmetry of the model, which is not included in $Pnma$, and therefore, should be broken in reality.
%% Indeed, the accidental degeneracy of surface states at $k_x=\pi$ is lifted by adding the following term to the BdG Hamiltonian,
%% \begin{subequations}
%% \begin{gather}
%%   \Delta H(k_x,k_y)=\epsilon(\tilde{U}^{G_a}(k_x,k_y,\pi)+\tilde{U}^{G_a}(k_x,k_y,\pi)^\dagger)\tau_z,\label{eq:additional_term}\\
%% H_{\mathrm{BdG}}(k_x,k_y,\pi)\to  H_{\mathrm{BdG}}(k_x,k_y,\pi)+\Delta H(k_x,k_y),
%% \end{gather}
%% \end{subequations}
%% which is compatible with time-reversal symmetry, particle-hole symmetry, and the glide symmetry $\tilde{U}_{\mathrm{BdG}}^{G_a}(k_x,k_y,\pi)$.
%% %@@-------Old version
We take $\epsilon=0.015$ to calculate the surface states. The result is shown in Figs.~\ref{fig:Ga_odd}~(c) and \ref{fig:Ga_odd}~(d).
The surface states show a characteristic feature of nontrivial surface states with $\theta_4^{(a)}(\pi)=2$~\cite{Shiozaki2016_SM}.

\subsubsection{$\hat{G}_a$-even TNCS}
Next, we show the result of $\hat{G}_a$-even superconductivity $B_{2u}$ and $B_{3u}$.
Figures~\ref{fig:Ga_even}(a) and \ref{fig:Ga_even}(b) show the $(010)$ surface states of $B_{2u}$ and $B_{3u}$ at $k_z=\pi$, which are consistent with $\mathbb{Z}_2$ nontrivial TNCS with $\nu_\pm^{(a)}(\pi)=\pm1$.
%The gapless surface states are pinned to $k_x=\pi$, where edge states with positive- and negative-glide eigenvalues are highlited by red and blue, respectively.
In contrast to glide-odd TNCS, the gapless surface states at $k_x=\pi$ are protected by nontrivial $\mathbb{Z}_2$ number $\nu_\pm^{(a)}(\pi)=1$, and therefore, they are not lifted by small perturbations. Actually, addition of the term \Eq{eq:additional_term}, which preserves the glide symmetry also in this case, does not affect the spectrum around $k_x=\pi$, as shown in Figures~\ref{fig:Ga_even}(c) and \ref{fig:Ga_even}(d).
% Thus, the obtained edge states are protected by $\nu_\pm^{(a)}(\pi)=1$ in consistent with our formulas.
\begin{figure}
  \centering
%  \begin{tabular}{cccc}
%    (a) $B_{2u}$ &(b) $B_{3u}$ &(c) $B_{2u}(+\Delta H)$ &(d) $B_{3u}(+\Delta H)$ \\
%    \includegraphics[width=45mm]{B2u_Ga.pdf}&\includegraphics[width=45mm]{B3u_Ga.pdf}&\includegraphics[width=45mm]{B2u_lifted_Ga.pdf}&\includegraphics[width=45mm]{B3u_lifted_Ga.pdf}\\
%  \end{tabular}
  \includegraphics[width=180mm]{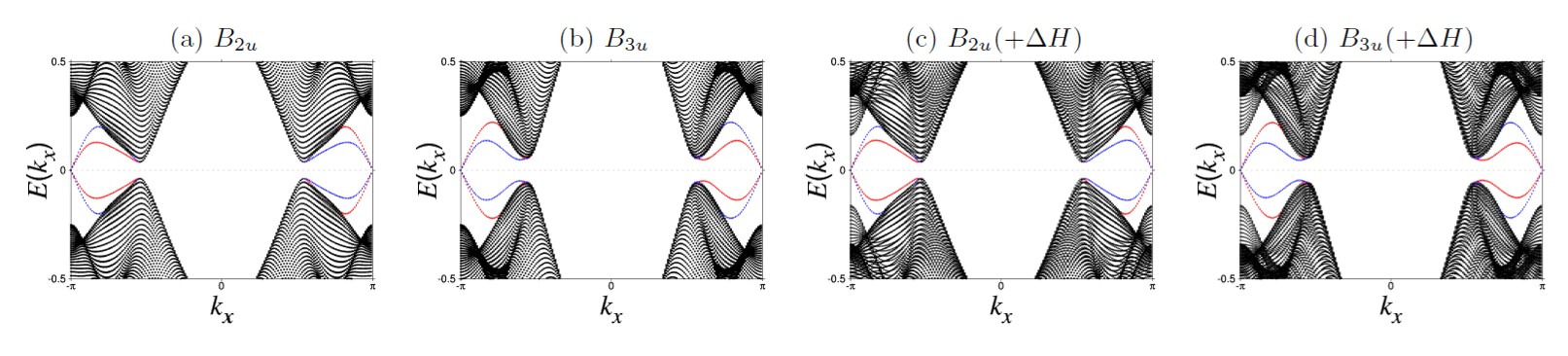}
  \caption{$(010)$ surface states of $\hat{G}_a$-even superconductivity with $k_z=\pi$.
    We take $L_x=200$ and $L_y=50$.
    Surface states with positive- and negative-glide eigenvalues are highlighted by red and blue, respectively.}
  \label{fig:Ga_even}
\end{figure}

\subsection{Surface states of $\hat{G}_n$-TNCS}
Here, we show the results of the model calculations for surface states protected by $\hat{G}_n$-topological invariants.
We adopt a slab geometry, where lattice sites are given by
\begin{subequations}
\begin{gather}
  \Set{\bm{R}=n\bm{a}+m\bm{b}+l\hat{c}|\ (n,l)\in\mathbb{Z}^2,\quad0\le m\le L_b},\\
\bm{a}=\hat{y}+\hat{z},\quad\bm{b}=\hat{z},\quad \hat{c}=\hat{x}.
\end{gather}
\end{subequations}
Translation operators satisfy
\begin{equation}
  \hat{T}_{\bm{a}}^{L_a}=\hat{1},\quad \hat{T}_{\hat{c}}^{L_c}=\hat{1}.
\end{equation}
It is easily confirmed that $\hat{G}_n$ is preserved on the surfaces $\Set{\bm{R}|m=0}$ and $\Set{\bm{R}|m=L_b}$.

\subsubsection{$\hat{G}_n$-odd TNCS}
First, we show the results of $\hat{G}_n$-odd superconductivity $A_u$ and $B_{3u}$.
Figures~\ref{fig:Gn_odd}(a) and \ref{fig:Gn_odd}(b) show the $(0\bar{1}1)$ surface states of $A_u$ and $B_{3u}$ at $k_c=\pi$, respectively.
\begin{figure}
  \centering
%  \begin{tabular}{cccc}
%    (a) $A_u$ &(b) $B_{3u}$ &(c) $A_u(+\Delta H+\Delta H')$ &(d) $B_{3u}(+\Delta H+\Delta H')$ \\
%    \includegraphics[width=45mm]{Au_Gn.pdf}&\includegraphics[width=45mm]{B3u_Gn.pdf}&\includegraphics[width=45mm]{Au_lifted_Gn.pdf}&\includegraphics[width=45mm]{B3u_lifted_Gn.pdf}\\
%  \end{tabular}
  \includegraphics[width=180mm]{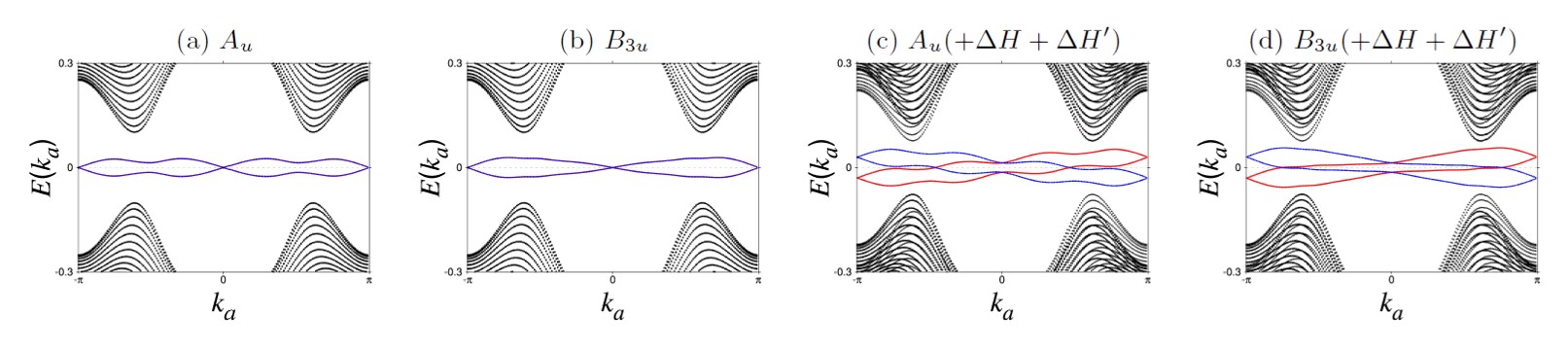}
  \caption{$(0\bar{1}1)$ surface states of $\hat{G}_n$-odd superconductivity at $k_c=\pi$.
    We take $L_a=200$ and $L_b=50$.
  Surface states with positive- and negative-glide eigenvalues are highlighted by red and blue, respectively.}
  \label{fig:Gn_odd}
\end{figure}
Surface states with positive and negative glide eigenvalues are highlighted by red and blue, respectively.
Topological surface states consistent with nontrivial $\mathbb{Z}_4$ invariant $\theta_4^{(n)}(\pi)=2$ appear in the figures.
Although zero-energy surface states are pinned to $k_a=0$ and $k_a=\pi$,
these are the artifact of the model.
Actually, there is an accidental internal symmetry $[H_{\mathrm{BdG}}(k_c=\pi),\sigma_z]=0$, and the fourfold degeneracy at $k_a=0,\pi$ follows from the following relations, by making use of the discussion in \Ref{Niu2017_SM},
\begin{subequations}
\begin{gather}
  (\Theta\sigma_z)^2=-1,\\
  \{\tilde{U}^{G_n}_{\mathrm{BdG}}(\bm{k}),\sigma_z\}=0.
\end{gather}
\end{subequations}
Indeed, the pinning can be removed by adding the following two terms to $H_{\mathrm{BdG}}(k_c=\pi)$, breaking the accidental symmetry $\sigma_z$.
The first one is similar to \Eq{eq:additional_term},
\begin{equation}
  \Delta H(\bm{k})=\epsilon(\tilde{U}^{G_n}(\bm{k})+\tilde{U}^{G_n}(\bm{k})^\dagger)\tau_z,\quad k_c=\pi,\label{eq:additional_term2}
\end{equation}
which lifts the accidental degeneracy at $k_a=\pi$.
This term vanishes at $k_a=0$ due to the anti-Hermitian property of $\tilde{U}^{G_a}(k_a=0,k_c=\pi)$. However, we have another term preserving the symmetry, for example,
\begin{gather}
  \Delta H'(\bm{k})=\epsilon s_x\sigma_y\eta_z\tau_x,\quad k_c=\pi,\label{eq:add_terms1}
\end{gather}
which is also compatible with the $\hat{G}_n$, $\hat{\Theta}$, and $\hat{C}$ symmetries.
By this term, the accidental degeneracy at $k_a=\pi$ is also lifted.
%We notice that \Eq{eq:add_terms1} also lifts the degeneracy at $k_a=\pi$. However, the induced gap is much smaller than that of \Eq{eq:additional_term2} for a fixed $\epsilon$. Here, we take both of the two terms into account for visibility, taking $\epsilon=0.015$.
The resulting surface states at $k_c=\pi$ are shown in \Fig{fig:Gn_odd}(c) and \Fig{fig:Gn_odd}(d), illustrating the characteristic surface states of $\mathbb{Z}_4$ nontrivial $\theta_4^{(n)}(\pi)=2$ state.
Here, we take $\epsilon=0.015$.
%, in agreement with the formulas given in \Table{tab:results}.

\subsubsection{$\hat{G}_n$-even TNCS}
Next, we show the results for $\hat{G}_n$-even superconductivity $B_{1u}$ and $B_{2u}$.
Figure~\ref{fig:Gn_even}(a) and \ref{fig:Gn_even}(b) show the $(0\bar{1}1)$ surface states of $B_{1u}$ and $B_{2u}$ at $k_c=\pi$, which are consistent with $\mathbb{Z}_2$ nontrivial TNCS with $\nu_\pm^{(n)}(\pi)=\pm1$.
\begin{figure}
  \centering
%  \begin{tabular}{cccc}
%    (a) $B_{1u}$ &(b) $B_{2u}$ &(c) $B_{1u}(+\Delta H+\Delta H'')$ &(d) $B_{2u}(+\Delta H+\Delta H'')$ \\
%    \includegraphics[width=45mm]{B1u_Gn.pdf}&\includegraphics[width=45mm]{B2u_Gn.pdf}&\includegraphics[width=45mm]{B1u_lifted_Gn.pdf}&\includegraphics[width=45mm]{B2u_lifted_Gn.pdf}\\
%  \end{tabular}
  \includegraphics[width=180mm]{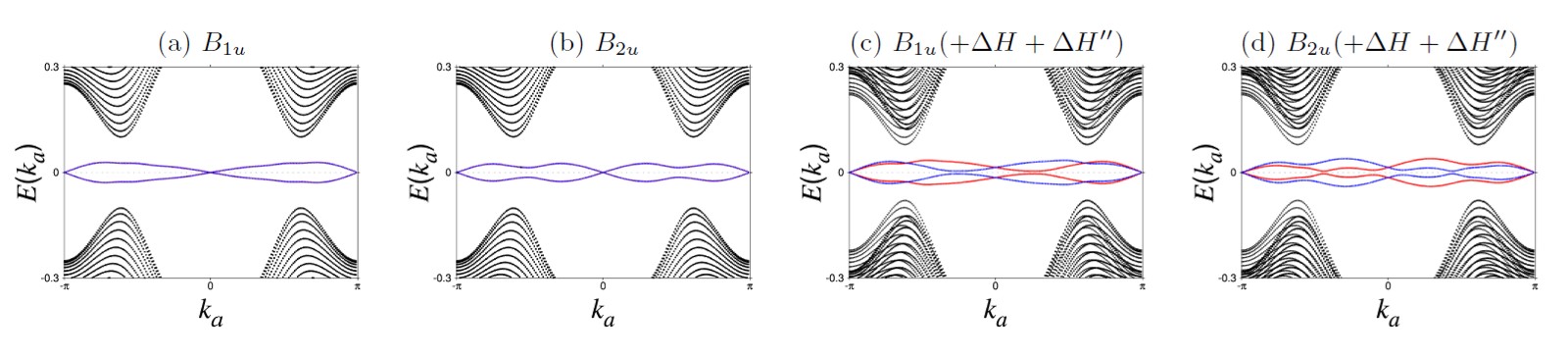}
  \caption{$(0\bar{1}1)$ surface states of $\hat{G}_n$-even superconductivity at $k_c=\pi$.
    We take $L_a=200$ and $L_b=50$.
    Surface states with positive- and negative-glide eigenvalues are highlighted by red and blue, respectively.}
  \label{fig:Gn_even}
\end{figure}
The pinning of gapless surface states at $k_a=0$ is again owing to the accidental symmetry $\sigma_z$.
It can be lifted as in the $\hat{G}_n$-odd superconductivity, by adding \Eq{eq:additional_term2} and
\begin{equation}
  \Delta H''(\bm{k})=\epsilon s_y\sigma_y\eta_z\tau_x,\quad k_c=\pi,\label{eq:add_terms2}
\end{equation}
instead of \Eq{eq:add_terms1}.
Then, zero-energy surface states are lifted from $k_a=0$ as shown in Figs.~\ref{fig:Gn_even}(c) and (d), while the pinning at $k_a=\pi$ remains intact.
The results are characteristic properties of glide-even TNCS with $\nu_\pm^{(n)}(\pi)=1$.

\subsection{Relationship to Shiozaki's toy model}
Here we discuss the connection of our model with Shiozaki's toy model: time-reversal symmetric 2D spin-triplet superconductivity with a single (Kramers-degenerate) Fermi surface, as a minimal model realizing $\theta_4=1$ state~\cite{Shiozaki2016_SM}.
Naively thinking, it seems that the nontrivial topology $\theta_4=2$ of our model at $k_c=\pi$ plane, namely $H_{\mathrm{BdG}}(k_a,k_b,\pi)$, could be directly understood as just two stacked copies of the Shiozaki's toy model, because $H_{\mathrm{BdG}}(k_a,k_b,\pi)$ describes spin-triplet superconductivity with two Fermi surfaces. However, this is not the case, when $H_{\mathrm{BdG}}(k_a,k_b,\pi)$ is regarded as a part of a 3D system, as we discuss below.

We first stress that the 2D toy model proposed by Shiozaki \textit{et al.} is essentially a 1D class D topological superconductivity. 
On the $k_a=0$ line (in our notation), Shiozaki's model consists of a Kitaev chain in each glide eigen-sector. 
Therefore, nontrivial topology is ensured by the presence of the Fermi surface crossing the $k_a=0$ line. On the other hand, the cylinder Fermi surfaces are on the $k_a=\pi$ line, and not on the $k_a=0$ line. It is by no means obvious whether the cylinder Fermi surfaces can be adiabatically deformed into two stacks of the Shiozaki's toy model. Indeed, this is \textit{impossible} as long as the screw symmetry is preserved. This point can be clearly seen in Table I in the main text: $\mathbb{Z}_4$ invariant of the $Pnma$ system at the Brillouin zone face is determined solely by the Fermi surfaces crossing the $k_a=\pi$ line, and presence/absence of the Fermi surfaces at the $k_a=0$ line does not matter. 
Thus, our model offers a new and realistic platform to realize nontrivial $\mathbb{Z}_4$ invariants in three dimensional systems.

\section{Surface directions to observe topological surface states}
\label{sec:options}
{In this section, we elucidate the boundary directions required to observe symmetry-protected surface states associated with glide topological invariants.
It is often stated that we have to take boundary direction preserving $\hat{G}=\Set{M_c|\bm{a}/2+\hat{c}/2}$, that is $\bm{a}$- and $\hat{c}$-normal surfaces.
However, we here show that the topological surface states appear not only on that surface but also on many other surface directions.
This is because there is another glide symmetry $\hat{G}'$ associated with $\hat{G}$, defined by
\begin{equation}
  \hat{G}'\equiv\Set{E|m\bm{a}-n\bm{b}}\hat{G}=\Set{M_c|(2m+1)\bm{a}/2-n\bm{b}+\hat{c}/2},
\end{equation}
with $m$ and $n$ being arbitrary integer.
The new glide operation $\hat{G}'$ is preserved on the surface normal to $(2m+1)\bm{a}-2n\bm{b}$ and $\hat{c}$.
In the following, we assume $2m+1$ and $2n$ are mutually prime without loss of generality, since otherwise we can find another pair of $(m',n')$ to give the same surface direction.
(Actually, we can write $2m+1=(2m'+1)d$ and $2n=2n'd$, with $d\in 2\mathbb{Z}+1$ the greatest common divisor.)
It can be shown that we can retake the set of basic lattice translation vectors so as to include $\bm{a}'=(2m+1)\bm{a}-2n\bm{b}$, without Brillouin zone folding.
Indeed, when we adopt another lattice translation vector $\bm{b}'=\alpha\bm{a}+\beta\bm{b}$, the volume of the unit cell is
\begin{equation}
  |\bm{a}',\bm{b}',\hat{c}|=|\bm{a},\bm{b},\hat{c}|\begin{vmatrix}2m+1&\alpha&0\\-2n&\beta&0\\0&0&1\end{vmatrix}=|\bm{a},\bm{b},\hat{c}|\left\{(2m+1)\beta+2n\alpha\right\}.
\end{equation}
We can find a pair of integer $(\alpha,\beta)$ such that $(2m+1)\beta+2n\alpha=1$, since $(2m+1)$ and $2n$ are mutually prime. Thereby, $|\bm{a},\bm{b},\hat{c}|=|\bm{a}',\bm{b}',\hat{c}|$ holds, and the volume of the unit cell does not change. The first Brillouin zone of the new basis is given by $-\pi<k_a',k_b',k_c\le\pi$.}

{Now, let us apply the formulas for topological invariants to the newly introduced glide symmetry $\hat{G}'=\Set{M_c|\bm{a}'/2+\hat{c}/2}$.
The point is that the two TRIM $\Gamma_i$ $(i=1,2)$, which determine topological invariants, are shared by $\hat{G}$ and $\hat{G}'$. Actually, $\Gamma_i$ $(i=1,2)$ for $\hat{G}'$ are the TRIM on the line
\begin{equation}
  k_c=k_a'=(2m+1)k_a-2nk_b=\pi.
\end{equation}
Clearly, they are $(k_a,\,k_b)=(\pi,0),(\pi,\pi)$, which coincide with $\Gamma_i$ $(i=1,2)$ for $\hat{G}$.
%, since $2\bm{\alpha}'\equiv2\bm{b}'\times\hat{c}/\bm{a}'\cdot\bm{b}'\times\hat{c}$ is a reciprocal lattice vector.
Thus, topological invariants for $\hat{G}$ and $\hat{G}'$ are identical, and bulk-boundary correspondence ensures the topological surface states on the $\hat{G}'$-preserving surface, when topological invariants enriched by $\hat{G}$ is nontrivial.}

For example, let us consider the $n$-glide symmetry of UCoGe.
We can retake the glide operation as
\begin{equation}
  %  \hat{G}_n'\equiv\Set{E|n\hat{y}-(m+1)\hat{z}}\hat{G}_n=\Set{M_x|\hat{x}/2+(2n+1)\hat{y}/2-(2m+1)\hat{z}/2},
    \hat{G}_n'\equiv\Set{E|m(\hat{y}+\hat{z})-n\hat{z}}\hat{G}_n=\Set{M_x|\hat{x}/2+(2m+1)\hat{y}/2+(2m-2n+1)\hat{z}/2},
\end{equation}
with $m$ and $n$ being arbitrary integer.
The glide topological invariants defined by $\hat{G}_n'$ are exactly the same as those by $\hat{G}_n$.
Thus, topologically-protected surface states also appear on the $(0,2m-2n+1,-(2m+1))$ surface. In the same way, it is also shown that the $(2n,-(2m+1),0)$ surface 
%states exist, associated with another choice of the $a$-glide,
hosts topologically-protected surface states associated with TNCS by considering the $a$-glide symmetry
\begin{equation}
  \hat{G}_a'\equiv\Set{E|m\hat{x}-n\hat{y}}\hat{G}_a=\Set{M_z|(2m+1)\hat{x}/2-n\hat{y}+\hat{z}/2}.
\end{equation}
%These facts may reduce experimental difficulty.
Thus, we have many options for the surface direction. Experimental difficulty to observe the surface states may be reduced by this fact.

\section{Strong topological indices of $\mathrm{UCoGe}$}
In this section, we complete the topological classification of UCoGe.
%the glide-preserving surfaces.
In particular, we clarify strong topological indices.
% of $\mathrm{UCoGe}$.
For convenience, we show generators of glide-odd and glide-even topological phases obtained by $K$-theory~\cite{Yanase2017_SM} in \Table{tab:generators} and \Table{tab:generators2}, respectively.
\begin{table}
  \centering
  \caption{Table of the generators of the glide-odd topological phases from \Ref{Yanase2017_SM}.}
  \label{tab:generators}
  \begingroup
  \renewcommand{\arraystretch}{1.5}
  \tabcolsep =2.5mm
  \begin{tabular}{cccc}\\\hline\hline
    &$W$&$\theta_4(0)$&$\theta_4(\pi)$\\\hline
    $\mathcal{H}_1$&$1$&$1$&$0$\\
    $\mathcal{H}_2$&$0$&$0$&$2$\\
    $\mathcal{H}_3$&$0$&$1$&$1$\\\hline\hline
  \end{tabular}
  \endgroup
\end{table}
\begin{table}
  \centering
  \caption{Table of the generators of the glide-even topological phases from Ref.~\cite{Yanase2017_SM}.}
  \label{tab:generators2}
  \begingroup
  \renewcommand{\arraystretch}{1.5}
  \tabcolsep =2.5mm
  \begin{tabular}{ccccc}\\\hline\hline
    &$\nu_+(0)$&$\nu_-(0)$&$\nu_+(\pi)$&$\nu_-(\pi)$\\\hline
    $\mathcal{H}_1$&$1$&$0$&$1$&$0$\\
    $\mathcal{H}_2$&$0$&$0$&$1$&$1$\\
    $\mathcal{H}_3$&$1$&$1$&$1$&$1$\\\hline\hline
  \end{tabular}
  \endgroup
\end{table}
Topological indices $\{\mathbb{Z}^W,\mathbb{Z}_2^{\mathrm{strong}},\mathbb{Z}_4^{\mathrm{weak}}\}$ of glide-odd superconductivity and $[\mathbb{Z}_2^{\mathrm{CS}_T},\mathbb{Z}_2^{\mathrm{strong}},\mathbb{Z}_2^{\mathrm{weak}}]$ of glide-even superconductivity are given by $\{n,m,l\}$ and $[n,m,l]$, respectively, when the BdG Hamiltonian has topological invariants equivalent to those of the Hamiltonian $(\oplus\mathcal{H}_1)^n(\oplus\mathcal{H}_2)^m(\oplus\mathcal{H}_3)^l$.
% in \Table{tab:generators} or \ref{tab:generators2}, respectively.
Here, $\mathbb{Z}^W$ corresponds to the usual three-dimensional winding number, while $\mathbb{Z}_2^{\mathrm{CS}_T}$ represents the usual one-dimensional class DIII $\mathbb{Z}_2$ invariant on the line $C_{\mathrm{AII}}(\Gamma_c)$~\cite{SatoAndo2017_SM,Yanase2017_SM}.
The strong glide topological indices is represented by $\mathbb{Z}_2^{\mathrm{strong}}$, while weak glide indices are represented by $\mathbb{Z}_4^{\mathrm{weak}}$ and $\mathbb{Z}_2^{\mathrm{weak}}$ for glide-odd and -even superconductivity, respectively~\cite{Yanase2017_SM}.

Here, we evaluate topological invariants by
considering only the cylinder FSs, and later we show irrelevant influence of the other FSs.
%ignoring the $\Gamma$-FS, and then discuss its influence. We assume that $X$- and $Y$- FSs can be adiabatically removed, since they are off TRIM, where typical $d$-vectors vanish.
%Thus, only we have to do is to determine the topological invariants of the cylinder FSs.

First, we consider $A_u$ representation. 
We can show that the winding number is an even integer $W\in2\mathbb{Z}$ by using the formula of \Ref{Sato2010_oddparity_SM}.
Thus, we take $W=4n$ or $W=4n+2$, where $n$ is an arbitrary integer.
Since there is no FS at $k_x=0$, we {naturally obtain}
\begin{equation}
\theta_4^{(n)}(0)=0,\label{eq:theta4n}
\end{equation}
%{\sout{is physically expected for $n$-glide.}}
for the $\mathbb{Z}_4$ invariant by the $n$-glide symmetry. %{@@@@Fact?}
On the other hand, it is natural to assume
\begin{equation}
  \theta_4^{(a)}(0)=2,\label{eq:theta4a}
\end{equation}
for the $a$-glide symmetry, considering the weak $k_z$ dependence of the cylinder FSs.
A numerical analysis of the model studied in previous sections consistently gives \Eqs{eq:theta4n}{eq:theta4a}\cite{Yoshida2017_SM}. %@@@@\textcolor{red}{LATER, calculate edge states.}
Thus, topological indices of the $A_u$ phase are given by
\begin{equation}
  (W;\,\theta_4^{(a)}(0),\theta_4^{(a)}(\pi);\,\theta_4^{(n)}(0),\theta_4^{(n)}(\pi))=(4n;2,2;0,2)\quad \text{or}\quad (4n+2;2,2;0,2),
\end{equation}
which are decomposed into the generators as in \Table{tab:TopoNums}.
Thus, the $A_u$ representation has nontrivial strong glide $\mathbb{Z}_2$ index;
%for $\hat{G}_a$ or $\hat{G}_n$, corresponding to $W=4n+2$ or $W=4n$, respectively.
The strong glide $\mathbb{Z}_2$ index for $\hat{G}_n$ ($\hat{G}_a$) is nontrivial when $W=4n$ ($W=4n+2$).%@@@\textcolor{red}{Grammer?}

\begin{table}
  \centering
  \caption{Topological indices of UCoGe. The first and the second row correspond to the two possibilities of the winding number, $W=4n$ and $W=4n+2$, respectively. Curly braces represent glide-odd indices $\{\mathbb{Z}^W,\mathbb{Z}_2^{\mathrm{strong}},\mathbb{Z}_4^{\mathrm{weak}}\}$, while square braces represent glide-even indices $[\mathbb{Z}_2^{\mathrm{CS}_T},\mathbb{Z}_2^{\mathrm{strong}},\mathbb{Z}_2^{\mathrm{weak}}]$.}
  \label{tab:TopoNums}
  \begingroup
  \renewcommand{\arraystretch}{1.5}
  \tabcolsep =2.5mm
  \begin{tabular}{ccc}\\\hline\hline
    &$\hat{G}_a$&$\hat{G}_n$\\\hline
    \multirow{2}{*}{$A_u$}&$\{4n,0,2\}$&$\{4n,1,0\}$\\
    &$\{4n+2,1,0\}$&$\{4n+2,0,2\}$\\
    $B_{1u}$&$\{0,0,2\}$&$[0,1,0]$\\
    $B_{2u}$&$[0,0,1]$&$[0,1,0]$\\
    $B_{3u}$&$[0,0,1]$&$\{0,1,0\}$\\\hline\hline
  \end{tabular}
  \endgroup
\end{table}

The glide-odd $\mathbb{Z}_4$ indices %$\theta_4^(a,n)(0)$ and $\theta_4^(a,n)(\pi)$
of $B_{1u}$ and $B_{3u}$ representations are the same as those of $A_u$ representation. On the other hand, the winding number $W=0$ follows from $\hat{M}_y$-even order parameters. Thus, the glide-odd indices are given by $(W,\theta_4^{(a)}(0),\theta_4^{(a)}(\pi))=(0,2,2)$ for $B_{1u}$ and $(W,\theta_4^{(n)}(0),\theta_4^{(n)}(\pi))=(0,0,2)$ for $B_{3u}$.
As for the glide-even indices, we expect
\begin{equation}
  \nu_\pm^{(n)}(0)=0,\quad\text{and}\quad\nu_\pm^{(a)}(0)=1,
\end{equation}
%{\sout{for}}@@
by the same reasons as those for glide-odd indices~\cite{Yoshida2017_SM}. %@@@
%The same glide-even indices are expected for the $B_{2u}$ representation as well.
The glide-even indices for the $B_{2u}$ representation are the same as those for $B_{1u}$ and $B_{3u}$ representations.
The glide topological invariants at ZF are
\begin{equation}
  (\nu_\pm^{(n)}(\pi),\,\nu_\pm^{(a)}(\pi))=(1,1),
\end{equation}
as given in the main text.
Thus, the glide-even indices of $\hat{G}_a$- and $\hat{G}_n$-even superconductivity are given by
\begin{equation}
  (\nu_\pm^{(a)}(0),\,\nu_\pm^{(a)}(\pi))=(1,1),\quad\text{and}\quad(\nu_\pm^{(n)}(0),\,\nu_\pm^{(n)}(\pi))=(0,1).
\end{equation}
The decomposition into generators leads to \Table{tab:TopoNums}. It should be noticed that $B_{1u}$, $B_{2u}$, and $B_{3u}$ representations are strong glide TNCS for $\hat{G}_n$.

\subsection{Influence of the tiny $\Gamma$-FS}
We here discuss the influence of the tiny FS around $\Gamma$ ($\Gamma$-FS).
Note that topological invariants of glide-even TNCS are not affected by the $\Gamma$-FS, since they are defined at ZF. Therefore, we discuss only glide-odd topological invariants.

First, we consider the $A_u$ representation.
The $\Gamma$-FS likely carries 3D winding number $W=\pm1$, as in the $B$-phase of ${}^3\mathrm{He}$.
%~\cite{Volovik2003_SM}.
The FS is also expected to carry $\theta_4(0)=\pm1$ or $\mp1$, since a single FS is placed on the line where $\mathbb{Z}_2$ part of $\theta_4(0)$ is defined~\cite{Shiozaki2016_SM}.
When $(W,\theta_4(0))=(\pm1,\pm1)$ is added, $\mathbb{Z}^W$ changes by $\pm1$ and other indices remain unchanged. On the other hand, addition of $(W,\theta_4(0))=(\pm1,\mp1)$ changes $\mathbb{Z}_2^{\mathrm{strong}}$ and $\mathbb{Z}_2^{\mathrm{weak}}$ as $\{\mathbb{Z}_2^{\mathrm{strong}},\mathbb{Z}_2^{\mathrm{weak}}\}=\{1,0\}\leftrightarrow\{0,2\}$. In any case, either one of strong indices for $\hat{G}_a$ and $\hat{G}_n$ are nontrivial, and $A_u$ representation is classified into strong glide TNCS, while it is also a strong TSC specified by the nontrivial winding number.

Next, we consider $\hat{G}_a$-odd $B_{1u}$ superconductivity and $\hat{G}_n$-odd $B_{3u}$ superconductivity. When we assume gapful excitation, the winding number should be trivial in this case, since they are $\hat{M}_y$-even superconductivity. On the other hand, $\theta_4(0)$ must be either $\pm1$, as in the case of $A_u$ representation.
However, the configuration $(W,\theta_4(0),\theta_4(\pi))=(0,\pm1,2)$ breaks the constraint $\theta_4(0)+\theta_4(\pi)\equiv W\ (\mathrm{mod}\,2)$ for gapful TNCS~\cite{Yanase2017_SM}. This indicates that $B_{1u}$ and $B_{3u}$ states must be gapless~\cite{Yanase2017_SM}. Indeed, point nodes are expected to appear on the $\Gamma$-FS, as naively expected, for example, from the $d$-vector of $B_{1u}$ representation $\bm{d}(\bm{k})\sim(k_y,k_x,0)$.
Similarly, $B_{2u}$ state is expected to host point nodes on the tiny $\Gamma$-FS.
The nodal excitations are not harmful to the topological surface states protected by the glide $\mathbb{Z}_4$ or $\mathbb{Z}_2$ invariants on the ZF, which are discussed in the main text. This is because gapless bulk states around $\Gamma$ do not hybridize with the surface states with $k_z =\pi$, at least in the clean limit. The signature of the surface states protected by nontrivial weak indices $\theta_4(\pi)$ may be robust in this situation.

%% @@@@@@@@
%% , where typical spin-triplet order parameters vanish. Thus, it is naively expected that all the FSs may be adiabatically removed from $k_x=0$ and $k_z=0$, giving trivial topological invariants (We discuss influence of tiny FS around $\Gamma$ in Supplemental Materials\cite{}). In addition, symmetry requires $W=0$ for $B_{1u}$, $B_{2u}$, and $B_{3u}$ superconductivity. Thus, $B_{1u}$, $B_{2u}$, and $B_{3u}$ superconductivity are expeccted to be TNCS with strong glide $\mathbb{Z}_2$ index\cite{}, while strong glide index of $A_u$ superconductivity depends on the value of $W$.
%% @@@@@@@@@@@@@@

\subsection{Influence of the $X$- and $Y$-FS}
The FSs near $X$ point ($X$-FS) and $Y$ point ($Y$-FS) do not affect topological properties when the excitation is gapful. This is because these FSs do not enclose the TRIM and is {naturally} removed without closing the gap. %{@@@Fact?}
Thus, we have only to consider the influence of possible gapless excitation on the FSs.
The $Y$-FS does not affect the {M\"obius topological surface states} on the glide invariant planes $k_x=\pi$ and $k_z=\pi$ because of its position.
On the other hand, the $X$-FS crosses the $k_x=\pi$ plane. {Even in the presence of gapless excitation on the $k_x=\pi$ plane, {\textit{zero-energy}} topological surface states avoid hybridization to the gapless bulk state when we choose an appropriate surface direction by using the option discussed in \Sec{sec:options}}.
%@@@case by case? Ga-type: OK Gn-type: Impossible.@@@@@
\section{Possibility and influence of excitation nodes}
%% Here we briefly discuss the possibility of excitation nodes, and their influence to the topological properties.
%% As we discussed in the previous section, presence of the tiny FS leads to point nodes on it for $B_{1u}$, $B_{2u}$ and $B_{3u}$ representations.
%% However, it is quite unclear whether this tiny FS really exists in the high-pressure regime.
%% We note that even in the presence of the FS, gapless bulk states around $\Gamma$ do not hybridize with the edge states with $k_z=\pi$, at least in the clean limit. We expect the signature of the surface states protected by nontrivial weak indices $\theta_4(\pi)$ may be still observed even in this situation.
In the previous section, we discussed the influence of nodal excitations on the $\Gamma$-, $X$-, and $Y$-FSs.
More significant possibility of excitation nodes is the line nodes at ZF predicted by Norman's theorem~\cite{Norman1995_SM,Kobayashi2016_SM}.
The Norman's theorem states that line nodes exist at the ZF of mirror- or glide-odd SCs with screw symmetry~\cite{Norman1995_SM}. Accordingly, line nodes are predicted as illustrated in Table~V.
However, it is also known that Norman's theorem does not hold when the gap function and the spin-orbit splitting of the Fermi surfaces are of the same order in magnitude~\cite{Kobayashi2016_SM}.
Therefore, we may obtain gapful excitation if FSs under high pressure have small splitting.
For instance, in the cylinder FSs of Fig.~1, the splitting at $S-X-S$ line and at $k_y=\pi$ plane may be sufficiently small to achieve gapful excitation. Actually, the splitting is estimated to be $\sim\SI{0.01}{\electronvolt}$~\cite{Fujimori2015_SM,Fujimori2016_SM}. Assuming the mass renormalization factor $1/z\sim100$ {as a typical value}, renormalized splitting is about $\SI{0.1}{\milli\electronvolt}$, which {is smaller than} the magnitude of gap function expected from the transition temperature $\sim\SI{0.5}{\kelvin}$~\cite{Slooten2009_SM}.
{We leave quantitative estimation of the splitting as a future work.}

%Furthermore, we expect signiture of $\mathbb{Z}_4$ topological surface states may be observed even in the presence of excitation nodes.
Furthermore, the $Z_4$ invariant is well-defined and corresponding topological surface states may be robust even in the presence of excitation nodes.
This is because Norman's line nodes \textit{preserve band gap}, although they make excitation gapless.
We show in \Fig{fig:Norman_gapless} an example of the surface spectrum in the presence of the Norman's line nodes.
Figures~\ref{fig:Norman_gapless}~(a) and \ref{fig:Norman_gapless}~(b) show the $(0\bar{1}1)$ surface states of $A_u$ superconductivity.
Parameters are the same as those of \Fig{fig:Gn_odd}~(c), with $\Delta H(\bm{k})$ replaced by $5\Delta H(\bm{k})$.
Figure~\ref{fig:Norman_gapless}~(a) shows the whole surface spectrum, while \Fig{fig:Norman_gapless}~(b) shows surface states with positive glide eigenvalue.
Clearly, the band gap is preserved in each glide eigen-sector, and therefore, glide topological invariants are still well-defined by assuming the ``curved chemical potential''~\cite{Wang2015_FeSeTe_SM}. Bulk-boundary correspondence leads to surface states, which do not hybridize with bulk states with opposite glide eigenvalue.
\begin{figure}[htbp]
  \centering
  \begin{tabular}{cc}
    (a) $A_u+5\Delta H+\Delta H'$ &(b) $A_u+5\Delta H+\Delta H'$ \\
    \includegraphics[width=45mm]{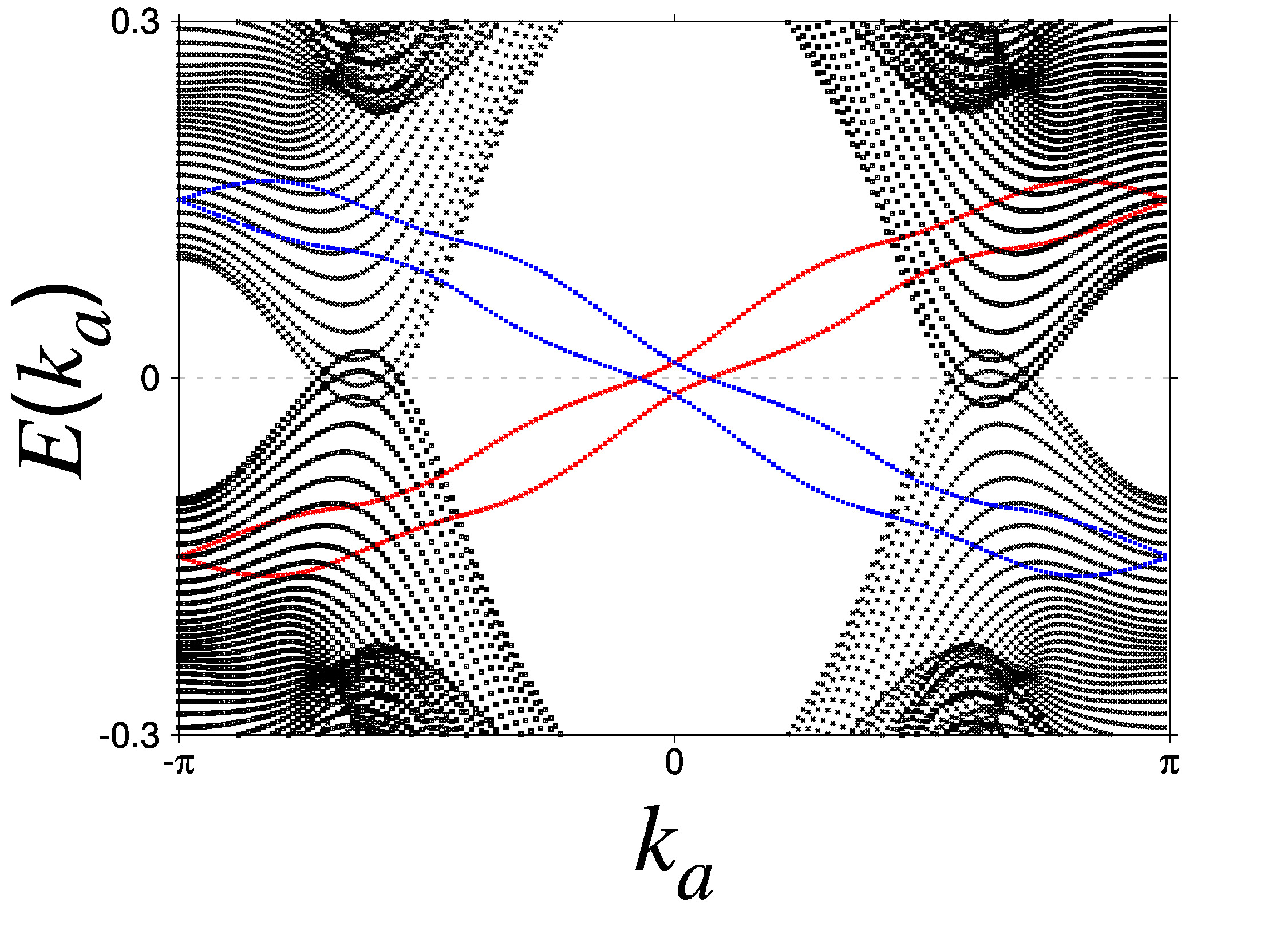}&  \includegraphics[width=45mm]{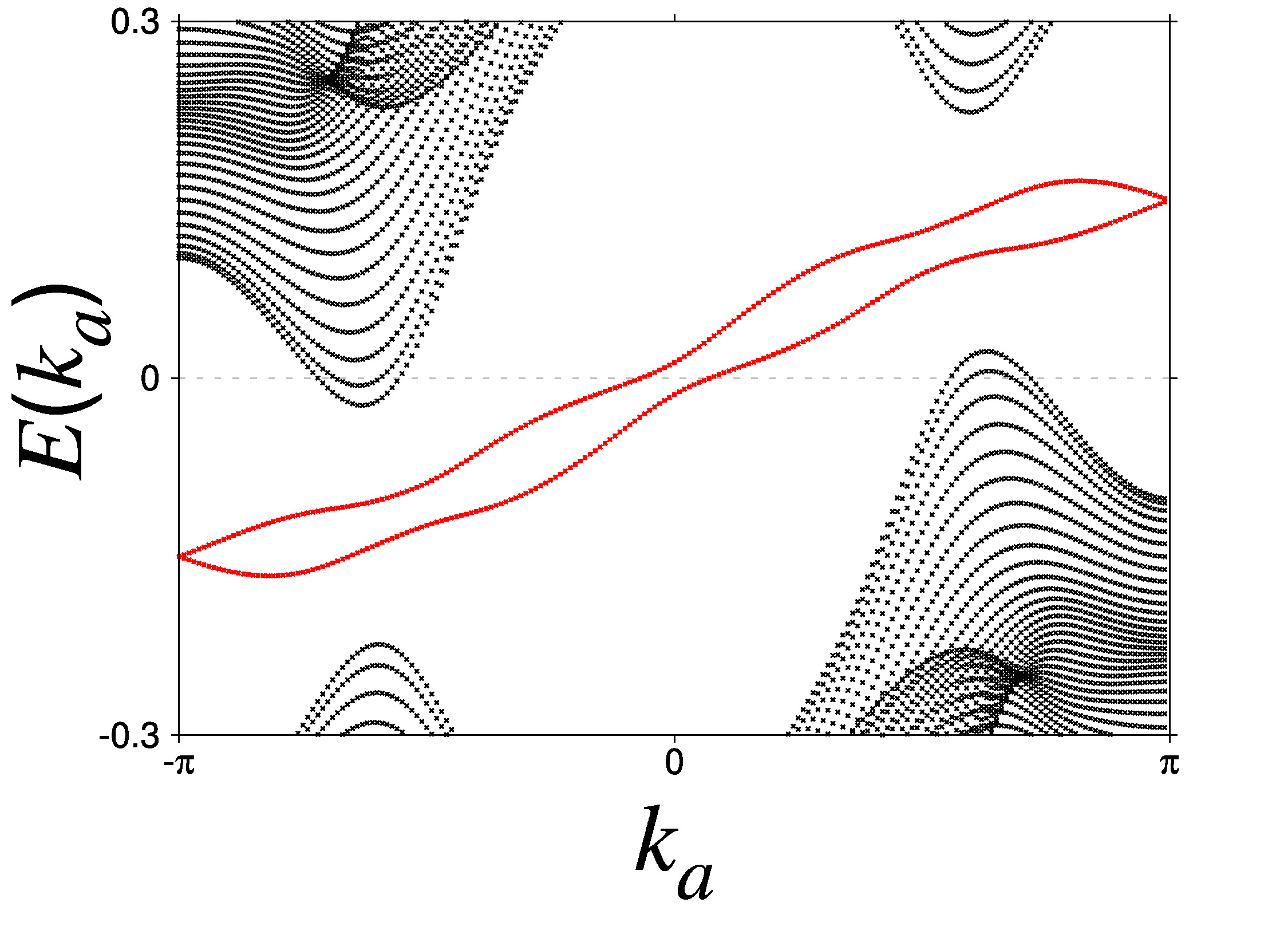}\\
  \end{tabular}
  \caption{Surface spectrum in the presence of Norman's line node. The left panel (a) shows the whole surface spectrum on the $(0\bar{1}1)$ surface at $k_x=\pi$, while the right panel (b) shows surface states with positive glide eigenvalue.}
  \label{fig:Norman_gapless}
\end{figure}

\end{document}